# An Ontology-Based Approach to the Optimization of

# Non-Binary (2, v)-Regular LDPC Codes

Chao Chen

chenchaoxidian@gmail.com

A non-binary (2,v)-regular LDPC code is defined by a parity-check matrix with column weight 2 and row weight v. In this report, we give an ontology-based approach to the optimization for this class of codes. All possible inter-connected cycle patterns that lead to low symbol-weight codewords are identified to put together the ontology. The optimization goal is to improve the distance property of equivalent binary images. Using the proposed method, the estimation and optimization of bit-distance spectrum becomes easily handleable. Three codes in the CCSDS recommendation are analyzed and several codes with good minimum bit-distance are designed.

Low bit-weight codewords in binary image code are most likely induced by low symbol-weight codewords in the original code. Therefore, to improve the bit-distance property, we first need to eliminate low symbol-weight codewords.

In [1], it was shown that for non-binary (2, v)-regular LDPC codes, low symbol-weight codewords are caused by short cycles and inter-connected short cycles in the Tanner graph. These subgraphs of the Tanner graph correspond to different submatrices of the parity-check matrix. Since the submatrix corresponding to a cycle is a square matrix, we can make it to be of full rank by properly choosing non-zero field elements. In this way, low symbol-weight codewords associated with short cycles can be avoided. If the corresponding submatrix is made to be of full rank, the cycle is said to be "cancelled" in [1]. In contrast, an inter-connected cycle corresponds to a rectangular submatrix, where the number of rows is strictly less than that of columns. Therefore, for inter-connected short cycles, low symbol-weight codewords are always caused, regardless of the selection of non-zero field elements.

With cycle cancellation, the symbol-distance property is improved, which can potentially improve the bit-distance property. But the codes with the same minimum symbol-distance can have different minimum bit-distances. To obtain codes with better bit-distance property, we adopt the following method. The cycle cancellation is performed in a random manner, so a large number of codes with the same minimum symbol-distance can be obtained. From these codes, we select one or more codes by estimating and comparing their bit-distance spectrum. In general, distance estimation is not an easy task and is usually time-consuming. Below, we develop an ontology-based approach for non-binary (2, v)-regular LDPC codes, and consequently we can very quickly estimate the bit-distance spectrum.

As mentioned before, short cycles and inter-connected short cycles correspond to submatrices of the parity-check matrix. If the non-zero entries have been assigned, low symbol-weight codewords can be obtained by solving the system of homogeneous linear equations defined by these submatrices. This can be efficiently performed because there are a small number of unknowns. Now the key is how to identify these submatrices, or equivalently the corresponding subgraphs.

For a given Tanner graph, short cycles can be easily identified. However, it is more difficult to identify inter-connected cycles. To address this problem, we following the methodology in [2] and develop an ontology, which consists of all possible patterns of inter-connected cycles that lead to low symbol-weight codewords. The ontology is illustrated in Fig. 1, where each pattern is denoted by an uppercase letter followed a tuple, e.g., $A-(n_1, n_2, n_3)$. The $i$-th element $n_i$ of the tuple represents the number of variable nodes in the edge $i$. For these inter-connected cycles, the number of variable nodes is given by $w = \sum_i n_i$, strictly larger than that of check nodes. For some codes, the Tanner graphs have only cycles of lengths divisible by 4, e.g., the codes defined on even-girth cages. Therefore, we categorize all inter-connected cycles into two types. An inter-connected cycle is called Type-I if all cycles involved have lengths divisible by 4; otherwise, it is called Type-II.

Based on the ontology, we create a database, as shown in Table I. The database enumerates all possible inter-connected cycles that lead to codewords of symbol weight up to $\left\lceil \dfrac{3}{4} g_T \right\rceil + 3$, where $g_T = 8, 10, \cdots, 16$ is the girth of the Tanner graph. For Tanner graphs with $g_T > 16$, codewords of symbol weight up to $\left\lceil \dfrac{3}{4} g_T \right\rceil + 3$ are only caused by the $A-(n_1, n_2, n_3)$ pattern. Since the reader can easily write out the configurations, we do not include them in Table I.

The value of the database lies in that with it, we know what kinds of inter-connected cycles we should search for. Once all short cycles have been obtained, the intended inter-connected short cycles can be easily identified.

For cycle cancellation, we follow the method of [1]. Conditioned that more short cycles can be cancelled, the non-zero field elements are replaced randomly in a row manner. Some good candidate rows have been tabulated in [1].

| Configurations | Examples |
|:---:|:---:|
| 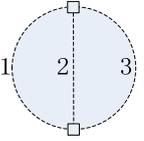 A $-(n_1, n_2, n_3)$ | 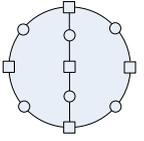 A $-(2,2,2)$ |
| 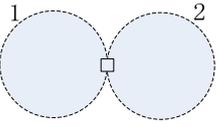 B $-(n_1, n_2)$ | 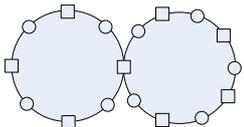 B $-(4,5)$ |
| 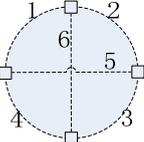 C $-(n_1, n_2, n_3, n_4, n_5, n_6)$ | 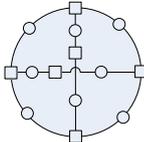 C $-(1,1,1,1,2,2)$ |
| 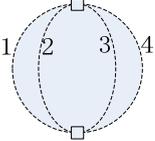 D $-(n_1, n_2, n_3, n_4)$ | 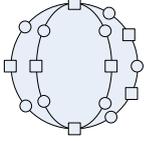 D $-(2,2,2,3)$ |
| 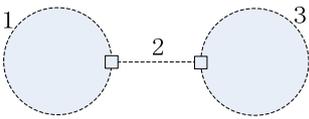 E $-(n_1, n_2, n_3)$ | 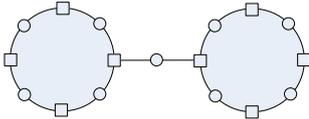 E $-(4,1,4)$ |
| 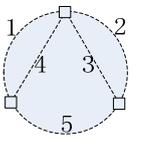 F $-(n_1, n_2, n_3, n_4, n_5)$ | 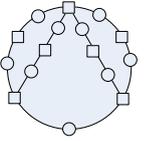 F $-(2,2,2,2,1)$ |
| 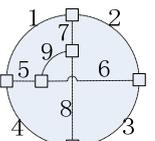 G $-(n_1, n_2, n_3, n_4, n_5, n_6, n_7, n_8, n_9)$ | 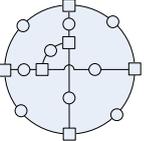 G $-(1,1,1,1,1,1,1,1,1)$ |

Fig. 1. The ontology of inter-connected cycles

TABLE I
A DATA BASE OF INTER-CONNECTED SHORT CYCLES

| $g_T$ | $w$ | Type-I | Type-II |
|---|---|---|---|
| 8 | 6 | A-(2,2,2) | |
| | 7 | A-(1,3,3) | A-(2,2,3) |
| | 8 | A-(2,2,4), B-(4,4), C-(1,1,1,1,2,2), D-(2,2,2,2) | A-(1,3,4), A-(2,3,3) |
| | 9 | A-(1,3,5), A-(3,3,3), C-(1,1,2,2,2,1), E-(4,1,4), F-(2,3,1,2,1), G-(1,1,1,1,1,1,1,1,1,1) | A-(1,4,4), A-(2,2,5), A-(2,3,4), B-(4,5), C-(1,1,1,1,2,3), C-(1,2,1,2,1,2), D-(2,2,2,3), F-(2,2,2,2,1) |
| 10 | 8 | | A-(2,3,3) |
| | 9 | A-(3,3,3) | A-(1,4,4), A-(2,3,4) |
| | 10 | A-(2,4,4) | A-(1,4,5), A-(2,3,5), A-(3,3,4), B-(5,5), C-(2,2,2,2,1,1) |
| | 11 | A-(1,5,5), A-(3,3,5) | A-(1,4,6), A-(2,3,6), A-(2,4,5), A-(3,4,4), B-(5,6), C-(2,2,2,2,1,2), C-(3,2,2,2,1,1), C-(1,3,2,2,1,2), D-(2,3,3,3), E-(5,1,5), F-(2,3,2,3,1) |
| 12 | 9 | A-(3,3,3) | |
| | 10 | A-(2,4,4) | A-(3,3,4) |
| | 11 | A-(1,5,5), A-(3,3,5) | A-(2,4,5), A-(3,4,4) |
| | 12 | A-(2,4,6), A-(4,4,4), B-(6,6), C-(2,2,2,2,2,2), C-(1,2,1,2,3,3), D-(3,3,3,3) | A-(1,5,6), A-(2,5,5), A-(3,3,6), A-(3,4,5) |
| 14 | 11 | | A-(3,4,4) |
| | 12 | A-(4,4,4) | A-(2,5,5), A-(3,4,5) |
| | 13 | A-(3,5,5) | A-(1,6,6), A-(2,5,6), A-(3,4,6), A-(4,4,5) |
| | 14 | A-(2,6,6), A-(4,4,6) | A-(2,5,7), A-(3,4,7), A-(3,5,6), A-(1,6,7), A-(4,5,5), B-(7,7), C-(2,2,2,2,3,3), C-(3,3,3,3,1,1) |
| 16 | 12 | A-(4,4,4) | |
| | 13 | A-(3,5,5) | A-(4,4,5) |
| | 14 | A-(2,6,6), A-(4,4,6) | A-(3,5,6), A-(4,5,5) |
| | 15 | A-(1,7,7), A-(3,5,7), A-(3,6,6), A-(5,5,5) | A-(2,6,7), A-(4,4,7), A-(4,5,6) |

Below we give three code design examples.

**Example 1**. We analyze three (16,8) codes in the CCSDS recommendation [3] and design a new (16,8) code. All these codes are (2,4)-regular LDPC codes over $\mathrm{GF}(2^8)$. The first code is designed by DLR, while the second and third codes are designed by JPL.

Up on some column permutations of the parity-check matrices of the first three codes, the obtained codes can be seen as defined on the same Tanner graph. We denote the three codes (after column permutation) by $C_1, C_2$ and $C_3$, respectively. Our designed code is also defined on this Tanner graph and denoted by $C_4$. The incidence matrix is described in Fig. 2, where the black squares correspond to non-zero entries of the parity-check matrix and white squares correspond to zero entries.

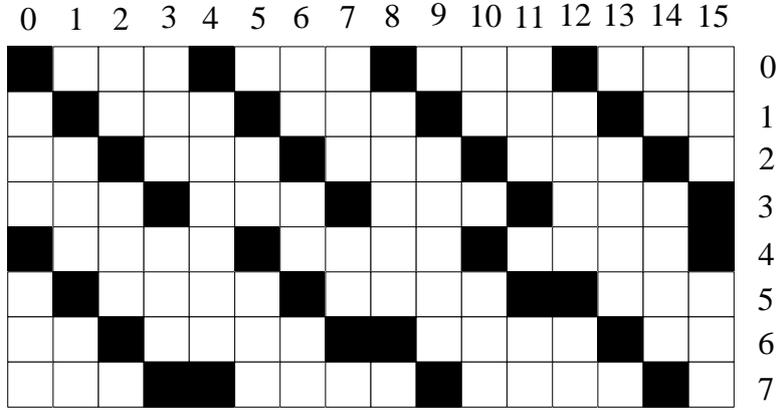

Fig. 2. The incidence matrix of the Tanner graph.

In a compact form, we record the positions of non-zero entries row by row as follows, where the numbers denote the column coordinates of non-zero entries of each row.

$$
\begin{array}{cccc}
0 & 4 & 8 & 12 \\
1 & 5 & 9 & 13 \\
2 & 6 & 10 & 14 \\
3 & 7 & 11 & 15 \\
0 & 5 & 10 & 15 \\
1 & 6 & 11 & 12 \\
2 & 7 & 8 & 13 \\
3 & 4 & 9 & 14
\end{array}
$$

The cycle distribution of the Tanner graph is given by $36x^8 + 96x^{12} + 72x^{16}$. So the girth is 8. We list these short cycles in the appendix.

Using these cycles, we identify inter-connected short cycles. The number of the corresponding patterns is given in Table II.

| Inter-connected short cycles | Number |
|---|---|
| A-(2,2,2) | 48 |
| A-(1,3,3) | 288 |
| A-(2,2,4) | 288 |
| B-(4,4) | 144 |
| C-(1,1,1,1,2,2) | 144 |
| D-(2,2,2,2) | 12 |
| A-(1,3,5) | 432 |
| A-(3,3,3) | 96 |
| C-(1,1,2,2,2,1) | 192 |
| E-(4,1,4) | 144 |
| F-(2,3,1,2,1) | 576 |
| G-(1,1,1,1,1,1,1,1,1) | 16 |

In the appendix, we also give the column indices (variable nodes) contained in each of these inter-connected short cycles. If we chose the columns of the parity-check matrix according to these indices and remove all-zero rows, we obtain the submatrix corresponding to the inter-connected short cycle. The submatrix is a rectangular matrix, for which the row number is strictly less than the column number.

The non-zero entries of the parity-check matrices of $C_1$, $C_2$, $C_3$, and $C_4$ are given as follows.

Each entry is represented by the power of a primitive element of $\mathrm{GF}(2^8)$, which is generated using the primitive polynomial $f(x) = 1 + x^2 + x^3 + x^4 + x^8$.

The non-zero entries for $C_1$ is as follows.

$$
\begin{array}{rrrr}
182 & 8 & 173 & 0 \\
89 & 0 & 9 & 81 \\
0 & 173 & 182 & 8 \\
8 & 182 & 173 & 0 \\
88 & 80 & 0 & 8 \\
169 & 0 & 127 & 40 \\
169 & 128 & 40 & 0 \\
8 & 80 & 88 & 0
\end{array}
$$

The non-zero entries for $C_2$ is as follows.

$$
\begin{array}{rrrr}
183 & 173 & 0 & 8 \\
0 & 88 & 8 & 80 \\
0 & 167 & 40 & 127
\end{array}
$$

$$\begin{matrix} 0 & 182 & 8 & 173 \\ 0 & 89 & 9 & 81 \\ 8 & 0 & 173 & 182 \\ 173 & 8 & 183 & 0 \\ 8 & 0 & 80 & 88 \end{matrix}$$

The non-zero entries for $C_3$ is as follows.

$$\begin{matrix} 89 & 9 & 81 & 0 \\ 89 & 9 & 81 & 0 \\ 89 & 9 & 81 & 0 \\ 89 & 9 & 81 & 0 \\ 0 & 80 & 8 & 88 \\ 0 & 80 & 8 & 88 \\ 0 & 80 & 8 & 88 \\ 0 & 80 & 8 & 88 \end{matrix}$$

The non-zero entries for $C_4$ is as follows.

$$\begin{matrix} 183 & 173 & 0 & 8 \\ 183 & 173 & 0 & 8 \\ 183 & 173 & 0 & 8 \\ 183 & 173 & 0 & 8 \\ 0 & 8 & 183 & 173 \\ 0 & 8 & 183 & 173 \\ 0 & 8 & 183 & 173 \\ 0 & 8 & 183 & 173 \end{matrix}$$

Once the non-zero entries have been given, using the submatrices corresponding to the identified short cycles and inter-connected short cycles, we can obtain the corresponding codewords by solving the system of homogeneous linear equations. Then we obtain the binary image codewords and count the low bit-weight codewords.

Based on the above method, the truncated bit-weight distribution is estimated as follows.

| bit-weight | $C_1$ | $C_2$ | $C_3$ | $C_4$ |
|---|---|---|---|---|
| 13 | 1 | 0 | 0 | 0 |
| 14 | 15 | 17 | 0 | 0 |
| 15 | 36 | 53 | 60 | 8 |
| 16 | 185 | 177 | 200 | 172 |
| 17 | 658 | 633 | 644 | 664 |
| 18 | 2057 | 2085 | 2114 | 2124 |
| 19 | 6192 | 6455 | 6376 | 6016 |
| 20 | 17727 | 17431 | 17910 | 17220 |
| 21 | 45506 | 45502 | 45600 | 46448 |
| 22 | 108955 | 109184 | 109000 | 107625 |
| 23 | 243146 | 243414 | 241721 | 241509 |
| 24 | 506546 | 506392 | 504134 | 501404 |
| 25 | 986635 | 988874 | 984918 | 980077 |
| 26 | 1808367 | 1807894 | 1799083 | 1800023 |
| 27 | 3109255 | 3109827 | 3108288 | 3093225 |
| 28 | 5030103 | 5035670 | 5018033 | 5025987 |
| 29 | 7680254 | 7680940 | 7673802 | 7676147 |
| 30 | 11052341 | 11056947 | 11042705 | 11053100 |
| 31 | 15016653 | 15017324 | 15003560 | 15021757 |
| 32 | 19270316 | 19280797 | 19288889 | 19284488 |
| 33 | 23393313 | 23398266 | 23390688 | 23405712 |
| 34 | 26851274 | 26845631 | 26856922 | 26842327 |
| 35 | 29146246 | 29147805 | 29173765 | 29171187 |
| 36 | 29943682 | 29939207 | 29950522 | 29998260 |
| 37 | 29114396 | 29108287 | 29144685 | 29138822 |
| 38 | 26804600 | 26788157 | 26822086 | 26795809 |
| 39 | 23355340 | 23340757 | 23371298 | 23341200 |
| 40 | 19253919 | 19242460 | 19262314 | 19259378 |
| 41 | 15019536 | 15013874 | 15017672 | 15011417 |
| 42 | 11084458 | 11088072 | 11089034 | 11065626 |
| 43 | 7732975 | 7736664 | 7723527 | 7728712 |
| 44 | 5100432 | 5107100 | 5092961 | 5100509 |
| 45 | 3178501 | 3182982 | 3171617 | 3174307 |
| 46 | 1867550 | 1871528 | 1861817 | 1865832 |
| 47 | 1034100 | 1038460 | 1029251 | 1035178 |
| 48 | 539895 | 542907 | 538069 | 540190 |
| 49 | 266060 | 267216 | 263542 | 267066 |
| 50 | 122788 | 123773 | 121287 | 121034 |
| 51 | 53209 | 53839 | 53102 | 53088 |
| 52 | 21807 | 22056 | 21180 | 22504 |
| 53 | 8308 | 8421 | 8424 | 8492 |
| 54 | 2909 | 3055 | 2952 | 3140 |
| 55 | 945 | 1022 | 1108 | 904 |
| 56 | 311 | 327 | 296 | 400 |
| 57 | 88 | 93 | 104 | 48 |
| 58 | 22 | 33 | 8 | 8 |
| 59 | 7 | 12 | 8 | 0 |
| 60 | 1 | 0 | 4 | 16 |

We now list the obtained minimum bit-weight codewords for each code.

For $C_1$, the minimum bit-weight is 13. There is one codeword with such bit-weight.

The coordinates of non-zero elements of the codeword are given by

$$0 \quad 3 \quad 4 \quad 7 \quad 8 \quad 15$$

The non-zero elements of the codeword are given by

|   |   |   |   |   |   |
|---|---|---|---|---|---|
| 6 | 1 | 89 | 138 | 224 | 181 |

The corresponding binary image codeword is given by the coordinates of each '1':

| 6 | 25 | 32 | 37 | 38 | 39 | 56 | 61 | 65 | 68 | 120 | 124 | 125 |
|---|----|----|----|----|----|----|----|----|----|-----|-----|-----|

For code $C_2$, the minimum bit-weight is 14, there are 17 codewords with such bit-weight.

The coordinates of the non-zero elements of these codewords are given by

| | | | | | | | | |
|---|---|---|---|---|---|---|---|---|
| | | 0 | 4 | 5 | 8 | 9 | 13 | |
| | | 0 | 4 | 5 | 8 | 9 | 13 | |
| | | 0 | 2 | 4 | 8 | 10 | 14 | |
| | | 0 | 2 | 6 | 8 | 10 | 12 | |
| | | 0 | 5 | 7 | 8 | 13 | 15 | |
| | | 1 | 5 | 7 | 11 | 13 | 15 | |
| | | 2 | 6 | 7 | 8 | 11 | 12 | |
| | | 3 | 4 | 7 | 8 | 9 | 13 | |
| | | 3 | 4 | 7 | 8 | 9 | 13 | |
| | | 3 | 4 | 7 | 8 | 9 | 13 | |
| | | 3 | 4 | 7 | 8 | 9 | 13 | |
| | | 3 | 4 | 7 | 8 | 11 | 12 | |
| | 1 | 4 | 5 | 6 | 9 | 10 | 12 | 14 |
| | 1 | 4 | 5 | 6 | 9 | 10 | 12 | 14 |
| 0 | 1 | 2 | 5 | 6 | 8 | 10 | 12 | 13 |
| 0 | 1 | 2 | 5 | 6 | 8 | 10 | 12 | 13 |
| 1 | 3 | 5 | 6 | 9 | 10 | 11 | 14 | 15 |

The non-zero elements of these codewords are given by

| | | | | | | | | |
|---|---|---|---|---|---|---|---|---|
| | | 36 | 51 | 202 | 102 | 226 | 30 | |
| | | 37 | 52 | 203 | 103 | 227 | 31 | |
| | | 14 | 39 | 191 | 29 | 5 | 103 | |
| | | 36 | 5 | 183 | 250 | 27 | 1 | |
| | | 225 | 139 | 103 | 153 | 147 | 112 | |
| | | 4 | 247 | 29 | 94 | 37 | 0 | |
| | | 103 | 191 | 147 | 15 | 66 | 7 | |
| | | 29 | 25 | 102 | 198 | 72 | 0 | |
| | | 30 | 26 | 103 | 199 | 73 | 1 | |
| | | 31 | 27 | 104 | 200 | 74 | 2 | |
| | | 55 | 51 | 128 | 224 | 98 | 26 | |
| | | 7 | 15 | 112 | 192 | 34 | 25 | |
| | 3 | 30 | 191 | 2 | 5 | 16 | 195 | 5 |
| | 4 | 31 | 192 | 3 | 6 | 17 | 196 | 6 |
| 100 | 102 | 1 | 3 | 50 | 50 | 28 | 0 | 1 |
| 101 | 103 | 2 | 4 | 51 | 51 | 29 | 1 | 2 |
| 1 | 6 | 3 | 4 | 55 | 0 | 224 | 224 | 240 |

The corresponding binary image codewords are given by the coordinates of each '1':

| | | | | | | | | | | | | | |
|---|---|---|---|---|---|---|---|---|---|---|---|---|---|
| 0 | 2 | 5 | 33 | 35 | 44 | 45 | 46 | 66 | 70 | 75 | 78 | 109 | 110 |
| 1 | 3 | 6 | 34 | 36 | 45 | 46 | 47 | 67 | 71 | 76 | 79 | 110 | 111 |
| 0 | 1 | 4 | 16 | 18 | 20 | 21 | 32 | 38 | 68 | 69 | 85 | 115 | 119 |
| 0 | 2 | 5 | 21 | 50 | 54 | 55 | 66 | 67 | 69 | 70 | 82 | 83 | 97 |
| 2 | 5 | 41 | 46 | 59 | 63 | 65 | 68 | 71 | 104 | 107 | 109 | 120 | 127 |
| 12 | 40 | 41 | 47 | 60 | 61 | 88 | 92 | 93 | 94 | 105 | 107 | 110 | 120 |
| 19 | 23 | 48 | 54 | 56 | 59 | 61 | 65 | 66 | 69 | 88 | 93 | 94 | 103 |
| 28 | 29 | 32 | 33 | 58 | 62 | 64 | 65 | 66 | 72 | 74 | 77 | 78 | 104 |
| 29 | 30 | 33 | 34 | 59 | 63 | 65 | 66 | 67 | 73 | 75 | 78 | 79 | 105 |
| 30 | 31 | 34 | 35 | 56 | 58 | 59 | 66 | 67 | 68 | 72 | 75 | 79 | 106 |
| 29 | 31 | 33 | 35 | 56 | 58 | 63 | 65 | 68 | 72 | 73 | 78 | 105 | 106 |
| 31 | 33 | 34 | 37 | 56 | 63 | 65 | 71 | 89 | 90 | 91 | 94 | 96 | 97 |
| 11 | 37 | 38 | 40 | 46 | 50 | 77 | 82 | 83 | 86 | 98 | 101 | 102 | 117 |
| 12 | 38 | 39 | 41 | 47 | 51 | 78 | 83 | 84 | 87 | 99 | 102 | 103 | 118 |
| 0 | 4 | 10 | 14 | 17 | 43 | 48 | 50 | 64 | 66 | 83 | 84 | 96 | 105 |
| 1 | 5 | 11 | 15 | 18 | 44 | 49 | 51 | 65 | 67 | 84 | 85 | 97 | 106 |
| 9 | 30 | 43 | 52 | 77 | 79 | 80 | 89 | 92 | 113 | 116 | 122 | 123 | 125 |

For code $C_3$, the minimum bit-weight is 15, there are 60 codewords with such bit-weight.

The coordinates of the non-zero elements of these codewords are given by

| | | | | | |
|---|---|---|---|---|---|
| 0 | 2 | 4 | 7 | 14 | 15 |
| 0 | 2 | 5 | 6 | 12 | 13 |
| 1 | 3 | 4 | 5 | 12 | 15 |
| 1 | 3 | 6 | 7 | 13 | 14 |
| 0 | 3 | 4 | 10 | 14 | 15 |
| 0 | 3 | 4 | 10 | 14 | 15 |
| 0 | 4 | 5 | 8 | 9 | 13 |
| 0 | 4 | 5 | 8 | 9 | 13 |
| 0 | 4 | 5 | 8 | 9 | 13 |
| 0 | 2 | 4 | 8 | 10 | 14 |
| 0 | 2 | 6 | 8 | 10 | 12 |
| 0 | 1 | 5 | 11 | 12 | 15 |
| 0 | 1 | 5 | 11 | 12 | 15 |
| 1 | 5 | 6 | 9 | 10 | 14 |
| 1 | 5 | 6 | 9 | 10 | 14 |
| 1 | 5 | 6 | 9 | 10 | 14 |
| 1 | 3 | 5 | 9 | 11 | 15 |
| 1 | 3 | 7 | 9 | 11 | 13 |
| 1 | 2 | 6 | 8 | 12 | 13 |
| 1 | 2 | 6 | 8 | 12 | 13 |
| 2 | 6 | 7 | 10 | 11 | 15 |
| 2 | 6 | 7 | 10 | 11 | 15 |
| 2 | 6 | 7 | 10 | 11 | 15 |

| | | | | | | | | |
|---|---|---|---|---|---|---|---|---|
| | | 2 | 3 | 7 | 9 | 13 | 14 | |
| | | 2 | 3 | 7 | 9 | 13 | 14 | |
| | | 3 | 4 | 7 | 8 | 11 | 12 | |
| | | 3 | 4 | 7 | 8 | 11 | 12 | |
| | | 3 | 4 | 7 | 8 | 11 | 12 | |
| | 0 | 3 | 4 | 5 | 9 | 11 | 12 | 15 |
| | 0 | 3 | 4 | 5 | 9 | 11 | 12 | 15 |
| | 0 | 3 | 4 | 7 | 8 | 9 | 13 | 15 |
| | 0 | 1 | 4 | 5 | 9 | 10 | 12 | 14 |
| | 0 | 4 | 6 | 10 | 11 | 12 | 14 | 15 |
| | 0 | 4 | 6 | 10 | 11 | 12 | 14 | 15 |
| | 0 | 1 | 5 | 6 | 8 | 10 | 12 | 13 |
| | 0 | 1 | 5 | 6 | 8 | 10 | 12 | 13 |
| | 1 | 2 | 5 | 6 | 10 | 11 | 13 | 15 |
| | 1 | 5 | 7 | 8 | 11 | 12 | 13 | 15 |
| | 1 | 5 | 7 | 8 | 11 | 12 | 13 | 15 |
| | 1 | 2 | 6 | 7 | 9 | 11 | 13 | 14 |
| | 1 | 2 | 6 | 7 | 9 | 11 | 13 | 14 |
| | 2 | 3 | 6 | 7 | 8 | 11 | 12 | 14 |
| | 2 | 4 | 6 | 8 | 9 | 12 | 13 | 14 |
| | 2 | 4 | 6 | 8 | 9 | 12 | 13 | 14 |
| | 2 | 3 | 4 | 7 | 8 | 10 | 14 | 15 |
| | 2 | 3 | 4 | 7 | 8 | 10 | 14 | 15 |
| | 3 | 5 | 7 | 9 | 10 | 13 | 14 | 15 |
| | 3 | 5 | 7 | 9 | 10 | 13 | 14 | 15 |
| 0 | 1 | 3 | 4 | 5 | 9 | 11 | 12 | 15 |
| 0 | 2 | 3 | 4 | 7 | 8 | 10 | 14 | 15 |
| 0 | 3 | 4 | 6 | 10 | 11 | 12 | 14 | 15 |
| 0 | 2 | 4 | 5 | 8 | 9 | 10 | 13 | 14 |
| 0 | 2 | 6 | 7 | 8 | 10 | 11 | 12 | 15 |
| 0 | 1 | 2 | 5 | 6 | 8 | 10 | 12 | 13 |
| 0 | 1 | 5 | 7 | 8 | 11 | 12 | 13 | 15 |
| 1 | 3 | 5 | 6 | 9 | 10 | 11 | 14 | 15 |
| 1 | 3 | 4 | 7 | 8 | 9 | 11 | 12 | 13 |
| 1 | 2 | 3 | 6 | 7 | 9 | 11 | 13 | 14 |
| 1 | 2 | 4 | 6 | 8 | 9 | 12 | 13 | 14 |
| 2 | 3 | 5 | 7 | 9 | 10 | 13 | 14 | 15 |

The non-zero elements of these codewords are given by

| | | | | | |
|---|---|---|---|---|---|
| 200 | 183 | 25 | 103 | 17 | 112 |
| 183 | 200 | 103 | 25 | 17 | 112 |
| 200 | 183 | 103 | 25 | 112 | 17 |
| 183 | 200 | 103 | 25 | 17 | 112 |
| 230 | 103 | 55 | 224 | 50 | 192 |
| 182 | 55 | 7 | 176 | 2 | 144 |

| | | | | | | | | |
|---|---|---|---|---|---|---|---|---|
| | 181 | 29 | 101 | 130 | 101 | 50 | | |
| | 182 | 30 | 102 | 131 | 102 | 51 | | |
| | 183 | 31 | 103 | 132 | 103 | 52 | | |
| | 11 | 37 | 112 | 29 | 3 | 104 | | |
| | 37 | 11 | 112 | 3 | 29 | 104 | | |
| | 55 | 182 | 7 | 176 | 144 | 2 | | |
| | 103 | 230 | 55 | 224 | 192 | 50 | | |
| | 181 | 29 | 101 | 130 | 101 | 50 | | |
| | 182 | 30 | 102 | 131 | 102 | 51 | | |
| | 183 | 31 | 103 | 132 | 103 | 52 | | |
| | 11 | 37 | 112 | 29 | 3 | 104 | | |
| | 37 | 11 | 112 | 3 | 29 | 104 | | |
| | 103 | 230 | 55 | 224 | 50 | 192 | | |
| | 55 | 182 | 7 | 176 | 2 | 144 | | |
| | 181 | 29 | 101 | 130 | 101 | 50 | | |
| | 182 | 30 | 102 | 131 | 102 | 51 | | |
| | 183 | 31 | 103 | 132 | 103 | 52 | | |
| | 103 | 230 | 55 | 224 | 50 | 192 | | |
| | 55 | 182 | 7 | 176 | 2 | 144 | | |
| | 181 | 101 | 29 | 101 | 130 | 50 | | |
| | 182 | 102 | 30 | 102 | 131 | 51 | | |
| | 183 | 103 | 31 | 103 | 132 | 52 | | |
| 5 | 139 | 30 | 100 | 28 | 182 | 102 | 5 | |
| 6 | 140 | 31 | 101 | 29 | 183 | 103 | 6 | |
| 140 | 139 | 6 | 138 | 50 | 225 | 51 | 52 | |
| 139 | 140 | 138 | 6 | 50 | 225 | 52 | 51 | |
| 3 | 199 | 6 | 29 | 225 | 51 | 191 | 51 | |
| 4 | 200 | 7 | 30 | 226 | 52 | 192 | 52 | |
| 139 | 5 | 30 | 100 | 182 | 28 | 5 | 102 | |
| 140 | 6 | 31 | 101 | 183 | 29 | 6 | 103 | |
| 139 | 140 | 138 | 6 | 50 | 225 | 52 | 51 | |
| 3 | 199 | 6 | 225 | 29 | 51 | 51 | 191 | |
| 4 | 200 | 7 | 226 | 30 | 52 | 52 | 192 | |
| 139 | 5 | 30 | 100 | 182 | 28 | 5 | 102 | |
| 140 | 6 | 31 | 101 | 183 | 29 | 6 | 103 | |
| 139 | 140 | 138 | 6 | 225 | 50 | 51 | 52 | |
| 3 | 6 | 199 | 29 | 225 | 191 | 51 | 51 | |
| 4 | 7 | 200 | 30 | 226 | 192 | 52 | 52 | |
| 139 | 5 | 100 | 30 | 28 | 182 | 5 | 102 | |
| 140 | 6 | 101 | 31 | 29 | 183 | 6 | 103 | |
| 3 | 6 | 199 | 29 | 225 | 191 | 51 | 51 | |
| 4 | 7 | 200 | 30 | 226 | 192 | 52 | 52 | |
| 53 | 66 | 29 | 0 | 140 | 4 | 1 | 4 | 52 |
| 66 | 29 | 53 | 140 | 0 | 4 | 1 | 52 | 4 |

| 55 | 27 | 2 | 52 | 3 | 15 | 6 | 2 | 138 |
| 31 | 1 | 25 | 2 | 195 | 26 | 6 | 7 | 55 |
| 1 | 31 | 25 | 2 | 6 | 195 | 26 | 55 | 7 |
| 29 | 53 | 66 | 0 | 140 | 1 | 4 | 52 | 4 |
| 27 | 55 | 2 | 52 | 15 | 3 | 138 | 6 | 2 |
| 31 | 1 | 25 | 2 | 195 | 26 | 6 | 7 | 55 |
| 1 | 31 | 2 | 25 | 26 | 6 | 195 | 7 | 55 |
| 29 | 53 | 66 | 0 | 140 | 1 | 4 | 52 | 4 |
| 27 | 55 | 52 | 2 | 3 | 15 | 2 | 138 | 6 |
| 27 | 55 | 52 | 2 | 3 | 15 | 2 | 138 | 6 |

The corresponding binary image codewords are given by the coordinates of each '1':

| 2 | 3 | 4 | 18 | 22 | 23 | 32 | 33 | 59 | 63 | 115 | 116 | 119 | 120 | 127 |
| 2 | 6 | 7 | 18 | 19 | 20 | 43 | 47 | 48 | 49 | 99 | 100 | 103 | 104 | 111 |
| 10 | 11 | 12 | 26 | 30 | 31 | 35 | 39 | 40 | 41 | 96 | 103 | 123 | 124 | 127 |
| 10 | 14 | 15 | 26 | 27 | 28 | 51 | 55 | 56 | 57 | 107 | 108 | 111 | 112 | 119 |
| 2 | 4 | 5 | 6 | 7 | 27 | 31 | 37 | 39 | 81 | 84 | 112 | 114 | 121 | 127 |
| 1 | 5 | 6 | 29 | 31 | 39 | 80 | 81 | 85 | 86 | 87 | 114 | 123 | 125 | 127 |
| 0 | 4 | 5 | 36 | 37 | 41 | 45 | 65 | 66 | 67 | 69 | 73 | 77 | 104 | 106 |
| 1 | 5 | 6 | 37 | 38 | 42 | 46 | 66 | 67 | 68 | 70 | 74 | 78 | 105 | 107 |
| 2 | 6 | 7 | 38 | 39 | 43 | 47 | 67 | 68 | 69 | 71 | 75 | 79 | 106 | 108 |
| 3 | 5 | 6 | 7 | 17 | 19 | 22 | 32 | 39 | 68 | 69 | 83 | 112 | 114 | 115 |
| 1 | 3 | 6 | 19 | 21 | 22 | 23 | 48 | 55 | 67 | 84 | 85 | 96 | 98 | 99 |
| 5 | 7 | 9 | 13 | 14 | 47 | 88 | 89 | 93 | 94 | 95 | 99 | 101 | 103 | 122 |
| 3 | 7 | 10 | 12 | 13 | 14 | 15 | 45 | 47 | 89 | 92 | 97 | 103 | 120 | 122 |
| 8 | 12 | 13 | 44 | 45 | 49 | 53 | 73 | 74 | 75 | 77 | 81 | 85 | 112 | 114 |
| 9 | 13 | 14 | 45 | 46 | 50 | 54 | 74 | 75 | 76 | 78 | 82 | 86 | 113 | 115 |
| 10 | 14 | 15 | 46 | 47 | 51 | 55 | 75 | 76 | 77 | 79 | 83 | 87 | 114 | 116 |
| 11 | 13 | 14 | 15 | 25 | 27 | 30 | 40 | 47 | 76 | 77 | 91 | 120 | 122 | 123 |
| 9 | 11 | 14 | 27 | 29 | 30 | 31 | 56 | 63 | 75 | 92 | 93 | 104 | 106 | 107 |
| 11 | 15 | 18 | 20 | 21 | 22 | 23 | 53 | 55 | 65 | 68 | 96 | 98 | 105 | 111 |
| 13 | 15 | 17 | 21 | 22 | 55 | 64 | 65 | 69 | 70 | 71 | 98 | 107 | 109 | 111 |
| 16 | 20 | 21 | 52 | 53 | 57 | 61 | 81 | 82 | 83 | 85 | 89 | 93 | 120 | 122 |
| 17 | 21 | 22 | 53 | 54 | 58 | 62 | 82 | 83 | 84 | 86 | 90 | 94 | 121 | 123 |
| 18 | 22 | 23 | 54 | 55 | 59 | 63 | 83 | 84 | 85 | 87 | 91 | 95 | 122 | 124 |
| 19 | 23 | 26 | 28 | 29 | 30 | 31 | 61 | 63 | 73 | 76 | 104 | 106 | 113 | 119 |
| 21 | 23 | 25 | 29 | 30 | 63 | 72 | 73 | 77 | 78 | 79 | 106 | 115 | 117 | 119 |
| 24 | 28 | 29 | 33 | 37 | 60 | 61 | 65 | 69 | 89 | 90 | 91 | 93 | 96 | 98 |
| 25 | 29 | 30 | 34 | 38 | 61 | 62 | 66 | 70 | 90 | 91 | 92 | 94 | 97 | 99 |
| 26 | 30 | 31 | 35 | 39 | 62 | 63 | 67 | 71 | 91 | 92 | 93 | 95 | 98 | 100 |
| 5 | 25 | 30 | 37 | 38 | 40 | 44 | 75 | 76 | 89 | 93 | 94 | 98 | 102 | 125 |
| 6 | 26 | 31 | 38 | 39 | 41 | 45 | 76 | 77 | 90 | 94 | 95 | 99 | 103 | 126 |
| 2 | 7 | 25 | 30 | 38 | 56 | 61 | 64 | 66 | 74 | 77 | 105 | 107 | 122 | 124 |
| 1 | 6 | 10 | 15 | 32 | 37 | 46 | 72 | 74 | 82 | 85 | 98 | 100 | 113 | 115 |
| 3 | 33 | 34 | 35 | 54 | 84 | 85 | 90 | 93 | 97 | 99 | 112 | 118 | 121 | 123 |

| | | | | | | | | | | | | | |
|---|---|---|---|---|---|---|---|---|---|---|---|---|---|
| 4 | 34 | 35 | 36 | 55 | 85 | 86 | 91 | 94 | 98 | 100 | 113 | 119 | 122 | 124 |
| 1 | 6 | 13 | 45 | 46 | 48 | 52 | 65 | 69 | 70 | 83 | 84 | 101 | 106 | 110 |
| 2 | 7 | 14 | 46 | 47 | 49 | 53 | 66 | 70 | 71 | 84 | 85 | 102 | 107 | 111 |
| 9 | 14 | 18 | 23 | 40 | 45 | 54 | 80 | 82 | 90 | 93 | 106 | 108 | 121 | 123 |
| 11 | 41 | 42 | 43 | 62 | 66 | 69 | 92 | 93 | 97 | 99 | 105 | 107 | 120 | 126 |
| 12 | 42 | 43 | 44 | 63 | 67 | 70 | 93 | 94 | 98 | 100 | 106 | 108 | 121 | 127 |
| 9 | 14 | 21 | 53 | 54 | 56 | 60 | 73 | 77 | 78 | 91 | 92 | 109 | 114 | 118 |
| 10 | 15 | 22 | 54 | 55 | 57 | 61 | 74 | 78 | 79 | 92 | 93 | 110 | 115 | 119 |
| 17 | 22 | 26 | 31 | 48 | 53 | 62 | 66 | 69 | 88 | 90 | 97 | 99 | 114 | 116 |
| 19 | 38 | 49 | 50 | 51 | 68 | 69 | 74 | 77 | 96 | 102 | 105 | 107 | 113 | 115 |
| 20 | 39 | 50 | 51 | 52 | 69 | 70 | 75 | 78 | 97 | 103 | 106 | 108 | 114 | 116 |
| 17 | 22 | 29 | 32 | 36 | 61 | 62 | 67 | 68 | 81 | 85 | 86 | 117 | 122 | 126 |
| 18 | 23 | 30 | 33 | 37 | 62 | 63 | 68 | 69 | 82 | 86 | 87 | 118 | 123 | 127 |
| 27 | 46 | 57 | 58 | 59 | 76 | 77 | 82 | 85 | 104 | 110 | 113 | 115 | 121 | 123 |
| 28 | 47 | 58 | 59 | 60 | 77 | 78 | 83 | 86 | 105 | 111 | 114 | 116 | 122 | 124 |
| 3 | 5 | 8 | 13 | 14 | 28 | 29 | 32 | 42 | 47 | 76 | 89 | 100 | 122 | 124 |
| 0 | 5 | 6 | 20 | 21 | 27 | 29 | 34 | 39 | 56 | 68 | 81 | 114 | 116 | 124 |
| 5 | 7 | 26 | 27 | 34 | 50 | 52 | 83 | 89 | 90 | 93 | 102 | 114 | 120 | 125 |
| 6 | 7 | 17 | 32 | 33 | 42 | 66 | 69 | 70 | 73 | 74 | 86 | 111 | 117 | 119 |
| 1 | 22 | 23 | 48 | 49 | 58 | 70 | 82 | 85 | 86 | 89 | 90 | 101 | 103 | 127 |
| 4 | 5 | 11 | 13 | 16 | 21 | 22 | 40 | 50 | 55 | 65 | 84 | 98 | 100 | 108 |
| 2 | 3 | 13 | 15 | 42 | 58 | 60 | 65 | 66 | 69 | 91 | 96 | 101 | 110 | 122 |
| 14 | 15 | 25 | 40 | 41 | 50 | 74 | 77 | 78 | 81 | 82 | 94 | 119 | 125 | 127 |
| 9 | 30 | 31 | 34 | 56 | 57 | 65 | 66 | 78 | 90 | 93 | 94 | 103 | 109 | 111 |
| 12 | 13 | 19 | 21 | 24 | 29 | 30 | 48 | 58 | 63 | 73 | 92 | 106 | 108 | 116 |
| 10 | 11 | 21 | 23 | 34 | 36 | 50 | 67 | 73 | 74 | 77 | 98 | 104 | 109 | 118 |
| 18 | 19 | 29 | 31 | 42 | 44 | 58 | 75 | 81 | 82 | 85 | 106 | 112 | 117 | 126 |

For code $C_4$, the minimum bit-weight is 15, there are 8 codewords with such bit-weight.

The coordinates of the non-zero elements of these codewords are given by

| | | | | | | | |
|---|---|---|---|---|---|---|---|
| 0 | 3 | 4 | 5 | 7 | 8 | 13 | 15 |
| 0 | 1 | 4 | 5 | 6 | 9 | 12 | 14 |
| 0 | 4 | 6 | 10 | 11 | 12 | 14 | 15 |
| 1 | 2 | 5 | 6 | 7 | 10 | 13 | 15 |
| 1 | 5 | 7 | 8 | 11 | 12 | 13 | 15 |
| 2 | 3 | 4 | 6 | 7 | 11 | 12 | 14 |
| 2 | 4 | 6 | 8 | 9 | 12 | 13 | 14 |
| 3 | 5 | 7 | 9 | 10 | 13 | 14 | 15 |

The non-zero elements of these codewords are given by

| | | | | | | | |
|---|---|---|---|---|---|---|---|
| 1 | 0 | 247 | 142 | 6 | 2 | 52 | 16 |
| 0 | 1 | 6 | 247 | 142 | 2 | 16 | 52 |
| 2 | 52 | 16 | 1 | 0 | 6 | 142 | 247 |
| 0 | 1 | 6 | 247 | 142 | 2 | 16 | 52 |
| 2 | 52 | 16 | 0 | 1   | 247 | 6 | 142 |
| 0 | 1  | 142| 6 | 247 | 2   | 52| 16  |
| 2 | 16 | 52 | 1 | 0   | 142 | 247| 6  |
| 2 | 16 | 52 | 1 | 0   | 142 | 247| 6  |

The corresponding binary image codewords are given by the coordinates of each '1':

| 1  | 24 | 32 | 33 | 39 | 41 | 43 | 45 | 62 | 66 | 106 | 108 | 122 | 123 | 126 |
|----|----|----|----|----|----|----|----|----|----|-----|-----|-----|-----|-----|
| 0  | 9  | 38 | 40 | 41 | 47 | 49 | 51 | 53 | 74 | 98  | 99  | 102 | 114 | 116 |
| 2  | 34 | 36 | 50 | 51 | 54 | 81 | 88 | 102| 113| 115 | 117 | 120 | 121 | 127 |
| 8  | 17 | 46 | 48 | 49 | 55 | 57 | 59 | 61 | 82 | 106 | 107 | 110 | 122 | 124 |
| 10 | 42 | 44 | 58 | 59 | 62 | 64 | 89 | 96 | 97 | 103 | 110 | 121 | 123 | 125 |
| 16 | 25 | 33 | 35 | 37 | 54 | 56 | 57 | 63 | 90 | 98  | 100 | 114 | 115 | 118 |
| 18 | 34 | 35 | 38 | 50 | 52 | 65 | 72 | 97 | 99 | 101 | 104 | 105 | 111 | 118 |
| 26 | 42 | 43 | 46 | 58 | 60 | 73 | 80 | 105| 107| 109 | 112 | 113 | 119 | 126 |

**Example 2:** We design a (52,26) LDPC code over $GF(2^8)$. The positions of non-zero entries of the parity-check matrix are given row-by-row as follows.

| 0  | 13 | 26 | 39 |
|----|----|----|----|
| 1  | 14 | 27 | 40 |
| 2  | 15 | 28 | 41 |
| 3  | 16 | 29 | 42 |
| 4  | 17 | 30 | 43 |
| 5  | 18 | 31 | 44 |
| 6  | 19 | 32 | 45 |
| 7  | 20 | 33 | 46 |
| 8  | 21 | 34 | 47 |
| 9  | 22 | 35 | 48 |
| 10 | 23 | 36 | 49 |
| 11 | 24 | 37 | 50 |
| 12 | 25 | 38 | 51 |
| 0  | 14 | 30 | 45 |
| 1  | 15 | 31 | 46 |
| 2  | 16 | 32 | 47 |
| 3  | 17 | 33 | 48 |
| 4  | 18 | 34 | 49 |
| 5  | 19 | 35 | 50 |
| 6  | 20 | 36 | 51 |
| 7  | 21 | 37 | 39 |
| 8  | 22 | 38 | 40 |
| 9  | 23 | 26 | 41 |
| 10 | 24 | 27 | 42 |
| 11 | 25 | 28 | 43 |
| 12 | 13 | 29 | 44 |

The cycle distribution of the corresponding Tanner graph is given by $234x^{12} + 702x^{16} + 5616x^{20} + 21060x^{24} + \cdots$. So the girth is 12.

Using the identified cycles, we identify inter-connected short cycles. The number of the corresponding patterns is given in Table III.

TABLE III.
THE NUMBER OF INTER-CONNECTED SHORT CYCLES

| Inter-connected short cycles | Number |
| --- | --- |
| A-(3,3,3) | 468 |
| A-(2,4,4) | 2808 |
| A-(1,5,5) | 8424 |
| A-(3,3,5) | 2808 |
| A-(2,4,6) | 14976 |
| A-(4,4,4) | 936 |
| B-(6,6) | 5616 |
| C-(2,2,2,2,2,2) | 468 |
| C-(1,2,1,2,3,3) | 2808 |
| D-(3,3,3,3) | 117 |

The non-zero entries of the parity-check matrix are given as follows.

8 173 183 0

81 0 89 9

183 0 172 8

167 127 40 0

169 40 127 0

169 40 128 0

8 173 183 0

8 182 173 0

172 0 182 8

0 80 8 88

40 127 0 169

0 8 183 173

0 80 88 8

0 172 8 183

88 8 80 0

80 88 0 8

0 80 8 88

81 89 9 0

167 127 40 0

182 172 8 0

0 169 40 127

40 167 127 0

$$169 \ 40 \ 0 \ 128$$
$$40 \ 167 \ 127 \ 0$$
$$127 \ 40 \ 0 \ 167$$
$$183 \ 173 \ 8 \ 0$$

Using the identified short cycles and inter-connected short cycles, the bit-weight distribution is estimated as $4x^{22} + 46x^{23} + 108x^{24} + 322x^{25} + 695x^{26} + 1540x^{27} + \cdots$. So the code has an estimated minimum bit-distance 22.

**Example 3**: We design a (160,80) LDPC code over $\mathrm{GF}(2^8)$. The positions of non-zero entries of the parity-check matrix are given row-by-row as follows.

| | | | |
|---|---|---|---|
| 0 | 1 | 2 | 3 |
| 0 | 4 | 5 | 6 |
| 1 | 7 | 8 | 9 |
| 2 | 10 | 11 | 12 |
| 3 | 13 | 14 | 15 |
| 4 | 16 | 17 | 18 |
| 5 | 19 | 20 | 21 |
| 6 | 22 | 23 | 24 |
| 7 | 25 | 26 | 27 |
| 8 | 28 | 29 | 30 |
| 9 | 31 | 32 | 33 |
| 10 | 34 | 35 | 36 |
| 11 | 37 | 38 | 39 |
| 12 | 40 | 41 | 42 |
| 13 | 43 | 44 | 45 |
| 14 | 46 | 47 | 48 |
| 15 | 49 | 50 | 51 |
| 16 | 52 | 53 | 54 |
| 17 | 55 | 56 | 57 |
| 18 | 58 | 59 | 60 |
| 19 | 61 | 62 | 63 |
| 20 | 64 | 65 | 66 |
| 21 | 67 | 68 | 69 |
| 22 | 70 | 71 | 72 |
| 23 | 73 | 74 | 75 |
| 24 | 76 | 77 | 78 |
| 25 | 79 | 80 | 81 |
| 26 | 82 | 83 | 84 |
| 27 | 85 | 86 | 87 |
| 28 | 88 | 89 | 90 |
| 29 | 91 | 92 | 93 |
| 30 | 94 | 95 | 96 |
| 31 | 97 | 98 | 99 |
| 32 | 100 | 101 | 102 |
| 33 | 103 | 104 | 105 |
| 34 | 106 | 107 | 108 |
| 35 | 109 | 110 | 111 |
| 36 | 112 | 113 | 114 |
| 37 | 115 | 116 | 117 |
| 38 | 118 | 119 | 120 |
| 39 | 121 | 122 | 123 |
| 40 | 124 | 125 | 126 |
| 41 | 127 | 128 | 129 |
| 42 | 130 | 131 | 132 |
| 43 | 133 | 134 | 135 |
| 44 | 136 | 137 | 138 |
| 45 | 139 | 140 | 141 |
| 46 | 142 | 143 | 144 |
| 47 | 145 | 146 | 147 |
| 48 | 148 | 149 | 150 |
| 49 | 151 | 152 | 153 |
| 50 | 154 | 155 | 156 |
| 51 | 157 | 158 | 159 |
| 52 | 88 | 119 | 153 |
| 53 | 79 | 127 | 134 |
| 54 | 104 | 106 | 147 |
| 55 | 91 | 121 | 159 |
| 56 | 83 | 126 | 139 |
| 57 | 102 | 114 | 150 |
| 58 | 95 | 115 | 154 |
| 59 | 98 | 111 | 144 |
| 60 | 85 | 131 | 138 |
| 61 | 82 | 118 | 142 |
| 62 | 96 | 113 | 133 |
| 63 | 105 | 130 | 157 |
| 64 | 89 | 110 | 141 |
| 65 | 87 | 122 | 146 |
| 66 | 101 | 128 | 155 |
| 67 | 97 | 125 | 151 |
| 68 | 81 | 116 | 149 |
| 69 | 92 | 107 | 137 |
| 70 | 100 | 120 | 136 |
| 71 | 94 | 124 | 145 |
| 72 | 80 | 109 | 158 |
| 73 | 103 | 117 | 140 |
| 74 | 86 | 112 | 152 |
| 75 | 93 | 129 | 143 |
| 76 | 84 | 108 | 156 |
| 77 | 99 | 123 | 135 |
| 78 | 90 | 132 | 148 |

The cycle distribution of the Tanner graph is given by $1620x^{16} + 5184x^{20} + 43200x^{24} + \cdots$. So the girth is 16. Using the identified cycles, we identify inter-connected short cycles. The number of the corresponding patterns is given in Table IV.

TABLE IV.
THE NUMBER OF INTER-CONNECTED SHORT CYCLES

| Inter-connected short cycles | Number |
|---|---|
| A-(4,4,4) | 4320 |
| A-(3,5,5) | 25920 |
| A-(2,6,6) | 77760 |
| A-(4,4,6) | 25920 |

The non-zero entries of the parity-check matrix are given as follows.

| | | | |
|---:|---:|---:|---:|
| 80 | 88 | 8 | 0 |
| 172 | 8 | 0 | 183 |
| 0 | 128 | 40 | 169 |
| 40 | 127 | 0 | 169 |
| 40 | 0 | 169 | 127 |
| 0 | 8 | 80 | 88 |
| 169 | 40 | 128 | 0 |
| 183 | 172 | 8 | 0 |
| 0 | 8 | 172 | 183 |
| 0 | 183 | 8 | 173 |
| 173 | 0 | 183 | 8 |
| 0 | 169 | 40 | 127 |
| 172 | 182 | 8 | 0 |
| 40 | 127 | 0 | 167 |
| 0 | 8 | 172 | 183 |
| 0 | 182 | 172 | 8 |
| 80 | 88 | 8 | 0 |
| 0 | 167 | 40 | 127 |
| 183 | 0 | 173 | 8 |
| 127 | 40 | 169 | 0 |
| 0 | 80 | 8 | 88 |
| 0 | 8 | 80 | 88 |
| 8 | 183 | 172 | 0 |
| 0 | 40 | 167 | 127 |
| 81 | 9 | 0 | 89 |
| 40 | 0 | 169 | 127 |
| 183 | 172 | 0 | 8 |
| 0 | 182 | 8 | 173 |
| 40 | 127 | 169 | 0 |
| 172 | 183 | 8 | 0 |
| 80 | 88 | 8 | 0 |
| 0 | 40 | 128 | 169 |
| 183 | 8 | 172 | 0 |
| 0 | 80 | 88 | 8 |
| 0 | 8 | 183 | 173 |
| 172 | 183 | 8 | 0 |
| 0 | 173 | 183 | 8 |
| 8 | 183 | 172 | 0 |
| 0 | 9 | 81 | 89 |
| 80 | 88 | 8 | 0 |
| 0 | 172 | 182 | 8 |
| 0 | 182 | 173 | 8 |
| 8 | 182 | 0 | 173 |
| 183 | 0 | 8 | 172 |
| 40 | 0 | 128 | 169 |
| 0 | 173 | 182 | 8 |
| 183 | 0 | 172 | 8 |
| 88 | 0 | 8 | 80 |
| 0 | 81 | 9 | 89 |
| 0 | 89 | 81 | 9 |
| 127 | 40 | 0 | 169 |
| 183 | 172 | 8 | 0 |
| 182 | 8 | 172 | 0 |
| 183 | 0 | 172 | 8 |
| 80 | 8 | 88 | 0 |
| 0 | 80 | 8 | 88 |
| 172 | 183 | 8 | 0 |
| 8 | 0 | 172 | 183 |
| 127 | 169 | 40 | 0 |
| 182 | 172 | 8 | 0 |
| 127 | 169 | 0 | 40 |
| 8 | 182 | 172 | 0 |
| 183 | 172 | 8 | 0 |
| 173 | 183 | 8 | 0 |
| 182 | 172 | 8 | 0 |
| 80 | 88 | 0 | 8 |
| 183 | 8 | 0 | 173 |
| 167 | 40 | 0 | 127 |
| 183 | 172 | 8 | 0 |
| 0 | 169 | 127 | 40 |
| 88  | 0   | 80  | 8   |
| 169 | 40  | 127 | 0   |
| 183 | 172 | 8   | 0   |
| 172 | 183 | 8   | 0   |
| 0   | 8   | 173 | 183 |
| 183 | 0   | 8   | 172 |
| 81  | 0   | 89  | 9   |
| 172 | 0   | 182 | 8   |
| 183 | 172 | 8   | 0   |
| 182 | 8   | 173 | 0   |

Using the identified short cycles and inter-connected short cycles, the bit-weight distribution is estimated as $21x^{30} + 54x^{31} + 183x^{32} + 480x^{33} + \cdots$. So the code has an estimated minimum bit-distance 30.

APPENDIX:
IDENTIFIED SHORT CYCLES AND COLUMN INDICES CONTAINED IN
INTER-CONNCTED SHORT CYCLES OF EXAMPLE 1

We first need to describe the format. For example, a length-8 cycle is given by

$$0\ 4\ 15\ 3\ 3\ 7\ 4\ 0$$

where the first number is the column index (corresponding to the a variable node), the second number is the row index (corresponding to a check node), the third number is the column index, and so forth. So the cycle can be illustrated in Fig. 3.

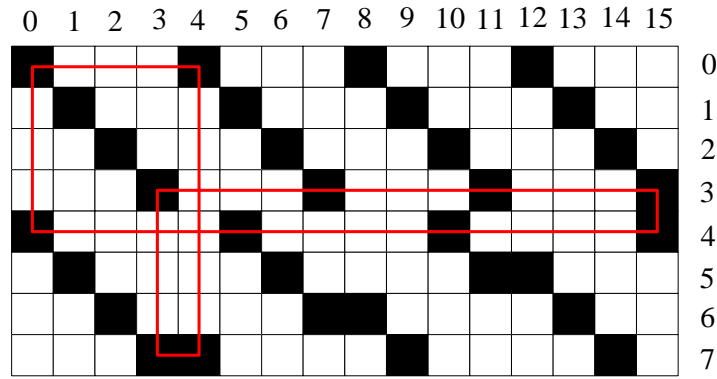

Fig. 3. A length-8 cycle

Now we list the cycles for the Tanner graph.

There are 36 length-8 cycles:

$$0\ 4\ 15\ 3\ 3\ 7\ 4\ 0$$
$$0\ 4\ 5\ 1\ 9\ 7\ 4\ 0$$
$$0\ 4\ 10\ 2\ 14\ 7\ 4\ 0$$
$$0\ 4\ 10\ 2\ 2\ 6\ 8\ 0$$

```
0 4 15 3 7 6 8 0
0 4 5 1 13 6 8 0
0 4 5 1 1 5 12 0
0 4 10 2 6 5 12 0
0 4 15 3 11 5 12 0
1 5 6 2 10 4 5 1
1 5 11 3 15 4 5 1
1 5 11 3 3 7 9 1
1 5 12 0 4 7 9 1
1 5 6 2 14 7 9 1
1 5 6 2 2 6 13 1
1 5 11 3 7 6 13 1
1 5 12 0 8 6 13 1
2 6 7 3 11 5 6 2
2 6 8 0 12 5 6 2
2 6 13 1 5 4 10 2
2 6 7 3 15 4 10 2
2 6 7 3 3 7 14 2
2 6 8 0 4 7 14 2
2 6 13 1 9 7 14 2
3 7 4 0 8 6 7 3
3 7 9 1 13 6 7 3
3 7 14 2 6 5 11 3
3 7 4 0 12 5 11 3
3 7 9 1 5 4 15 3
3 7 14 2 10 4 15 3
4 7 9 1 13 6 8 0
4 7 14 2 6 5 12 0
5 4 10 2 14 7 9 1
5 4 15 3 7 6 13 1
6 5 11 3 15 4 10 2
7 6 8 0 12 5 11 3
```

There are 96 length-12 cycles:

```
0 4 5 1 13 6 7 3 3 7 4 0
0 4 10 2 2 6 7 3 3 7 4 0
0 4 5 1 1 5 11 3 3 7 4 0
0 4 10 2 6 5 11 3 3 7 4 0
0 4 10 2 6 5 1 1 9 7 4 0
0 4 15 3 11 5 1 1 9 7 4 0
0 4 10 2 2 6 13 1 9 7 4 0
0 4 15 3 7 6 13 1 9 7 4 0
0 4 5 1 13 6 2 2 14 7 4 0
0 4 15 3 7 6 2 2 14 7 4 0
```

0 4 5 1 1 5 6 2 14 7 4 0
0 4 15 3 11 5 6 2 14 7 4 0
0 4 5 1 1 5 6 2 2 6 8 0
0 4 15 3 11 5 6 2 2 6 8 0
0 4 5 1 9 7 14 2 2 6 8 0
0 4 15 3 3 7 14 2 2 6 8 0
0 4 5 1 9 7 3 3 7 6 8 0
0 4 10 2 14 7 3 3 7 6 8 0
0 4 5 1 1 5 11 3 7 6 8 0
0 4 10 2 6 5 11 3 7 6 8 0
0 4 10 2 6 5 1 1 13 6 8 0
0 4 15 3 11 5 1 1 13 6 8 0
0 4 10 2 14 7 9 1 13 6 8 0
0 4 15 3 3 7 9 1 13 6 8 0
0 4 10 2 14 7 9 1 1 5 12 0
0 4 15 3 3 7 9 1 1 5 12 0
0 4 10 2 2 6 13 1 1 5 12 0
0 4 15 3 7 6 13 1 1 5 12 0
0 4 5 1 13 6 2 2 6 5 12 0
0 4 15 3 7 6 2 2 6 5 12 0
0 4 5 1 9 7 14 2 6 5 12 0
0 4 15 3 3 7 14 2 6 5 12 0
0 4 5 1 9 7 3 3 11 5 12 0
0 4 10 2 14 7 3 3 11 5 12 0
0 4 5 1 13 6 7 3 11 5 12 0
0 4 10 2 2 6 7 3 11 5 12 0
1 5 11 3 7 6 2 2 10 4 5 1
1 5 12 0 8 6 2 2 10 4 5 1
1 5 11 3 3 7 14 2 10 4 5 1
1 5 12 0 4 7 14 2 10 4 5 1
1 5 6 2 14 7 3 3 15 4 5 1
1 5 12 0 4 7 3 3 15 4 5 1
1 5 6 2 2 6 7 3 15 4 5 1
1 5 12 0 8 6 7 3 15 4 5 1
1 5 6 2 2 6 7 3 3 7 9 1
1 5 12 0 8 6 7 3 3 7 9 1
1 5 6 2 10 4 15 3 3 7 9 1
1 5 6 2 2 6 8 0 4 7 9 1
1 5 11 3 7 6 8 0 4 7 9 1
1 5 11 3 7 6 2 2 14 7 9 1
1 5 12 0 8 6 2 2 14 7 9 1
1 5 11 3 15 4 10 2 14 7 9 1
1 5 11 3 15 4 10 2 2 6 13 1
1 5 11 3 3 7 14 2 2 6 13 1

1 5 12 0 4 7 14 2 2 6 13 1
1 5 6 2 14 7 3 3 7 6 13 1
1 5 12 0 4 7 3 3 7 6 13 1
1 5 6 2 10 4 15 3 7 6 13 1
1 5 6 2 14 7 4 0 8 6 13 1
1 5 11 3 3 7 4 0 8 6 13 1
2 6 8 0 4 7 3 3 11 5 6 2
2 6 13 1 9 7 3 3 11 5 6 2
2 6 13 1 5 4 15 3 11 5 6 2
2 6 7 3 3 7 4 0 12 5 6 2
2 6 13 1 9 7 4 0 12 5 6 2
2 6 7 3 3 7 9 1 5 4 10 2
2 6 8 0 4 7 9 1 5 4 10 2
2 6 8 0 4 7 3 3 15 4 10 2
2 6 13 1 9 7 3 3 15 4 10 2
2 6 8 0 12 5 11 3 15 4 10 2
2 6 8 0 12 5 11 3 3 7 14 2
2 6 13 1 5 4 15 3 3 7 14 2
2 6 7 3 11 5 12 0 4 7 14 2
2 6 7 3 15 4 5 1 9 7 14 2
3 7 14 2 6 5 12 0 8 6 7 3
3 7 14 2 10 4 5 1 13 6 7 3
3 7 9 1 5 4 10 2 6 5 11 3
3 7 9 1 13 6 8 0 12 5 11 3
3 7 4 0 8 6 13 1 5 4 15 3
3 7 4 0 12 5 6 2 10 4 15 3
4 7 14 2 6 5 11 3 7 6 8 0
4 7 9 1 5 4 15 3 7 6 8 0
4 7 14 2 10 4 15 3 7 6 8 0
4 7 14 2 10 4 5 1 13 6 8 0
4 7 9 1 5 4 10 2 6 5 12 0
4 7 9 1 13 6 7 3 11 5 12 0
4 7 9 1 5 4 15 3 11 5 12 0
4 7 14 2 10 4 15 3 11 5 12 0
5 4 15 3 11 5 6 2 14 7 9 1
5 4 10 2 6 5 11 3 7 6 13 1
5 4 10 2 6 5 12 0 8 6 13 1
5 4 15 3 11 5 12 0 8 6 13 1
6 5 12 0 8 6 7 3 15 4 10 2
6 5 11 3 7 6 13 1 9 7 14 2
6 5 12 0 8 6 13 1 9 7 14 2
7 6 13 1 9 7 14 2 10 4 15 3

There are 72 length-16 cycles:

```
        0 4 5 1 1 5 6 2 2 6 7 3 3 7 4 0
        0 4 10 2 6 5 1 1 13 6 7 3 3 7 4 0
        0 4 10 2 2 6 13 1 1 5 11 3 3 7 4 0
        0 4 5 1 13 6 2 2 6 5 11 3 3 7 4 0
        0 4 15 3 7 6 2 2 6 5 1 1 9 7 4 0
        0 4 10 2 2 6 7 3 11 5 1 1 9 7 4 0
        0 4 15 3 11 5 6 2 2 6 13 1 9 7 4 0
        0 4 10 2 6 5 11 3 7 6 13 1 9 7 4 0
        0 4 5 1 1 5 11 3 7 6 2 2 14 7 4 0
        0 4 15 3 11 5 1 1 13 6 2 2 14 7 4 0
        0 4 15 3 7 6 13 1 1 5 6 2 14 7 4 0
        0 4 5 1 13 6 7 3 11 5 6 2 14 7 4 0
        0 4 15 3 3 7 9 1 1 5 6 2 2 6 8 0
        0 4 5 1 9 7 3 3 11 5 6 2 2 6 8 0
        0 4 5 1 1 5 11 3 3 7 14 2 2 6 8 0
        0 4 15 3 11 5 1 1 9 7 14 2 2 6 8 0
        0 4 10 2 6 5 1 1 9 7 3 3 7 6 8 0
        0 4 5 1 1 5 6 2 14 7 3 3 7 6 8 0
        0 4 10 2 14 7 9 1 1 5 11 3 7 6 8 0
        0 4 5 1 9 7 14 2 6 5 11 3 7 6 8 0
        0 4 15 3 3 7 14 2 6 5 1 1 13 6 8 0
        0 4 10 2 14 7 3 3 11 5 1 1 13 6 8 0
        0 4 10 2 6 5 11 3 3 7 9 1 13 6 8 0
        0 4 15 3 11 5 6 2 14 7 9 1 13 6 8 0
        0 4 10 2 2 6 7 3 3 7 9 1 1 5 12 0
        0 4 15 3 7 6 2 2 14 7 9 1 1 5 12 0
        0 4 15 3 3 7 14 2 2 6 13 1 1 5 12 0
        0 4 10 2 14 7 3 3 7 6 13 1 1 5 12 0
        0 4 5 1 9 7 3 3 7 6 2 2 6 5 12 0
        0 4 15 3 3 7 9 1 13 6 2 2 6 5 12 0
        0 4 5 1 13 6 7 3 3 7 14 2 6 5 12 0
        0 4 15 3 7 6 13 1 9 7 14 2 6 5 12 0
        0 4 10 2 2 6 13 1 9 7 3 3 11 5 12 0
        0 4 5 1 13 6 2 2 14 7 3 3 11 5 12 0
        0 4 5 1 9 7 14 2 2 6 7 3 11 5 12 0
        0 4 10 2 14 7 9 1 1 3 6 7 3 11 5 12 0
        1 5 12 0 4 7 3 3 7 6 2 2 10 4 5 1
        1 5 11 3 3 7 4 0 8 6 2 2 10 4 5 1
        1 5 12 0 8 6 7 3 3 7 14 2 10 4 5 1
        1 5 11 3 7 6 8 0 4 7 14 2 10 4 5 1
        1 5 6 2 2 6 8 0 4 7 3 3 15 4 5 1
        1 5 12 0 8 6 2 2 14 7 3 3 15 4 5 1
        1 5 12 0 4 7 14 2 2 6 7 3 15 4 5 1
```

```
1 5 6 2 14 7 4 0 8 6 7 3 15 4 5 1
1 5 12 0 8 6 2 2 10 4 15 3 3 7 9 1
1 5 11 3 15 4 10 2 2 6 8 0 4 7 9 1
1 5 6 2 10 4 15 3 7 6 8 0 4 7 9 1
1 5 12 0 8 6 7 3 15 4 10 2 14 7 9 1
1 5 12 0 4 7 3 3 15 4 10 2 2 6 13 1
1 5 12 0 4 7 14 2 10 4 15 3 7 6 13 1
1 5 6 2 10 4 15 3 3 7 4 0 8 6 13 1
1 5 11 3 15 4 10 2 14 7 4 0 8 6 13 1
2 6 8 0 4 7 9 1 5 4 15 3 11 5 6 2
2 6 13 1 5 4 15 3 3 7 4 0 12 5 6 2
2 6 7 3 15 4 5 1 9 7 4 0 12 5 6 2
2 6 8 0 12 5 11 3 3 7 9 1 5 4 10 2
2 6 7 3 11 5 12 0 4 7 9 1 5 4 10 2
2 6 13 1 9 7 4 0 12 5 11 3 15 4 10 2
2 6 13 1 5 4 15 3 11 5 12 0 4 7 14 2
2 6 8 0 12 5 11 3 15 4 5 1 9 7 14 2
3 7 9 1 5 4 10 2 6 5 12 0 8 6 7 3
3 7 4 0 12 5 6 2 10 4 5 1 13 6 7 3
3 7 4 0 8 6 13 1 5 4 10 2 6 5 11 3
3 7 14 2 10 4 5 1 13 6 8 0 12 5 11 3
3 7 14 2 6 5 12 0 8 6 13 1 5 4 15 3
3 7 9 1 13 6 8 0 12 5 6 2 10 4 15 3
4 7 9 1 5 4 10 2 6 5 11 3 7 6 8 0
4 7 14 2 6 5 11 3 15 4 5 1 13 6 8 0
4 7 9 1 13 6 7 3 15 4 10 2 6 5 12 0
4 7 14 2 10 4 5 1 13 6 7 3 11 5 12 0
5 4 15 3 7 6 8 0 12 5 6 2 14 7 9 1
8 6 13 1 9 7 14 2 10 4 15 3 11 5 12 0
```

The column indices contained in each A-(2,2,2) inter-connected cycle (there are 48 A-(2,2,2) inter-connected cycles, each row below corresponds to a A-(2,2,2) inter-connected cycle):

| | | | | | |
|---|---|---|---|---|---|
| 0 | 3 | 4 | 5 | 9 | 15 |
| 0 | 3 | 4 | 10 | 14 | 15 |
| 0 | 3 | 4 | 7 | 8 | 15 |
| 0 | 3 | 4 | 11 | 12 | 15 |
| 0 | 4 | 5 | 9 | 10 | 14 |
| 0 | 4 | 5 | 8 | 9 | 13 |
| 0 | 1 | 4 | 5 | 9 | 12 |
| 0 | 2 | 4 | 8 | 10 | 14 |
| 0 | 4 | 6 | 10 | 12 | 14 |
| 0 | 2 | 7 | 8 | 10 | 15 |
| 0 | 2 | 5 | 8 | 10 | 13 |
| 0 | 2 | 6 | 8 | 10 | 12 |
| 0 | 5 | 7 | 8 | 13 | 15 |
| 0 | 7 | 8 | 11 | 12 | 15 |
| 0 | 1 | 5 | 8 | 12 | 13 |
| 0 | 1 | 5 | 6 | 10 | 12 |
| 0 | 1 | 5 | 11 | 12 | 15 |
| 0 | 6 | 10 | 11 | 12 | 15 |
| 1 | 5 | 6 | 10 | 11 | 15 |
| 1 | 5 | 6 | 9 | 10 | 14 |
| 1 | 2 | 5 | 6 | 10 | 13 |
| 1 | 3 | 5 | 9 | 11 | 15 |
| 1 | 5 | 7 | 11 | 13 | 15 |
| 1 | 3 | 4 | 9 | 11 | 12 |
| 1 | 3 | 6 | 9 | 11 | 14 |
| 1 | 3 | 7 | 9 | 11 | 13 |
| 1 | 4 | 6 | 9 | 12 | 14 |
| 1 | 4 | 8 | 9 | 12 | 13 |
| 1 | 2 | 6 | 9 | 13 | 14 |
| 1 | 2 | 6 | 7 | 11 | 13 |
| 1 | 2 | 6 | 8 | 12 | 13 |
| 1 | 7 | 8 | 11 | 12 | 13 |
| 2 | 6 | 7 | 8 | 11 | 12 |
| 2 | 6 | 7 | 10 | 11 | 15 |
| 2 | 3 | 6 | 7 | 11 | 14 |
| 2 | 4 | 6 | 8 | 12 | 14 |
| 2 | 5 | 7 | 10 | 13 | 15 |
| 2 | 5 | 9 | 10 | 13 | 14 |
| 2 | 3 | 7 | 10 | 14 | 15 |
| 2 | 3 | 4 | 7 | 8 | 14 |
| 2 | 3 | 7 | 9 | 13 | 14 |
| 2 | 4 | 8 | 9 | 13 | 14 |
| 3 | 4 | 7 | 8 | 9 | 13 |
| 3 | 4 | 7 | 8 | 11 | 12 |
| 3 | 5 | 7 | 9 | 13 | 15 |
| 3 | 4 | 6 | 11 | 12 | 14 |
| 3 | 6 | 10 | 11 | 14 | 15 |
| 3 | 5 | 9 | 10 | 14 | 15 |

The column indices contained in 288 A-(1,3,3) inter-connected cycles:

| | | | | | | |
|---|---|---|---|---|---|---|
| 0 | 2 | 3 | 4 | 8 | 10 | 15 |
| 0 | 3 | 4 | 5 | 8 | 13 | 15 |
| 0 | 1 | 3 | 4 | 5 | 12 | 15 |
| 0 | 3 | 4 | 6 | 10 | 12 | 15 |
| 0 | 1 | 3 | 4 | 5 | 11 | 15 |
| 0 | 1 | 3 | 4 | 9 | 11 | 15 |

| 0 | 1 | 3 | 4 | 9 | 12 | 15 |
|---|---|---|---|---|----|----|
| 0 | 2 | 3 | 4 | 7 | 10 | 15 |
| 0 | 2 | 3 | 4 | 7 | 14 | 15 |
| 0 | 2 | 3 | 4 | 8 | 14 | 15 |
| 0 | 3 | 4 | 7 | 9 | 13 | 15 |
| 0 | 3 | 4 | 6 | 11 | 14 | 15 |
| 0 | 3 | 4 | 8 | 9 | 13 | 15 |
| 0 | 3 | 4 | 6 | 12 | 14 | 15 |
| 0 | 3 | 4 | 5 | 7 | 13 | 15 |
| 0 | 3 | 4 | 6 | 10 | 11 | 15 |
| 0 | 2 | 4 | 5 | 8 | 9 | 10 |
| 0 | 4 | 5 | 7 | 8 | 9 | 15 |
| 0 | 4 | 5 | 6 | 9 | 10 | 12 |
| 0 | 4 | 5 | 9 | 11 | 12 | 15 |
| 0 | 1 | 4 | 5 | 6 | 9 | 10 |
| 0 | 1 | 4 | 5 | 9 | 11 | 15 |
| 0 | 1 | 3 | 4 | 5 | 9 | 11 |
| 0 | 1 | 4 | 5 | 6 | 9 | 14 |
| 0 | 2 | 4 | 5 | 9 | 10 | 13 |
| 0 | 2 | 4 | 5 | 8 | 9 | 14 |
| 0 | 2 | 4 | 5 | 9 | 13 | 14 |
| 0 | 3 | 4 | 5 | 7 | 8 | 9 |
| 0 | 3 | 4 | 5 | 7 | 9 | 13 |
| 0 | 3 | 4 | 5 | 9 | 11 | 12 |
| 0 | 4 | 5 | 6 | 9 | 12 | 14 |
| 0 | 4 | 5 | 7 | 9 | 13 | 15 |
| 0 | 4 | 7 | 8 | 10 | 14 | 15 |
| 0 | 4 | 5 | 8 | 10 | 13 | 14 |
| 0 | 1 | 4 | 5 | 10 | 12 | 14 |
| 0 | 4 | 10 | 11 | 12 | 14 | 15 |
| 0 | 1 | 4 | 5 | 6 | 10 | 14 |
| 0 | 1 | 4 | 9 | 10 | 12 | 14 |
| 0 | 1 | 4 | 6 | 9 | 10 | 14 |
| 0 | 2 | 4 | 5 | 10 | 13 | 14 |
| 0 | 2 | 4 | 7 | 10 | 14 | 15 |
| 0 | 2 | 3 | 4 | 7 | 10 | 14 |
| 0 | 2 | 4 | 9 | 10 | 13 | 14 |
| 0 | 3 | 4 | 7 | 8 | 10 | 14 |
| 0 | 3 | 4 | 6 | 10 | 11 | 14 |
| 0 | 3 | 4 | 10 | 11 | 12 | 14 |
| 0 | 4 | 8 | 9 | 10 | 13 | 14 |
| 0 | 4 | 6 | 10 | 11 | 14 | 15 |
| 0 | 1 | 2 | 5 | 8 | 10 | 12 |
| 0 | 2 | 8 | 10 | 11 | 12 | 15 |

| 0 | 1 | 2 | 5 | 6 | 8 | 10 |
|---|---|---|---|---|---|---|
| 0 | 1 | 2 | 6 | 8 | 10 | 13 |
| 0 | 1 | 2 | 8 | 10 | 12 | 13 |
| 0 | 2 | 6 | 7 | 8 | 10 | 11 |
| 0 | 2 | 3 | 7 | 8 | 10 | 14 |
| 0 | 2 | 8 | 9 | 10 | 13 | 14 |
| 0 | 2 | 3 | 4 | 7 | 8 | 10 |
| 0 | 2 | 3 | 8 | 10 | 14 | 15 |
| 0 | 2 | 4 | 8 | 9 | 10 | 13 |
| 0 | 2 | 5 | 8 | 9 | 10 | 14 |
| 0 | 2 | 6 | 8 | 10 | 11 | 15 |
| 0 | 2 | 7 | 8 | 10 | 11 | 12 |
| 0 | 1 | 5 | 7 | 8 | 12 | 15 |
| 0 | 6 | 7 | 8 | 10 | 12 | 15 |
| 0 | 1 | 5 | 7 | 8 | 11 | 15 |
| 0 | 1 | 7 | 8 | 11 | 13 | 15 |
| 0 | 1 | 7 | 8 | 12 | 13 | 15 |
| 0 | 2 | 6 | 7 | 8 | 11 | 15 |
| 0 | 2 | 6 | 7 | 8 | 12 | 15 |
| 0 | 2 | 3 | 7 | 8 | 14 | 15 |
| 0 | 2 | 4 | 7 | 8 | 14 | 15 |
| 0 | 3 | 7 | 8 | 9 | 13 | 15 |
| 0 | 3 | 5 | 7 | 8 | 9 | 15 |
| 0 | 3 | 7 | 8 | 10 | 14 | 15 |
| 0 | 4 | 7 | 8 | 9 | 13 | 15 |
| 0 | 6 | 7 | 8 | 10 | 11 | 15 |
| 0 | 5 | 6 | 8 | 10 | 12 | 13 |
| 0 | 5 | 8 | 11 | 12 | 13 | 15 |
| 0 | 1 | 5 | 6 | 8 | 10 | 13 |
| 0 | 1 | 5 | 8 | 11 | 13 | 15 |
| 0 | 1 | 2 | 5 | 6 | 8 | 13 |
| 0 | 1 | 5 | 7 | 8 | 11 | 13 |
| 0 | 2 | 5 | 6 | 8 | 12 | 13 |
| 0 | 2 | 4 | 5 | 8 | 13 | 14 |
| 0 | 2 | 5 | 8 | 9 | 13 | 14 |
| 0 | 3 | 4 | 5 | 7 | 8 | 13 |
| 0 | 3 | 5 | 7 | 8 | 9 | 13 |
| 0 | 3 | 5 | 8 | 9 | 13 | 15 |
| 0 | 5 | 8 | 9 | 10 | 13 | 14 |
| 0 | 5 | 7 | 8 | 11 | 12 | 13 |
| 0 | 1 | 3 | 5 | 9 | 11 | 12 |
| 0 | 1 | 5 | 6 | 9 | 12 | 14 |
| 0 | 1 | 2 | 5 | 6 | 12 | 13 |
| 0 | 1 | 5 | 7 | 11 | 12 | 13 |

| 0 | 1 | 2 | 5 | 6 | 8 | 12 |
|---|---|---|---|---|---|---|
| 0 | 1 | 2 | 5 | 10 | 12 | 13 |
| 0 | 1 | 3 | 4 | 5 | 11 | 12 |
| 0 | 1 | 3 | 5 | 9 | 12 | 15 |
| 0 | 1 | 4 | 5 | 6 | 12 | 14 |
| 0 | 1 | 5 | 9 | 10 | 12 | 14 |
| 0 | 1 | 5 | 7 | 12 | 13 | 15 |
| 0 | 1 | 5 | 7 | 8 | 11 | 12 |
| 0 | 1 | 4 | 6 | 9 | 10 | 12 |
| 0 | 1 | 6 | 9 | 10 | 12 | 14 |
| 0 | 1 | 2 | 6 | 10 | 12 | 13 |
| 0 | 1 | 6 | 8 | 10 | 12 | 13 |
| 0 | 2 | 6 | 7 | 10 | 11 | 12 |
| 0 | 2 | 5 | 6 | 10 | 12 | 13 |
| 0 | 2 | 6 | 7 | 10 | 12 | 15 |
| 0 | 3 | 6 | 10 | 11 | 12 | 14 |
| 0 | 3 | 4 | 6 | 10 | 11 | 12 |
| 0 | 3 | 6 | 10 | 12 | 14 | 15 |
| 0 | 5 | 6 | 9 | 10 | 12 | 14 |
| 0 | 6 | 7 | 8 | 10 | 11 | 12 |
| 0 | 1 | 3 | 9 | 11 | 12 | 15 |
| 0 | 1 | 4 | 9 | 11 | 12 | 15 |
| 0 | 1 | 7 | 11 | 12 | 13 | 15 |
| 0 | 1 | 8 | 11 | 12 | 13 | 15 |
| 0 | 2 | 6 | 7 | 11 | 12 | 15 |
| 0 | 2 | 6 | 8 | 11 | 12 | 15 |
| 0 | 2 | 7 | 10 | 11 | 12 | 15 |
| 0 | 3 | 6 | 11 | 12 | 14 | 15 |
| 0 | 3 | 5 | 9 | 11 | 12 | 15 |
| 0 | 3 | 10 | 11 | 12 | 14 | 15 |
| 0 | 4 | 6 | 11 | 12 | 14 | 15 |
| 0 | 5 | 7 | 11 | 12 | 13 | 15 |
| 1 | 3 | 5 | 6 | 9 | 10 | 11 |
| 1 | 4 | 5 | 6 | 9 | 10 | 12 |
| 1 | 5 | 6 | 7 | 10 | 11 | 13 |
| 1 | 5 | 6 | 8 | 10 | 12 | 13 |
| 1 | 2 | 5 | 6 | 7 | 10 | 11 |
| 1 | 2 | 5 | 6 | 8 | 10 | 12 |
| 1 | 2 | 5 | 6 | 7 | 10 | 15 |
| 1 | 3 | 5 | 6 | 10 | 11 | 14 |
| 1 | 3 | 5 | 6 | 9 | 10 | 15 |
| 1 | 3 | 5 | 6 | 10 | 14 | 15 |
| 1 | 4 | 5 | 6 | 10 | 12 | 14 |
| 1 | 5 | 6 | 7 | 10 | 13 | 15 |

| 1 | 4 | 5 | 9 | 11 | 12 | 15 |
|---|---|---|---|----|----|----|
| 1 | 5 | 6 | 9 | 11 | 14 | 15 |
| 1 | 2 | 5 | 6 | 11 | 13 | 15 |
| 1 | 5 | 8 | 11 | 12 | 13 | 15 |
| 1 | 2 | 5 | 6 | 7 | 11 | 15 |
| 1 | 2 | 5 | 10 | 11 | 13 | 15 |
| 1 | 2 | 5 | 7 | 10 | 11 | 15 |
| 1 | 3 | 5 | 6 | 11 | 14 | 15 |
| 1 | 3 | 4 | 5 | 11 | 12 | 15 |
| 1 | 3 | 5 | 10 | 11 | 14 | 15 |
| 1 | 5 | 9 | 10 | 11 | 14 | 15 |
| 1 | 5 | 7 | 8 | 11 | 12 | 15 |
| 1 | 2 | 3 | 6 | 9 | 11 | 13 |
| 1 | 3 | 8 | 9 | 11 | 12 | 13 |
| 1 | 2 | 3 | 6 | 7 | 9 | 11 |
| 1 | 2 | 3 | 7 | 9 | 11 | 14 |
| 1 | 2 | 3 | 9 | 11 | 13 | 14 |
| 1 | 3 | 4 | 7 | 8 | 9 | 11 |
| 1 | 3 | 9 | 10 | 11 | 14 | 15 |
| 1 | 3 | 4 | 8 | 9 | 11 | 13 |
| 1 | 3 | 5 | 9 | 10 | 11 | 14 |
| 1 | 3 | 6 | 9 | 10 | 11 | 15 |
| 1 | 3 | 7 | 8 | 9 | 11 | 12 |
| 1 | 2 | 4 | 6 | 9 | 12 | 13 |
| 1 | 4 | 7 | 9 | 11 | 12 | 13 |
| 1 | 2 | 4 | 6 | 8 | 9 | 12 |
| 1 | 2 | 4 | 8 | 9 | 12 | 14 |
| 1 | 2 | 4 | 9 | 12 | 13 | 14 |
| 1 | 3 | 4 | 7 | 8 | 9 | 12 |
| 1 | 3 | 4 | 7 | 9 | 12 | 13 |
| 1 | 3 | 4 | 5 | 9 | 12 | 15 |
| 1 | 4 | 5 | 9 | 10 | 12 | 14 |
| 1 | 4 | 7 | 8 | 9 | 11 | 12 |
| 1 | 6 | 7 | 9 | 11 | 13 | 14 |
| 1 | 6 | 8 | 9 | 12 | 13 | 14 |
| 1 | 2 | 6 | 7 | 9 | 11 | 14 |
| 1 | 2 | 6 | 8 | 9 | 12 | 14 |
| 1 | 2 | 3 | 6 | 7 | 9 | 14 |
| 1 | 2 | 4 | 6 | 8 | 9 | 14 |
| 1 | 3 | 6 | 7 | 9 | 13 | 14 |
| 1 | 3 | 5 | 6 | 9 | 14 | 15 |
| 1 | 3 | 6 | 9 | 10 | 14 | 15 |
| 1 | 4 | 6 | 8 | 9 | 13 | 14 |
| 1 | 6 | 9 | 10 | 11 | 14 | 15 |
| 1 | 2 | 6 | 7 | 10 | 13 | 15 |
| 1 | 2 | 3 | 6 | 7 | 13 | 14 |
| 1 | 2 | 4 | 6 | 8 | 13 | 14 |
| 1 | 2 | 3 | 6 | 7 | 9 | 13 |
| 1 | 2 | 3 | 6 | 11 | 13 | 14 |
| 1 | 2 | 4 | 6 | 8 | 9 | 13 |
| 1 | 2 | 4 | 6 | 12 | 13 | 14 |
| 1 | 2 | 5 | 6 | 7 | 13 | 15 |
| 1 | 2 | 6 | 10 | 11 | 13 | 15 |
| 1 | 2 | 5 | 7 | 10 | 11 | 13 |
| 1 | 2 | 7 | 10 | 11 | 13 | 15 |
| 1 | 2 | 3 | 7 | 11 | 13 | 14 |
| 1 | 2 | 7 | 9 | 11 | 13 | 14 |
| 1 | 3 | 4 | 7 | 8 | 11 | 13 |
| 1 | 3 | 6 | 7 | 11 | 13 | 14 |
| 1 | 3 | 4 | 7 | 11 | 12 | 13 |
| 1 | 4 | 7 | 8 | 9 | 11 | 13 |
| 1 | 6 | 7 | 10 | 11 | 13 | 15 |
| 1 | 2 | 5 | 8 | 10 | 12 | 13 |
| 1 | 2 | 4 | 8 | 12 | 13 | 14 |
| 1 | 2 | 8 | 9 | 12 | 13 | 14 |
| 1 | 3 | 4 | 7 | 8 | 12 | 13 |
| 1 | 3 | 7 | 8 | 9 | 12 | 13 |
| 1 | 3 | 4 | 8 | 11 | 12 | 13 |
| 1 | 4 | 6 | 8 | 12 | 13 | 14 |
| 1 | 5 | 7 | 8 | 12 | 13 | 15 |
| 2 | 5 | 6 | 7 | 10 | 11 | 13 |
| 2 | 4 | 6 | 7 | 8 | 11 | 14 |
| 2 | 6 | 7 | 9 | 11 | 13 | 14 |
| 2 | 3 | 4 | 6 | 7 | 8 | 11 |
| 2 | 3 | 6 | 7 | 9 | 11 | 13 |
| 2 | 3 | 4 | 6 | 7 | 11 | 12 |
| 2 | 4 | 6 | 7 | 11 | 12 | 14 |
| 2 | 5 | 6 | 7 | 11 | 13 | 15 |
| 2 | 5 | 6 | 8 | 10 | 12 | 13 |
| 2 | 6 | 7 | 8 | 10 | 12 | 15 |
| 2 | 3 | 6 | 7 | 8 | 12 | 14 |
| 2 | 6 | 8 | 9 | 12 | 13 | 14 |
| 2 | 3 | 4 | 6 | 7 | 8 | 12 |
| 2 | 3 | 6 | 8 | 11 | 12 | 14 |
| 2 | 3 | 4 | 6 | 8 | 11 | 12 |
| 2 | 4 | 6 | 8 | 9 | 12 | 13 |
| 2 | 6 | 8 | 10 | 11 | 12 | 15 |
| 2 | 3 | 5 | 7 | 10 | 13 | 14 |

| 2 | 4 | 5 | 8 | 10 | 13 | 14 |
|---|---|---|---|---|---|---|
| 2 | 3 | 5 | 7 | 9 | 10 | 13 |
| 2 | 3 | 5 | 9 | 10 | 13 | 15 |
| 2 | 3 | 5 | 10 | 13 | 14 | 15 |
| 2 | 4 | 5 | 8 | 9 | 10 | 13 |
| 2 | 5 | 6 | 10 | 11 | 13 | 15 |
| 2 | 4 | 7 | 8 | 10 | 14 | 15 |
| 2 | 7 | 9 | 10 | 13 | 14 | 15 |
| 2 | 3 | 4 | 7 | 8 | 10 | 15 |
| 2 | 3 | 7 | 9 | 10 | 13 | 15 |
| 2 | 3 | 5 | 7 | 9 | 10 | 15 |
| 2 | 5 | 7 | 9 | 10 | 14 | 15 |
| 2 | 7 | 8 | 10 | 11 | 12 | 15 |
| 2 | 3 | 4 | 7 | 11 | 12 | 14 |
| 2 | 3 | 5 | 7 | 9 | 14 | 15 |
| 2 | 3 | 4 | 6 | 7 | 12 | 14 |
| 2 | 3 | 5 | 7 | 9 | 10 | 14 |
| 2 | 3 | 5 | 7 | 13 | 14 | 15 |
| 2 | 3 | 7 | 8 | 11 | 12 | 14 |
| 2 | 3 | 4 | 6 | 8 | 11 | 14 |
| 2 | 3 | 4 | 8 | 11 | 12 | 14 |
| 2 | 3 | 4 | 8 | 10 | 14 | 15 |
| 2 | 4 | 5 | 8 | 9 | 10 | 14 |
| 2 | 4 | 7 | 8 | 11 | 12 | 14 |
| 2 | 3 | 6 | 9 | 11 | 13 | 14 |
| 2 | 3 | 5 | 9 | 13 | 14 | 15 |
| 2 | 3 | 9 | 10 | 13 | 14 | 15 |
| 2 | 4 | 6 | 9 | 12 | 13 | 14 |
| 2 | 5 | 7 | 9 | 13 | 14 | 15 |
| 3 | 4 | 6 | 7 | 8 | 11 | 14 |
| 3 | 4 | 5 | 7 | 8 | 9 | 15 |
| 3 | 4 | 7 | 8 | 10 | 14 | 15 |
| 3 | 4 | 6 | 7 | 8 | 12 | 14 |
| 3 | 4 | 5 | 7 | 8 | 13 | 15 |
| 3 | 6 | 7 | 9 | 11 | 13 | 14 |
| 3 | 4 | 7 | 9 | 11 | 12 | 13 |
| 3 | 7 | 9 | 10 | 13 | 14 | 15 |
| 3 | 5 | 7 | 9 | 10 | 13 | 14 |
| 3 | 7 | 8 | 9 | 11 | 12 | 13 |
| 3 | 5 | 6 | 9 | 11 | 14 | 15 |
| 3 | 5 | 6 | 9 | 10 | 11 | 14 |
| 3 | 6 | 7 | 8 | 11 | 12 | 14 |
| 3 | 4 | 5 | 9 | 11 | 12 | 15 |
| 3 | 4 | 10 | 11 | 12 | 14 | 15 |
| 3 | 4 | 8 | 9 | 11 | 12 | 13 |
| 3 | 4 | 6 | 10 | 11 | 12 | 15 |
| 3 | 4 | 5 | 8 | 9 | 13 | 15 |
| 3 | 5 | 6 | 9 | 10 | 11 | 15 |
| 3 | 4 | 6 | 10 | 12 | 14 | 15 |
| 3 | 5 | 7 | 10 | 13 | 14 | 15 |
| 4 | 6 | 8 | 9 | 12 | 13 | 14 |
| 4 | 5 | 8 | 9 | 10 | 13 | 14 |
| 4 | 5 | 7 | 8 | 9 | 13 | 15 |
| 4 | 7 | 8 | 9 | 11 | 12 | 13 |
| 4 | 5 | 6 | 9 | 10 | 12 | 14 |
| 4 | 6 | 10 | 11 | 12 | 14 | 15 |
| 4 | 6 | 7 | 8 | 11 | 12 | 14 |
| 5 | 7 | 9 | 10 | 13 | 14 | 15 |
| 5 | 6 | 9 | 10 | 11 | 14 | 15 |
| 5 | 6 | 7 | 10 | 11 | 13 | 15 |
| 5 | 7 | 8 | 11 | 12 | 13 | 15 |
| 6 | 7 | 8 | 10 | 11 | 12 | 15 |

The column indices contained in 288 A-(2,2,4) inter-connected cycles:

| | | | | | | | |
|---|---|---|---|---|---|---|---|
| 0 | 2 | 3 | 4 | 5 | 7 | 10 | 13 |
| 0 | 1 | 3 | 4 | 5 | 7 | 11 | 13 |
| 0 | 2 | 3 | 4 | 5 | 7 | 13 | 14 |
| 0 | 3 | 4 | 5 | 7 | 11 | 12 | 13 |
| 0 | 1 | 3 | 4 | 5 | 7 | 12 | 13 |
| 0 | 3 | 4 | 5 | 7 | 10 | 13 | 14 |
| 0 | 2 | 3 | 4 | 6 | 7 | 10 | 11 |
| 0 | 2 | 3 | 4 | 7 | 9 | 10 | 13 |
| 0 | 2 | 3 | 4 | 7 | 10 | 11 | 12 |
| 0 | 2 | 3 | 4 | 6 | 7 | 10 | 12 |
| 0 | 2 | 3 | 4 | 5 | 7 | 9 | 10 |
| 0 | 1 | 3 | 4 | 5 | 6 | 10 | 11 |
| 0 | 1 | 3 | 4 | 5 | 6 | 11 | 14 |
| 0 | 1 | 3 | 4 | 5 | 7 | 8 | 11 |
| 0 | 1 | 3 | 4 | 5 | 10 | 11 | 14 |
| 0 | 1 | 3 | 4 | 5 | 8 | 11 | 13 |
| 0 | 1 | 3 | 4 | 6 | 9 | 10 | 11 |
| 0 | 3 | 4 | 6 | 7 | 8 | 10 | 11 |
| 0 | 2 | 3 | 4 | 6 | 8 | 10 | 11 |
| 0 | 3 | 4 | 5 | 6 | 9 | 10 | 11 |
| 0 | 1 | 4 | 6 | 9 | 10 | 11 | 15 |
| 0 | 1 | 2 | 4 | 6 | 9 | 10 | 13 |
| 0 | 1 | 4 | 6 | 8 | 9 | 10 | 13 |
| 0 | 1 | 3 | 4 | 6 | 9 | 10 | 15 |

| 0 | 1 | 2 | 4 | 6 | 8 | 9 | 10 |
|---|---|---|---|---|---|---|---|
| 0 | 1 | 4 | 7 | 9 | 11 | 13 | 15 |
| 0 | 1 | 4 | 6 | 9 | 11 | 14 | 15 |
| 0 | 1 | 4 | 8 | 9 | 11 | 13 | 15 |
| 0 | 1 | 4 | 7 | 8 | 9 | 11 | 15 |
| 0 | 1 | 4 | 9 | 10 | 11 | 14 | 15 |
| 0 | 2 | 4 | 7 | 9 | 10 | 13 | 15 |
| 0 | 1 | 2 | 4 | 9 | 10 | 12 | 13 |
| 0 | 2 | 4 | 6 | 9 | 10 | 12 | 13 |
| 0 | 2 | 3 | 4 | 9 | 10 | 13 | 15 |
| 0 | 2 | 4 | 7 | 9 | 13 | 14 | 15 |
| 0 | 1 | 4 | 7 | 9 | 12 | 13 | 15 |
| 0 | 4 | 7 | 9 | 11 | 12 | 13 | 15 |
| 0 | 4 | 7 | 9 | 10 | 13 | 14 | 15 |
| 0 | 2 | 4 | 5 | 7 | 13 | 14 | 15 |
| 0 | 1 | 2 | 4 | 5 | 6 | 13 | 14 |
| 0 | 2 | 4 | 5 | 6 | 12 | 13 | 14 |
| 0 | 1 | 2 | 4 | 5 | 12 | 13 | 14 |
| 0 | 2 | 3 | 4 | 5 | 13 | 14 | 15 |
| 0 | 2 | 4 | 6 | 7 | 11 | 14 | 15 |
| 0 | 2 | 4 | 6 | 7 | 12 | 14 | 15 |
| 0 | 2 | 4 | 7 | 11 | 12 | 14 | 15 |
| 0 | 2 | 4 | 5 | 7 | 9 | 14 | 15 |
| 0 | 1 | 4 | 5 | 6 | 11 | 14 | 15 |
| 0 | 1 | 2 | 4 | 5 | 6 | 8 | 14 |
| 0 | 1 | 3 | 4 | 5 | 6 | 14 | 15 |
| 0 | 1 | 4 | 5 | 6 | 8 | 13 | 14 |
| 0 | 2 | 4 | 6 | 8 | 11 | 14 | 15 |
| 0 | 4 | 6 | 7 | 8 | 11 | 14 | 15 |
| 0 | 4 | 5 | 6 | 9 | 11 | 14 | 15 |
| 0 | 1 | 2 | 5 | 6 | 8 | 11 | 15 |
| 0 | 1 | 2 | 5 | 6 | 8 | 9 | 14 |
| 0 | 1 | 2 | 5 | 6 | 7 | 8 | 11 |
| 0 | 1 | 2 | 5 | 6 | 7 | 8 | 15 |
| 0 | 1 | 2 | 4 | 5 | 6 | 8 | 9 |
| 0 | 2 | 3 | 6 | 8 | 11 | 14 | 15 |
| 0 | 1 | 2 | 6 | 8 | 11 | 13 | 15 |
| 0 | 2 | 3 | 4 | 6 | 8 | 11 | 15 |
| 0 | 2 | 5 | 6 | 8 | 11 | 13 | 15 |
| 0 | 2 | 3 | 5 | 8 | 9 | 14 | 15 |
| 0 | 2 | 3 | 5 | 7 | 8 | 9 | 14 |
| 0 | 2 | 5 | 6 | 8 | 9 | 12 | 14 |
| 0 | 1 | 2 | 5 | 8 | 9 | 12 | 14 |
| 0 | 2 | 5 | 7 | 8 | 9 | 14 | 15 |

| 0 | 2 | 3 | 8 | 9 | 13 | 14 | 15 |
| 0 | 2 | 3 | 6 | 8 | 12 | 14 | 15 |
| 0 | 2 | 3 | 8 | 11 | 12 | 14 | 15 |
| 0 | 2 | 3 | 5 | 8 | 13 | 14 | 15 |
| 0 | 3 | 5 | 7 | 8 | 9 | 10 | 14 |
| 0 | 1 | 3 | 5 | 7 | 8 | 9 | 11 |
| 0 | 3 | 5 | 7 | 8 | 9 | 11 | 12 |
| 0 | 1 | 3 | 5 | 7 | 8 | 9 | 12 |
| 0 | 2 | 3 | 5 | 7 | 8 | 9 | 10 |
| 0 | 3 | 6 | 7 | 8 | 10 | 11 | 14 |
| 0 | 3 | 7 | 8 | 9 | 10 | 13 | 14 |
| 0 | 3 | 7 | 8 | 10 | 11 | 12 | 14 |
| 0 | 3 | 6 | 7 | 8 | 10 | 12 | 14 |
| 0 | 3 | 5 | 7 | 8 | 10 | 13 | 14 |
| 0 | 1 | 5 | 6 | 7 | 8 | 10 | 11 |
| 0 | 1 | 2 | 5 | 7 | 8 | 10 | 11 |
| 0 | 1 | 4 | 5 | 7 | 8 | 9 | 11 |
| 0 | 1 | 6 | 7 | 8 | 10 | 11 | 13 |
| 0 | 4 | 6 | 7 | 8 | 10 | 11 | 14 |
| 0 | 5 | 6 | 7 | 8 | 10 | 11 | 13 |
| 0 | 1 | 6 | 8 | 10 | 11 | 13 | 15 |
| 0 | 1 | 6 | 8 | 9 | 10 | 13 | 14 |
| 0 | 1 | 6 | 7 | 8 | 10 | 13 | 15 |
| 0 | 1 | 4 | 6 | 8 | 10 | 13 | 14 |
| 0 | 1 | 3 | 8 | 9 | 11 | 13 | 15 |
| 0 | 1 | 2 | 8 | 10 | 11 | 13 | 15 |
| 0 | 1 | 3 | 4 | 8 | 11 | 13 | 15 |
| 0 | 3 | 8 | 9 | 10 | 13 | 14 | 15 |
| 0 | 1 | 8 | 9 | 10 | 12 | 13 | 14 |
| 0 | 6 | 8 | 9 | 10 | 12 | 13 | 14 |
| 0 | 7 | 8 | 9 | 10 | 13 | 14 | 15 |
| 0 | 1 | 3 | 8 | 9 | 12 | 13 | 15 |
| 0 | 2 | 3 | 8 | 9 | 10 | 13 | 15 |
| 0 | 3 | 8 | 9 | 11 | 12 | 13 | 15 |
| 0 | 1 | 3 | 9 | 10 | 12 | 14 | 15 |
| 0 | 1 | 2 | 9 | 10 | 12 | 13 | 14 |
| 0 | 1 | 3 | 9 | 10 | 11 | 12 | 14 |
| 0 | 1 | 2 | 8 | 9 | 10 | 12 | 14 |
| 0 | 1 | 9 | 10 | 11 | 12 | 14 | 15 |
| 0 | 1 | 3 | 7 | 9 | 12 | 13 | 15 |
| 0 | 1 | 3 | 6 | 9 | 12 | 14 | 15 |
| 0 | 1 | 3 | 7 | 8 | 9 | 12 | 15 |
| 0 | 1 | 3 | 6 | 9 | 10 | 12 | 15 |
| 0 | 1 | 2 | 7 | 10 | 12 | 13 | 15 |
| 0 | 1 | 2 | 7 | 10 | 11 | 12 | 13 |
| 0 | 1 | 2 | 10 | 11 | 12 | 13 | 15 |
| 0 | 1 | 2 | 4 | 10 | 12 | 13 | 14 |
| 0 | 1 | 2 | 6 | 7 | 12 | 13 | 15 |
| 0 | 1 | 3 | 4 | 7 | 12 | 13 | 15 |
| 0 | 1 | 6 | 7 | 10 | 12 | 13 | 15 |
| 0 | 2 | 5 | 6 | 7 | 12 | 13 | 15 |
| 0 | 2 | 5 | 6 | 9 | 12 | 13 | 14 |
| 0 | 2 | 5 | 6 | 7 | 11 | 12 | 13 |
| 0 | 2 | 5 | 6 | 11 | 12 | 13 | 15 |
| 0 | 2 | 4 | 5 | 6 | 9 | 12 | 13 |
| 0 | 2 | 3 | 6 | 7 | 12 | 14 | 15 |
| 0 | 1 | 2 | 5 | 6 | 7 | 12 | 15 |
| 0 | 2 | 3 | 4 | 6 | 7 | 12 | 15 |
| 0 | 3 | 5 | 6 | 9 | 12 | 14 | 15 |
| 0 | 3 | 5 | 6 | 9 | 11 | 12 | 14 |
| 0 | 5 | 6 | 9 | 11 | 12 | 14 | 15 |
| 0 | 5 | 6 | 8 | 9 | 12 | 13 | 14 |
| 0 | 1 | 3 | 5 | 6 | 12 | 14 | 15 |
| 0 | 3 | 6 | 7 | 8 | 12 | 14 | 15 |
| 0 | 3 | 5 | 9 | 10 | 11 | 12 | 14 |
| 0 | 3 | 5 | 7 | 9 | 11 | 12 | 13 |
| 0 | 3 | 5 | 6 | 9 | 10 | 11 | 12 |
| 0 | 3 | 5 | 8 | 9 | 11 | 12 | 13 |
| 0 | 2 | 3 | 7 | 10 | 11 | 12 | 14 |
| 0 | 1 | 3 | 5 | 10 | 11 | 12 | 14 |
| 0 | 2 | 3 | 8 | 10 | 11 | 12 | 14 |
| 0 | 2 | 5 | 7 | 10 | 11 | 12 | 13 |
| 0 | 4 | 5 | 7 | 9 | 11 | 12 | 13 |
| 0 | 5 | 6 | 7 | 10 | 11 | 12 | 13 |
| 0 | 1 | 2 | 5 | 7 | 10 | 11 | 12 |
| 0 | 2 | 4 | 7 | 10 | 11 | 12 | 14 |
| 1 | 2 | 5 | 7 | 8 | 10 | 11 | 12 |
| 1 | 2 | 3 | 5 | 7 | 10 | 11 | 14 |
| 1 | 2 | 5 | 7 | 9 | 10 | 11 | 14 |
| 1 | 2 | 3 | 5 | 7 | 9 | 10 | 11 |
| 1 | 2 | 4 | 5 | 8 | 10 | 12 | 14 |
| 1 | 2 | 5 | 7 | 8 | 10 | 12 | 15 |
| 1 | 2 | 5 | 8 | 9 | 10 | 12 | 14 |
| 1 | 2 | 4 | 5 | 8 | 9 | 10 | 12 |
| 1 | 2 | 5 | 8 | 10 | 11 | 12 | 15 |
| 1 | 3 | 4 | 5 | 10 | 11 | 12 | 14 |
| 1 | 2 | 3 | 5 | 10 | 11 | 13 | 14 |
| 1 | 3 | 5 | 7 | 10 | 11 | 13 | 14 |

| 1 | 3 | 4 | 5 | 10 | 12 | 14 | 15 |
| 1 | 2 | 4 | 5 | 10 | 12 | 13 | 14 |
| 1 | 4 | 5 | 8 | 10 | 12 | 13 | 14 |
| 1 | 4 | 5 | 10 | 11 | 12 | 14 | 15 |
| 1 | 3 | 4 | 5 | 6 | 12 | 14 | 15 |
| 1 | 2 | 3 | 5 | 6 | 7 | 14 | 15 |
| 1 | 3 | 5 | 6 | 7 | 13 | 14 | 15 |
| 1 | 2 | 3 | 5 | 6 | 13 | 14 | 15 |
| 1 | 3 | 4 | 5 | 7 | 8 | 12 | 15 |
| 1 | 3 | 4 | 5 | 7 | 12 | 13 | 15 |
| 1 | 3 | 4 | 5 | 8 | 12 | 13 | 15 |
| 1 | 3 | 4 | 5 | 6 | 10 | 12 | 15 |
| 1 | 2 | 5 | 6 | 7 | 8 | 12 | 15 |
| 1 | 2 | 3 | 5 | 6 | 7 | 9 | 15 |
| 1 | 2 | 5 | 6 | 7 | 9 | 14 | 15 |
| 1 | 3 | 5 | 7 | 8 | 9 | 12 | 15 |
| 1 | 4 | 5 | 7 | 8 | 9 | 12 | 15 |
| 1 | 5 | 6 | 7 | 8 | 10 | 12 | 15 |
| 1 | 2 | 3 | 6 | 7 | 8 | 9 | 12 |
| 1 | 2 | 3 | 6 | 7 | 9 | 10 | 15 |
| 1 | 2 | 3 | 4 | 6 | 7 | 8 | 9 |
| 1 | 2 | 3 | 4 | 6 | 7 | 9 | 12 |
| 1 | 2 | 3 | 5 | 6 | 7 | 9 | 10 |
| 1 | 2 | 3 | 7 | 8 | 9 | 12 | 14 |
| 1 | 3 | 6 | 7 | 8 | 9 | 12 | 14 |
| 1 | 3 | 6 | 7 | 9 | 10 | 13 | 15 |
| 1 | 2 | 3 | 6 | 9 | 10 | 13 | 15 |
| 1 | 3 | 4 | 6 | 9 | 10 | 12 | 15 |
| 1 | 2 | 4 | 6 | 7 | 8 | 9 | 11 |
| 1 | 2 | 3 | 4 | 6 | 8 | 9 | 11 |
| 1 | 2 | 4 | 5 | 6 | 8 | 9 | 10 |
| 1 | 2 | 4 | 7 | 8 | 9 | 11 | 14 |
| 1 | 4 | 6 | 7 | 8 | 9 | 11 | 14 |
| 1 | 4 | 5 | 7 | 8 | 9 | 11 | 15 |
| 1 | 2 | 7 | 8 | 9 | 11 | 12 | 14 |
| 1 | 2 | 7 | 9 | 10 | 11 | 14 | 15 |
| 1 | 2 | 4 | 7 | 9 | 11 | 12 | 14 |
| 1 | 2 | 5 | 7 | 9 | 11 | 14 | 15 |
| 1 | 2 | 3 | 8 | 9 | 11 | 12 | 14 |
| 1 | 2 | 9 | 10 | 11 | 13 | 14 | 15 |
| 1 | 4 | 9 | 10 | 11 | 12 | 14 | 15 |
| 1 | 7 | 9 | 10 | 11 | 13 | 14 | 15 |
| 1 | 2 | 3 | 10 | 11 | 13 | 14 | 15 |
| 1 | 2 | 3 | 9 | 10 | 11 | 13 | 15 |

| 1 | 2 | 8 | 10 | 11 | 12 | 13 | 15 |
|---|---|---|----|----|----|----|----|
| 1 | 2 | 3 | 4 | 11 | 12 | 13 | 14 |
| 1 | 2 | 3 | 4 | 8 | 11 | 13 | 14 |
| 1 | 2 | 3 | 8 | 11 | 12 | 13 | 14 |
| 1 | 2 | 3 | 5 | 11 | 13 | 14 | 15 |
| 1 | 2 | 3 | 4 | 7 | 12 | 13 | 14 |
| 1 | 2 | 4 | 7 | 11 | 12 | 13 | 14 |
| 1 | 3 | 4 | 6 | 7 | 12 | 13 | 14 |
| 1 | 3 | 6 | 7 | 10 | 13 | 14 | 15 |
| 1 | 3 | 4 | 6 | 7 | 8 | 13 | 14 |
| 1 | 3 | 6 | 7 | 8 | 12 | 13 | 14 |
| 1 | 3 | 5 | 6 | 7 | 10 | 13 | 14 |
| 1 | 2 | 3 | 4 | 6 | 7 | 12 | 13 |
| 1 | 6 | 7 | 8 | 10 | 12 | 13 | 15 |
| 1 | 6 | 7 | 9 | 10 | 13 | 14 | 15 |
| 1 | 3 | 4 | 6 | 8 | 11 | 13 | 14 |
| 1 | 4 | 6 | 7 | 8 | 11 | 13 | 14 |
| 1 | 4 | 5 | 6 | 8 | 10 | 13 | 14 |
| 1 | 2 | 3 | 4 | 6 | 8 | 11 | 13 |
| 1 | 3 | 4 | 5 | 8 | 11 | 13 | 15 |
| 2 | 3 | 4 | 6 | 8 | 9 | 11 | 13 |
| 2 | 3 | 4 | 6 | 8 | 10 | 11 | 15 |
| 2 | 3 | 5 | 6 | 9 | 11 | 13 | 15 |
| 2 | 3 | 4 | 6 | 9 | 11 | 12 | 13 |
| 2 | 3 | 6 | 9 | 10 | 11 | 13 | 15 |
| 2 | 3 | 5 | 6 | 9 | 10 | 11 | 13 |
| 2 | 3 | 6 | 8 | 9 | 11 | 12 | 13 |
| 2 | 3 | 5 | 6 | 11 | 13 | 14 | 15 |
| 2 | 5 | 6 | 9 | 11 | 13 | 14 | 15 |
| 2 | 5 | 6 | 8 | 11 | 12 | 13 | 15 |
| 2 | 3 | 4 | 6 | 7 | 9 | 12 | 13 |
| 2 | 3 | 4 | 6 | 7 | 10 | 12 | 15 |
| 2 | 4 | 5 | 6 | 9 | 10 | 12 | 13 |
| 2 | 4 | 6 | 7 | 9 | 11 | 12 | 13 |
| 2 | 3 | 4 | 5 | 7 | 8 | 9 | 10 |
| 2 | 3 | 5 | 6 | 7 | 9 | 10 | 11 |
| 2 | 3 | 4 | 5 | 8 | 9 | 10 | 15 |
| 2 | 4 | 5 | 7 | 8 | 9 | 10 | 15 |
| 2 | 4 | 5 | 6 | 8 | 9 | 10 | 12 |
| 2 | 3 | 4 | 8 | 9 | 10 | 13 | 15 |
| 2 | 3 | 4 | 8 | 10 | 11 | 12 | 15 |
| 2 | 3 | 4 | 5 | 8 | 10 | 13 | 15 |
| 2 | 3 | 4 | 6 | 8 | 10 | 12 | 15 |
| 2 | 3 | 8 | 10 | 11 | 12 | 14 | 15 |
| 2 | 4 | 8 | 10 | 11 | 12 | 14 | 15 |
| 2 | 5 | 8 | 10 | 11 | 12 | 13 | 15 |
| 2 | 3 | 8 | 9 | 11 | 12 | 13 | 14 |
| 2 | 3 | 4 | 5 | 8 | 13 | 14 | 15 |
| 2 | 4 | 7 | 9 | 11 | 12 | 13 | 14 |
| 2 | 4 | 7 | 10 | 11 | 12 | 14 | 15 |
| 2 | 4 | 5 | 7 | 8 | 9 | 14 | 15 |
| 2 | 5 | 6 | 7 | 9 | 11 | 14 | 15 |
| 3 | 6 | 7 | 8 | 10 | 12 | 14 | 15 |
| 3 | 6 | 7 | 8 | 9 | 12 | 13 | 14 |
| 3 | 4 | 5 | 7 | 8 | 10 | 13 | 14 |
| 3 | 5 | 6 | 7 | 10 | 11 | 13 | 14 |
| 3 | 4 | 5 | 6 | 9 | 10 | 11 | 12 |
| 3 | 5 | 6 | 7 | 9 | 10 | 11 | 13 |
| 3 | 5 | 8 | 9 | 11 | 12 | 13 | 15 |
| 3 | 6 | 8 | 9 | 11 | 12 | 13 | 14 |
| 3 | 4 | 5 | 8 | 10 | 13 | 14 | 15 |
| 3 | 4 | 5 | 8 | 11 | 12 | 13 | 15 |
| 3 | 4 | 5 | 6 | 9 | 10 | 12 | 15 |
| 3 | 4 | 6 | 7 | 8 | 10 | 12 | 15 |
| 4 | 6 | 7 | 8 | 10 | 11 | 14 | 15 |
| 4 | 6 | 7 | 8 | 9 | 11 | 13 | 14 |
| 4 | 5 | 7 | 8 | 9 | 10 | 14 | 15 |
| 4 | 5 | 7 | 8 | 9 | 11 | 12 | 15 |
| 4 | 5 | 7 | 8 | 10 | 13 | 14 | 15 |
| 4 | 7 | 8 | 10 | 11 | 12 | 14 | 15 |
| 4 | 6 | 7 | 8 | 10 | 12 | 14 | 15 |
| 4 | 7 | 8 | 9 | 10 | 13 | 14 | 15 |
| 4 | 5 | 6 | 8 | 10 | 12 | 13 | 14 |
| 4 | 5 | 6 | 9 | 10 | 11 | 12 | 15 |
| 4 | 5 | 6 | 8 | 9 | 10 | 12 | 13 |
| 4 | 5 | 7 | 9 | 11 | 12 | 13 | 15 |
| 4 | 6 | 7 | 9 | 11 | 12 | 13 | 14 |
| 4 | 5 | 9 | 10 | 11 | 12 | 14 | 15 |
| 4 | 5 | 6 | 9 | 11 | 12 | 14 | 15 |
| 4 | 5 | 8 | 9 | 11 | 12 | 13 | 15 |
| 5 | 6 | 7 | 9 | 11 | 13 | 14 | 15 |
| 5 | 6 | 7 | 8 | 10 | 11 | 12 | 13 |
| 5 | 6 | 7 | 9 | 10 | 11 | 13 | 14 |
| 5 | 6 | 8 | 10 | 11 | 12 | 13 | 15 |
| 5 | 6 | 7 | 8 | 10 | 12 | 13 | 15 |
| 5 | 6 | 8 | 9 | 10 | 12 | 13 | 14 |
| 6 | 7 | 8 | 9 | 11 | 12 | 13 | 14 |
| 6 | 7 | 9 | 10 | 11 | 13 | 14 | 15 |

The column indices contained in 144 B-(4,4) inter-connected cycles:

| | | | | | | | |
|---|---|---|---|---|---|---|---|
| 0 | 1 | 3 | 4 | 5 | 6 | 10 | 15 |
| 0 | 1 | 3 | 4 | 6 | 9 | 14 | 15 |
| 0 | 1 | 3 | 4 | 7 | 11 | 13 | 15 |
| 0 | 1 | 3 | 4 | 8 | 12 | 13 | 15 |
| 0 | 2 | 3 | 4 | 6 | 7 | 11 | 15 |
| 0 | 2 | 3 | 4 | 6 | 8 | 12 | 15 |
| 0 | 2 | 3 | 4 | 5 | 10 | 13 | 15 |
| 0 | 2 | 3 | 4 | 9 | 13 | 14 | 15 |
| 0 | 1 | 2 | 4 | 5 | 6 | 9 | 13 |
| 0 | 1 | 4 | 5 | 7 | 9 | 11 | 13 |
| 0 | 2 | 4 | 5 | 6 | 8 | 9 | 12 |
| 0 | 2 | 4 | 5 | 7 | 9 | 10 | 15 |
| 0 | 2 | 3 | 4 | 5 | 7 | 9 | 14 |
| 0 | 3 | 4 | 5 | 6 | 9 | 11 | 14 |
| 0 | 4 | 5 | 6 | 9 | 10 | 11 | 15 |
| 0 | 4 | 5 | 7 | 8 | 9 | 11 | 12 |
| 0 | 1 | 4 | 5 | 10 | 11 | 14 | 15 |
| 0 | 1 | 3 | 4 | 9 | 10 | 11 | 14 |
| 0 | 1 | 2 | 4 | 6 | 10 | 13 | 14 |
| 0 | 1 | 4 | 8 | 10 | 12 | 13 | 14 |
| 0 | 2 | 4 | 6 | 7 | 10 | 11 | 14 |
| 0 | 3 | 4 | 7 | 9 | 10 | 13 | 14 |
| 0 | 4 | 5 | 7 | 10 | 13 | 14 | 15 |
| 0 | 4 | 7 | 8 | 10 | 11 | 12 | 14 |
| 0 | 1 | 2 | 5 | 8 | 10 | 11 | 15 |
| 0 | 1 | 2 | 4 | 8 | 9 | 10 | 12 |
| 0 | 1 | 2 | 6 | 8 | 9 | 10 | 14 |
| 0 | 1 | 2 | 7 | 8 | 10 | 11 | 13 |
| 0 | 2 | 3 | 7 | 8 | 9 | 10 | 13 |
| 0 | 2 | 3 | 6 | 8 | 10 | 11 | 14 |
| 0 | 2 | 3 | 4 | 8 | 10 | 11 | 12 |
| 0 | 2 | 3 | 5 | 8 | 9 | 10 | 15 |
| 0 | 1 | 5 | 6 | 7 | 8 | 10 | 15 |
| 0 | 1 | 3 | 7 | 8 | 9 | 11 | 15 |
| 0 | 1 | 4 | 7 | 8 | 9 | 12 | 15 |
| 0 | 1 | 2 | 6 | 7 | 8 | 13 | 15 |
| 0 | 2 | 7 | 8 | 9 | 13 | 14 | 15 |
| 0 | 3 | 6 | 7 | 8 | 11 | 14 | 15 |
| 0 | 4 | 6 | 7 | 8 | 12 | 14 | 15 |
| 0 | 5 | 7 | 8 | 9 | 10 | 14 | 15 |
| 0 | 1 | 3 | 5 | 8 | 9 | 11 | 13 |
| 0 | 1 | 5 | 6 | 8 | 9 | 13 | 14 |
| 0 | 2 | 5 | 6 | 7 | 8 | 11 | 13 |
| 0 | 2 | 3 | 5 | 7 | 8 | 13 | 14 |
| 0 | 3 | 4 | 5 | 8 | 11 | 12 | 13 |
| 0 | 3 | 5 | 8 | 10 | 13 | 14 | 15 |
| 0 | 4 | 5 | 6 | 8 | 12 | 13 | 14 |
| 0 | 5 | 6 | 8 | 10 | 11 | 13 | 15 |
| 0 | 1 | 2 | 5 | 6 | 7 | 11 | 12 |
| 0 | 1 | 2 | 5 | 7 | 10 | 12 | 15 |
| 0 | 1 | 2 | 4 | 5 | 8 | 12 | 14 |
| 0 | 1 | 2 | 5 | 9 | 12 | 13 | 14 |
| 0 | 1 | 3 | 4 | 5 | 7 | 8 | 12 |
| 0 | 1 | 3 | 5 | 7 | 9 | 12 | 13 |
| 0 | 1 | 3 | 5 | 6 | 11 | 12 | 14 |
| 0 | 1 | 3 | 5 | 10 | 12 | 14 | 15 |
| 0 | 1 | 3 | 6 | 9 | 10 | 11 | 12 |
| 0 | 1 | 6 | 7 | 10 | 11 | 12 | 13 |
| 0 | 2 | 3 | 6 | 7 | 10 | 12 | 14 |
| 0 | 2 | 6 | 9 | 10 | 12 | 13 | 14 |
| 0 | 3 | 4 | 6 | 7 | 8 | 10 | 12 |
| 0 | 3 | 5 | 6 | 9 | 10 | 12 | 15 |
| 0 | 4 | 6 | 8 | 9 | 10 | 12 | 13 |
| 0 | 5 | 6 | 7 | 10 | 12 | 13 | 15 |
| 0 | 1 | 6 | 9 | 11 | 12 | 14 | 15 |
| 0 | 1 | 2 | 6 | 11 | 12 | 13 | 15 |
| 0 | 2 | 5 | 10 | 11 | 12 | 13 | 15 |
| 0 | 2 | 3 | 7 | 11 | 12 | 14 | 15 |
| 0 | 2 | 4 | 8 | 11 | 12 | 14 | 15 |
| 0 | 3 | 7 | 9 | 11 | 12 | 13 | 15 |
| 0 | 4 | 8 | 9 | 11 | 12 | 13 | 15 |
| 0 | 5 | 9 | 10 | 11 | 12 | 14 | 15 |
| 1 | 2 | 3 | 5 | 6 | 7 | 10 | 14 |
| 1 | 2 | 4 | 5 | 6 | 8 | 10 | 14 |
| 1 | 3 | 5 | 6 | 7 | 9 | 10 | 13 |
| 1 | 3 | 4 | 5 | 6 | 10 | 11 | 12 |
| 1 | 4 | 5 | 6 | 8 | 9 | 10 | 13 |
| 1 | 5 | 6 | 7 | 8 | 10 | 11 | 12 |
| 1 | 2 | 5 | 6 | 8 | 11 | 12 | 15 |
| 1 | 2 | 3 | 5 | 7 | 11 | 14 | 15 |
| 1 | 2 | 5 | 9 | 11 | 13 | 14 | 15 |
| 1 | 3 | 4 | 5 | 7 | 8 | 11 | 15 |
| 1 | 4 | 5 | 8 | 9 | 11 | 13 | 15 |
| 1 | 4 | 5 | 6 | 11 | 12 | 14 | 15 |
| 1 | 2 | 3 | 6 | 8 | 9 | 11 | 12 |
| 1 | 2 | 3 | 5 | 9 | 10 | 11 | 13 |
| 1 | 2 | 3 | 7 | 9 | 10 | 11 | 15 |
| 1 | 2 | 3 | 4 | 8 | 9 | 11 | 14 |
| 1 | 2 | 4 | 6 | 7 | 9 | 11 | 12 |
| 1 | 2 | 4 | 5 | 9 | 10 | 12 | 13 |
| 1 | 2 | 3 | 4 | 7 | 9 | 12 | 14 |
| 1 | 3 | 4 | 9 | 10 | 12 | 14 | 15 |
| 1 | 4 | 5 | 7 | 9 | 12 | 13 | 15 |
| 1 | 4 | 6 | 9 | 10 | 11 | 12 | 15 |
| 1 | 2 | 6 | 7 | 9 | 10 | 14 | 15 |
| 1 | 3 | 4 | 6 | 7 | 8 | 9 | 14 |
| 1 | 5 | 6 | 7 | 9 | 13 | 14 | 15 |
| 1 | 6 | 7 | 8 | 9 | 11 | 12 | 14 |
| 1 | 2 | 3 | 4 | 6 | 7 | 8 | 13 |
| 1 | 2 | 3 | 4 | 6 | 11 | 12 | 13 |
| 1 | 2 | 3 | 5 | 6 | 9 | 13 | 15 |
| 1 | 2 | 3 | 6 | 10 | 13 | 14 | 15 |
| 1 | 2 | 4 | 7 | 8 | 11 | 13 | 14 |
| 1 | 3 | 7 | 10 | 11 | 13 | 14 | 15 |
| 1 | 4 | 6 | 7 | 11 | 12 | 13 | 14 |
| 1 | 5 | 7 | 9 | 10 | 11 | 13 | 14 |
| 1 | 2 | 7 | 8 | 10 | 12 | 13 | 15 |
| 1 | 2 | 3 | 7 | 8 | 12 | 13 | 14 |
| 1 | 3 | 6 | 8 | 11 | 12 | 13 | 14 |
| 1 | 3 | 5 | 8 | 9 | 12 | 13 | 15 |
| 1 | 5 | 8 | 9 | 10 | 12 | 13 | 14 |
| 1 | 6 | 8 | 10 | 11 | 12 | 13 | 15 |
| 2 | 3 | 5 | 6 | 7 | 9 | 11 | 15 |
| 2 | 4 | 6 | 7 | 8 | 9 | 11 | 13 |
| 2 | 5 | 6 | 7 | 9 | 10 | 11 | 14 |
| 2 | 3 | 6 | 7 | 8 | 9 | 12 | 13 |
| 2 | 3 | 6 | 8 | 10 | 12 | 14 | 15 |
| 2 | 5 | 6 | 8 | 9 | 10 | 12 | 14 |
| 2 | 5 | 6 | 7 | 8 | 12 | 13 | 15 |
| 2 | 3 | 4 | 5 | 7 | 8 | 10 | 13 |
| 2 | 3 | 5 | 6 | 10 | 11 | 13 | 14 |
| 2 | 4 | 5 | 6 | 10 | 12 | 13 | 14 |
| 2 | 5 | 7 | 8 | 10 | 11 | 12 | 13 |
| 2 | 3 | 4 | 7 | 10 | 11 | 12 | 15 |
| 2 | 4 | 7 | 8 | 9 | 10 | 13 | 15 |
| 2 | 4 | 6 | 7 | 10 | 12 | 14 | 15 |
| 2 | 3 | 4 | 5 | 8 | 9 | 14 | 15 |
| 2 | 4 | 5 | 7 | 8 | 13 | 14 | 15 |
| 2 | 4 | 6 | 8 | 10 | 11 | 14 | 15 |
| 2 | 3 | 4 | 9 | 11 | 12 | 13 | 14 |
| 2 | 6 | 9 | 10 | 11 | 13 | 14 | 15 |
| 2 | 7 | 8 | 9 | 11 | 12 | 13 | 14 |
| 3 | 4 | 5 | 7 | 8 | 9 | 10 | 14 |
| 3 | 4 | 6 | 7 | 8 | 10 | 11 | 15 |
| 3 | 4 | 6 | 7 | 9 | 12 | 13 | 14 |
| 3 | 6 | 7 | 9 | 10 | 11 | 13 | 15 |
| 3 | 4 | 6 | 8 | 9 | 11 | 13 | 14 |
| 3 | 5 | 6 | 7 | 11 | 13 | 14 | 15 |
| 3 | 4 | 5 | 9 | 10 | 11 | 12 | 14 |
| 3 | 4 | 5 | 7 | 11 | 12 | 13 | 15 |
| 3 | 4 | 5 | 6 | 9 | 12 | 14 | 15 |
| 3 | 5 | 7 | 8 | 9 | 11 | 12 | 15 |
| 3 | 4 | 8 | 9 | 10 | 13 | 14 | 15 |
| 3 | 7 | 8 | 10 | 11 | 12 | 14 | 15 |

The column indices contained in 144 C-(1,1,1,1,2,2) inter-connected cycles:

| | | | | | | | |
|---|---|---|---|---|---|---|---|
| 0 | 3 | 4 | 5 | 7 | 8 | 9 | 15 |
| 0 | 3 | 4 | 5 | 9 | 11 | 12 | 15 |
| 0 | 3 | 4 | 5 | 8 | 9 | 13 | 15 |
| 0 | 1 | 3 | 4 | 5 | 9 | 12 | 15 |
| 0 | 1 | 3 | 4 | 5 | 9 | 11 | 15 |
| 0 | 3 | 4 | 5 | 7 | 9 | 13 | 15 |
| 0 | 3 | 4 | 7 | 8 | 10 | 14 | 15 |
| 0 | 3 | 4 | 10 | 11 | 12 | 14 | 15 |
| 0 | 2 | 3 | 4 | 8 | 10 | 14 | 15 |
| 0 | 3 | 4 | 6 | 10 | 12 | 14 | 15 |
| 0 | 2 | 3 | 4 | 7 | 10 | 14 | 15 |
| 0 | 3 | 4 | 6 | 10 | 11 | 14 | 15 |
| 0 | 2 | 3 | 4 | 7 | 8 | 10 | 15 |
| 0 | 3 | 4 | 5 | 7 | 8 | 13 | 15 |
| 0 | 2 | 3 | 4 | 7 | 8 | 14 | 15 |
| 0 | 3 | 4 | 7 | 8 | 9 | 13 | 15 |
| 0 | 1 | 3 | 4 | 5 | 11 | 12 | 15 |
| 0 | 3 | 4 | 6 | 10 | 11 | 12 | 15 |
| 0 | 1 | 3 | 4 | 9 | 11 | 12 | 15 |
| 0 | 3 | 4 | 6 | 11 | 12 | 14 | 15 |
| 0 | 4 | 5 | 8 | 9 | 10 | 13 | 14 |
| 0 | 1 | 4 | 5 | 9 | 10 | 12 | 14 |
| 0 | 2 | 4 | 5 | 8 | 9 | 10 | 14 |
| 0 | 4 | 5 | 6 | 9 | 10 | 12 | 14 |
| 0 | 1 | 4 | 5 | 6 | 9 | 10 | 14 |
| 0 | 2 | 4 | 5 | 9 | 10 | 13 | 14 |
| 0 | 2 | 4 | 5 | 8 | 9 | 10 | 13 |
| 0 | 4 | 5 | 7 | 8 | 9 | 13 | 15 |
| 0 | 2 | 4 | 5 | 8 | 9 | 13 | 14 |

| 0 | 3 | 4 | 5 | 7 | 8 | 9 | 13 |
| 0 | 1 | 4 | 5 | 6 | 9 | 10 | 12 |
| 0 | 1 | 4 | 5 | 9 | 11 | 12 | 15 |
| 0 | 1 | 3 | 4 | 5 | 9 | 11 | 12 |
| 0 | 1 | 4 | 5 | 6 | 9 | 12 | 14 |
| 0 | 2 | 4 | 7 | 8 | 10 | 14 | 15 |
| 0 | 2 | 4 | 5 | 8 | 10 | 13 | 14 |
| 0 | 2 | 3 | 4 | 7 | 8 | 10 | 14 |
| 0 | 2 | 4 | 8 | 9 | 10 | 13 | 14 |
| 0 | 1 | 4 | 5 | 6 | 10 | 12 | 14 |
| 0 | 4 | 6 | 10 | 11 | 12 | 14 | 15 |
| 0 | 1 | 4 | 6 | 9 | 10 | 12 | 14 |
| 0 | 3 | 4 | 6 | 10 | 11 | 12 | 14 |
| 0 | 2 | 6 | 7 | 8 | 10 | 12 | 15 |
| 0 | 2 | 7 | 8 | 10 | 11 | 12 | 15 |
| 0 | 2 | 6 | 7 | 8 | 10 | 11 | 15 |
| 0 | 2 | 3 | 7 | 8 | 10 | 14 | 15 |
| 0 | 2 | 5 | 6 | 8 | 10 | 12 | 13 |
| 0 | 1 | 2 | 5 | 8 | 10 | 12 | 13 |
| 0 | 1 | 2 | 5 | 6 | 8 | 10 | 13 |
| 0 | 2 | 5 | 8 | 9 | 10 | 13 | 14 |
| 0 | 1 | 2 | 5 | 6 | 8 | 10 | 12 |
| 0 | 2 | 6 | 8 | 10 | 11 | 12 | 15 |
| 0 | 1 | 2 | 6 | 8 | 10 | 12 | 13 |
| 0 | 2 | 6 | 7 | 8 | 10 | 11 | 12 |
| 0 | 5 | 7 | 8 | 11 | 12 | 13 | 15 |
| 0 | 1 | 5 | 7 | 8 | 12 | 13 | 15 |
| 0 | 1 | 5 | 7 | 8 | 11 | 13 | 15 |
| 0 | 3 | 5 | 7 | 8 | 9 | 13 | 15 |
| 0 | 1 | 5 | 7 | 8 | 11 | 12 | 15 |
| 0 | 6 | 7 | 8 | 10 | 11 | 12 | 15 |
| 0 | 1 | 7 | 8 | 11 | 12 | 13 | 15 |
| 0 | 2 | 6 | 7 | 8 | 11 | 12 | 15 |
| 0 | 1 | 5 | 6 | 8 | 10 | 12 | 13 |
| 0 | 1 | 5 | 8 | 11 | 12 | 13 | 15 |
| 0 | 1 | 2 | 5 | 6 | 8 | 12 | 13 |
| 0 | 1 | 5 | 7 | 8 | 11 | 12 | 13 |
| 0 | 1 | 5 | 6 | 9 | 10 | 12 | 14 |
| 0 | 1 | 2 | 5 | 6 | 10 | 12 | 13 |
| 0 | 1 | 3 | 5 | 9 | 11 | 12 | 15 |
| 0 | 1 | 5 | 7 | 11 | 12 | 13 | 15 |
| 0 | 2 | 6 | 7 | 10 | 11 | 12 | 15 |
| 0 | 3 | 6 | 10 | 11 | 12 | 14 | 15 |
| 1 | 5 | 6 | 9 | 10 | 11 | 14 | 15 |

| 1 | 2 | 5 | 6 | 10 | 11 | 13 | 15 |
| 1 | 3 | 5 | 6 | 9 | 10 | 11 | 15 |
| 1 | 5 | 6 | 7 | 10 | 11 | 13 | 15 |
| 1 | 2 | 5 | 6 | 7 | 10 | 11 | 15 |
| 1 | 3 | 5 | 6 | 10 | 11 | 14 | 15 |
| 1 | 3 | 5 | 6 | 9 | 10 | 11 | 14 |
| 1 | 4 | 5 | 6 | 9 | 10 | 12 | 14 |
| 1 | 3 | 5 | 6 | 9 | 10 | 14 | 15 |
| 1 | 2 | 5 | 6 | 7 | 10 | 11 | 13 |
| 1 | 2 | 5 | 6 | 8 | 10 | 12 | 13 |
| 1 | 2 | 5 | 6 | 7 | 10 | 13 | 15 |
| 1 | 3 | 4 | 5 | 9 | 11 | 12 | 15 |
| 1 | 3 | 5 | 6 | 9 | 11 | 14 | 15 |
| 1 | 3 | 5 | 9 | 10 | 11 | 14 | 15 |
| 1 | 2 | 5 | 6 | 7 | 11 | 13 | 15 |
| 1 | 5 | 7 | 8 | 11 | 12 | 13 | 15 |
| 1 | 2 | 5 | 7 | 10 | 11 | 13 | 15 |
| 1 | 3 | 4 | 7 | 9 | 11 | 12 | 13 |
| 1 | 3 | 4 | 8 | 9 | 11 | 12 | 13 |
| 1 | 3 | 4 | 7 | 8 | 9 | 11 | 12 |
| 1 | 3 | 6 | 7 | 9 | 11 | 13 | 14 |
| 1 | 2 | 3 | 6 | 9 | 11 | 13 | 14 |
| 1 | 2 | 3 | 6 | 7 | 9 | 11 | 14 |
| 1 | 3 | 6 | 9 | 10 | 11 | 14 | 15 |
| 1 | 2 | 3 | 6 | 7 | 9 | 11 | 13 |
| 1 | 3 | 7 | 8 | 9 | 11 | 12 | 13 |
| 1 | 2 | 3 | 7 | 9 | 11 | 13 | 14 |
| 1 | 3 | 4 | 7 | 8 | 9 | 11 | 13 |
| 1 | 4 | 6 | 8 | 9 | 12 | 13 | 14 |
| 1 | 2 | 4 | 6 | 9 | 12 | 13 | 14 |
| 1 | 2 | 4 | 6 | 8 | 9 | 12 | 14 |
| 1 | 2 | 4 | 6 | 8 | 9 | 12 | 13 |
| 1 | 4 | 7 | 8 | 9 | 11 | 12 | 13 |
| 1 | 2 | 4 | 8 | 9 | 12 | 13 | 14 |
| 1 | 3 | 4 | 7 | 8 | 9 | 12 | 13 |
| 1 | 2 | 6 | 7 | 9 | 11 | 13 | 14 |
| 1 | 2 | 6 | 8 | 9 | 12 | 13 | 14 |
| 1 | 2 | 3 | 6 | 7 | 9 | 13 | 14 |
| 1 | 2 | 4 | 6 | 8 | 9 | 13 | 14 |
| 1 | 2 | 6 | 7 | 10 | 11 | 13 | 15 |
| 1 | 2 | 3 | 6 | 7 | 11 | 13 | 14 |
| 1 | 2 | 4 | 6 | 8 | 12 | 13 | 14 |
| 1 | 3 | 4 | 7 | 8 | 11 | 12 | 13 |
| 2 | 6 | 7 | 8 | 10 | 11 | 12 | 15 |
| 2 | 3 | 6 | 7 | 8 | 11 | 12 | 14 |
| 2 | 4 | 6 | 7 | 8 | 11 | 12 | 14 |
| 2 | 3 | 4 | 6 | 7 | 8 | 11 | 12 |
| 2 | 5 | 6 | 7 | 10 | 11 | 13 | 15 |
| 2 | 3 | 4 | 6 | 7 | 8 | 11 | 14 |
| 2 | 3 | 6 | 7 | 9 | 11 | 13 | 14 |
| 2 | 3 | 4 | 6 | 7 | 11 | 12 | 14 |
| 2 | 3 | 4 | 6 | 7 | 8 | 12 | 14 |
| 2 | 4 | 6 | 8 | 9 | 12 | 13 | 14 |
| 2 | 3 | 4 | 6 | 8 | 11 | 12 | 14 |
| 2 | 5 | 7 | 9 | 10 | 13 | 14 | 15 |
| 2 | 3 | 5 | 7 | 10 | 13 | 14 | 15 |
| 2 | 3 | 5 | 7 | 9 | 10 | 13 | 15 |
| 2 | 3 | 5 | 7 | 9 | 10 | 13 | 14 |
| 2 | 4 | 5 | 8 | 9 | 10 | 13 | 14 |
| 2 | 3 | 5 | 9 | 10 | 13 | 14 | 15 |
| 2 | 3 | 4 | 7 | 8 | 10 | 14 | 15 |
| 2 | 3 | 7 | 9 | 10 | 13 | 14 | 15 |
| 2 | 3 | 5 | 7 | 9 | 10 | 14 | 15 |
| 2 | 3 | 4 | 7 | 8 | 11 | 12 | 14 |
| 2 | 3 | 5 | 7 | 9 | 13 | 14 | 15 |
| 3 | 4 | 7 | 8 | 9 | 11 | 12 | 13 |
| 3 | 4 | 5 | 7 | 8 | 9 | 13 | 15 |
| 3 | 4 | 6 | 7 | 8 | 11 | 12 | 14 |
| 3 | 5 | 7 | 9 | 10 | 13 | 14 | 15 |
| 3 | 4 | 6 | 10 | 11 | 12 | 14 | 15 |
| 3 | 5 | 6 | 9 | 10 | 11 | 14 | 15 |

The column indices contained in 12 D-(2,2,2,2) inter-connected cycles:

| | | | | | | | |
|---|---|---|---|---|---|---|---|
| 0 | 3 | 4 | 5 | 9 | 10 | 14 | 15 |
| 0 | 3 | 4 | 7 | 8 | 11 | 12 | 15 |
| 0 | 1 | 4 | 5 | 8 | 9 | 12 | 13 |
| 0 | 2 | 4 | 6 | 8 | 10 | 12 | 14 |
| 0 | 2 | 5 | 7 | 8 | 10 | 13 | 15 |
| 0 | 1 | 5 | 6 | 10 | 11 | 12 | 15 |
| 1 | 2 | 5 | 6 | 9 | 10 | 13 | 14 |
| 1 | 3 | 5 | 7 | 9 | 11 | 13 | 15 |
| 1 | 3 | 4 | 6 | 9 | 11 | 12 | 14 |
| 1 | 2 | 6 | 7 | 8 | 11 | 12 | 13 |
| 2 | 3 | 6 | 7 | 10 | 11 | 14 | 15 |
| 2 | 3 | 4 | 7 | 8 | 9 | 13 | 14 |

The column indices contained in 432 A-(1,3,5) inter-connected cycles:

| | | | | | | | | |
|---|---|---|---|---|---|---|---|---|
| 0 | 1 | 3 | 4 | 5 | 6 | 7 | 10 | 13 |
| 0 | 1 | 2 | 3 | 4 | 5 | 6 | 7 | 13 |
| 0 | 2 | 3 | 4 | 5 | 6 | 7 | 11 | 13 |
| 0 | 1 | 2 | 3 | 4 | 5 | 6 | 7 | 10 |
| 0 | 1 | 2 | 3 | 4 | 7 | 9 | 10 | 11 |
| 0 | 1 | 2 | 3 | 4 | 7 | 9 | 10 | 12 |
| 0 | 1 | 2 | 3 | 4 | 6 | 7 | 10 | 13 |
| 0 | 1 | 2 | 3 | 4 | 7 | 10 | 11 | 13 |
| 0 | 1 | 2 | 3 | 4 | 5 | 6 | 7 | 11 |
| 0 | 1 | 2 | 3 | 4 | 5 | 10 | 11 | 13 |
| 0 | 1 | 2 | 3 | 4 | 5 | 7 | 11 | 14 |
| 0 | 1 | 2 | 3 | 4 | 5 | 8 | 11 | 14 |
| 0 | 1 | 2 | 3 | 4 | 6 | 10 | 11 | 13 |
| 0 | 1 | 3 | 4 | 6 | 7 | 10 | 11 | 13 |
| 0 | 2 | 3 | 4 | 5 | 6 | 10 | 11 | 13 |
| 0 | 1 | 2 | 4 | 6 | 7 | 9 | 10 | 11 |
| 0 | 1 | 2 | 4 | 6 | 7 | 9 | 10 | 15 |
| 0 | 1 | 3 | 4 | 6 | 7 | 8 | 9 | 10 |
| 0 | 1 | 3 | 4 | 6 | 7 | 9 | 10 | 13 |
| 0 | 1 | 2 | 4 | 6 | 7 | 9 | 11 | 15 |
| 0 | 1 | 2 | 4 | 7 | 9 | 10 | 11 | 15 |
| 0 | 1 | 2 | 4 | 8 | 9 | 11 | 14 | 15 |
| 0 | 1 | 2 | 4 | 9 | 11 | 13 | 14 | 15 |
| 0 | 1 | 2 | 3 | 4 | 9 | 10 | 11 | 13 |
| 0 | 1 | 2 | 4 | 7 | 9 | 10 | 11 | 13 |
| 0 | 2 | 3 | 4 | 9 | 10 | 11 | 12 | 13 |
| 0 | 2 | 4 | 6 | 9 | 10 | 11 | 13 | 15 |
| 0 | 1 | 4 | 6 | 7 | 9 | 13 | 14 | 15 |
| 0 | 1 | 2 | 4 | 6 | 7 | 9 | 13 | 15 |
| 0 | 2 | 4 | 6 | 7 | 9 | 11 | 13 | 15 |
| 0 | 4 | 6 | 7 | 9 | 10 | 11 | 13 | 15 |
| 0 | 1 | 2 | 4 | 5 | 11 | 13 | 14 | 15 |
| 0 | 1 | 2 | 4 | 5 | 7 | 11 | 13 | 14 |
| 0 | 2 | 3 | 4 | 5 | 6 | 11 | 13 | 14 |
| 0 | 2 | 3 | 4 | 5 | 11 | 12 | 13 | 14 |
| 0 | 1 | 2 | 4 | 5 | 7 | 11 | 14 | 15 |
| 0 | 1 | 2 | 4 | 7 | 9 | 12 | 14 | 15 |
| 0 | 1 | 2 | 4 | 6 | 7 | 9 | 14 | 15 |
| 0 | 1 | 2 | 4 | 6 | 7 | 13 | 14 | 15 |
| 0 | 1 | 2 | 4 | 7 | 11 | 13 | 14 | 15 |
| 0 | 1 | 2 | 4 | 5 | 6 | 7 | 11 | 14 |
| 0 | 1 | 2 | 3 | 4 | 5 | 6 | 7 | 14 |
| 0 | 1 | 3 | 4 | 5 | 6 | 7 | 8 | 14 |
| 0 | 1 | 2 | 4 | 6 | 11 | 13 | 14 | 15 |
| 0 | 1 | 4 | 6 | 7 | 11 | 13 | 14 | 15 |
| 0 | 2 | 4 | 6 | 9 | 11 | 13 | 14 | 15 |
| 0 | 4 | 5 | 6 | 7 | 11 | 13 | 14 | 15 |
| 0 | 1 | 2 | 3 | 4 | 5 | 6 | 7 | 8 |
| 0 | 1 | 2 | 3 | 5 | 6 | 8 | 11 | 14 |
| 0 | 1 | 2 | 3 | 5 | 6 | 8 | 9 | 15 |
| 0 | 1 | 2 | 3 | 6 | 8 | 9 | 11 | 15 |
| 0 | 1 | 2 | 6 | 8 | 9 | 11 | 14 | 15 |
| 0 | 2 | 3 | 5 | 6 | 8 | 9 | 11 | 15 |
| 0 | 2 | 4 | 6 | 8 | 9 | 11 | 13 | 15 |
| 0 | 1 | 2 | 5 | 8 | 9 | 11 | 14 | 15 |
| 0 | 1 | 2 | 3 | 5 | 8 | 9 | 11 | 14 |
| 0 | 2 | 3 | 5 | 6 | 8 | 9 | 11 | 14 |
| 0 | 2 | 5 | 7 | 8 | 9 | 11 | 12 | 14 |
| 0 | 1 | 2 | 3 | 5 | 8 | 11 | 14 | 15 |
| 0 | 1 | 2 | 3 | 8 | 9 | 11 | 14 | 15 |
| 0 | 1 | 2 | 3 | 6 | 8 | 9 | 14 | 15 |
| 0 | 1 | 2 | 3 | 6 | 8 | 13 | 14 | 15 |
| 0 | 1 | 2 | 3 | 8 | 12 | 13 | 14 | 15 |
| 0 | 1 | 3 | 5 | 6 | 7 | 8 | 9 | 10 |
| 0 | 1 | 3 | 5 | 6 | 7 | 8 | 9 | 14 |
| 0 | 2 | 3 | 5 | 6 | 7 | 8 | 9 | 11 |
| 0 | 2 | 3 | 5 | 6 | 7 | 8 | 9 | 12 |
| 0 | 1 | 3 | 5 | 6 | 7 | 8 | 10 | 14 |
| 0 | 1 | 3 | 7 | 8 | 9 | 10 | 11 | 14 |
| 0 | 1 | 3 | 6 | 7 | 8 | 9 | 10 | 14 |
| 0 | 1 | 3 | 7 | 8 | 10 | 11 | 13 | 14 |
| 0 | 1 | 3 | 7 | 8 | 10 | 12 | 13 | 14 |
| 0 | 1 | 2 | 3 | 5 | 7 | 8 | 11 | 14 |
| 0 | 1 | 2 | 4 | 5 | 7 | 8 | 11 | 14 |
| 0 | 1 | 3 | 5 | 6 | 7 | 8 | 11 | 14 |
| 0 | 1 | 3 | 6 | 7 | 8 | 9 | 10 | 11 |
| 0 | 1 | 6 | 7 | 8 | 9 | 10 | 11 | 14 |
| 0 | 3 | 6 | 7 | 8 | 9 | 10 | 11 | 13 |
| 0 | 4 | 6 | 7 | 8 | 9 | 10 | 11 | 13 |
| 0 | 5 | 6 | 7 | 8 | 9 | 10 | 11 | 14 |
| 0 | 1 | 3 | 4 | 6 | 7 | 8 | 10 | 13 |
| 0 | 1 | 3 | 6 | 7 | 8 | 9 | 10 | 13 |
| 0 | 1 | 3 | 6 | 8 | 10 | 11 | 13 | 14 |
| 0 | 1 | 3 | 6 | 8 | 10 | 13 | 14 | 15 |
| 0 | 1 | 2 | 4 | 8 | 11 | 13 | 14 | 15 |
| 0 | 1 | 2 | 8 | 9 | 11 | 13 | 14 | 15 |
| 0 | 1 | 3 | 6 | 8 | 11 | 13 | 14 | 15 |
| 0 | 1 | 3 | 8 | 10 | 11 | 13 | 14 | 15 |
| 0 | 1 | 3 | 8 | 9 | 10 | 11 | 13 | 14 |

| 0 | 1 | 7 | 8 | 9 | 10 | 11 | 13 | 14 |
|---|---|---|---|---|----|----|----|----|
| 0 | 3 | 6 | 8 | 9 | 10 | 11 | 13 | 14 |
| 0 | 6 | 8 | 9 | 10 | 11 | 13 | 14 | 15 |
| 0 | 7 | 8 | 9 | 10 | 11 | 12 | 13 | 14 |
| 0 | 1 | 3 | 6 | 8 | 9 | 13 | 14 | 15 |
| 0 | 1 | 2 | 3 | 6 | 8 | 9 | 13 | 15 |
| 0 | 2 | 3 | 6 | 8 | 9 | 12 | 13 | 15 |
| 0 | 3 | 6 | 8 | 9 | 10 | 11 | 13 | 15 |
| 0 | 1 | 2 | 7 | 9 | 10 | 12 | 14 | 15 |
| 0 | 1 | 2 | 3 | 7 | 9 | 10 | 12 | 14 |
| 0 | 1 | 3 | 7 | 9 | 10 | 12 | 13 | 14 |
| 0 | 1 | 7 | 8 | 9 | 10 | 11 | 12 | 14 |
| 0 | 1 | 2 | 3 | 6 | 8 | 9 | 12 | 15 |
| 0 | 1 | 2 | 3 | 7 | 9 | 10 | 12 | 15 |
| 0 | 1 | 2 | 3 | 7 | 9 | 12 | 14 | 15 |
| 0 | 1 | 2 | 3 | 9 | 12 | 13 | 14 | 15 |
| 0 | 1 | 2 | 3 | 7 | 9 | 10 | 12 | 13 |
| 0 | 1 | 2 | 3 | 4 | 10 | 11 | 12 | 13 |
| 0 | 1 | 2 | 3 | 10 | 12 | 13 | 14 | 15 |
| 0 | 1 | 2 | 3 | 7 | 12 | 13 | 14 | 15 |
| 0 | 1 | 2 | 7 | 9 | 12 | 13 | 14 | 15 |
| 0 | 1 | 3 | 7 | 10 | 12 | 13 | 14 | 15 |
| 0 | 1 | 4 | 6 | 7 | 12 | 13 | 14 | 15 |
| 0 | 2 | 3 | 5 | 6 | 7 | 9 | 12 | 13 |
| 0 | 2 | 3 | 5 | 6 | 11 | 12 | 13 | 14 |
| 0 | 2 | 3 | 4 | 5 | 6 | 11 | 12 | 13 |
| 0 | 2 | 3 | 5 | 6 | 9 | 12 | 13 | 15 |
| 0 | 1 | 2 | 4 | 6 | 7 | 9 | 12 | 15 |
| 0 | 1 | 2 | 6 | 7 | 9 | 12 | 14 | 15 |
| 0 | 2 | 3 | 6 | 7 | 9 | 12 | 13 | 15 |
| 0 | 2 | 3 | 5 | 6 | 7 | 9 | 12 | 15 |
| 0 | 2 | 5 | 6 | 7 | 9 | 11 | 12 | 14 |
| 0 | 2 | 3 | 5 | 6 | 7 | 9 | 12 | 14 |
| 0 | 3 | 5 | 6 | 7 | 9 | 12 | 13 | 14 |
| 0 | 5 | 6 | 7 | 8 | 9 | 11 | 12 | 14 |
| 0 | 1 | 2 | 3 | 6 | 12 | 13 | 14 | 15 |
| 0 | 1 | 3 | 6 | 8 | 12 | 13 | 14 | 15 |
| 0 | 2 | 3 | 6 | 9 | 12 | 13 | 14 | 15 |
| 0 | 3 | 5 | 6 | 7 | 12 | 13 | 14 | 15 |
| 0 | 2 | 3 | 5 | 6 | 7 | 9 | 11 | 12 |
| 0 | 2 | 3 | 5 | 6 | 8 | 9 | 11 | 12 |
| 0 | 2 | 3 | 5 | 9 | 10 | 11 | 12 | 13 |
| 0 | 2 | 3 | 5 | 7 | 9 | 11 | 12 | 14 |
| 0 | 2 | 3 | 5 | 9 | 11 | 12 | 13 | 14 |
| 0 | 1 | 3 | 7 | 10 | 11 | 12 | 13 | 14 |
| 0 | 1 | 3 | 8 | 10 | 11 | 12 | 13 | 14 |
| 0 | 2 | 3 | 5 | 10 | 11 | 12 | 13 | 14 |
| 0 | 2 | 3 | 9 | 10 | 11 | 12 | 13 | 14 |
| 0 | 2 | 3 | 5 | 7 | 11 | 12 | 13 | 14 |
| 0 | 2 | 5 | 7 | 9 | 11 | 12 | 13 | 14 |
| 0 | 3 | 5 | 6 | 7 | 11 | 12 | 13 | 14 |
| 0 | 4 | 5 | 6 | 7 | 11 | 12 | 13 | 14 |
| 0 | 1 | 2 | 3 | 7 | 9 | 10 | 11 | 12 |
| 0 | 1 | 2 | 4 | 7 | 9 | 10 | 11 | 12 |
| 0 | 2 | 3 | 7 | 9 | 10 | 11 | 12 | 13 |
| 0 | 2 | 5 | 7 | 9 | 10 | 11 | 12 | 14 |
| 0 | 1 | 2 | 4 | 5 | 7 | 9 | 10 | 11 |
| 0 | 1 | 2 | 4 | 5 | 7 | 10 | 11 | 14 |
| 1 | 2 | 3 | 4 | 5 | 7 | 8 | 10 | 11 |
| 1 | 2 | 3 | 4 | 5 | 7 | 10 | 11 | 12 |
| 1 | 2 | 3 | 4 | 5 | 7 | 8 | 10 | 12 |
| 1 | 2 | 3 | 4 | 5 | 8 | 10 | 11 | 12 |
| 1 | 2 | 3 | 5 | 8 | 9 | 10 | 12 | 15 |
| 1 | 2 | 3 | 5 | 8 | 10 | 12 | 14 | 15 |
| 0 | 1 | 2 | 3 | 5 | 8 | 10 | 11 | 14 |
| 0 | 1 | 3 | 5 | 8 | 10 | 11 | 13 | 14 |
| 1 | 2 | 3 | 4 | 5 | 8 | 10 | 11 | 14 |
| 1 | 3 | 5 | 7 | 8 | 10 | 11 | 12 | 14 |
| 1 | 2 | 4 | 5 | 7 | 10 | 12 | 14 | 15 |
| 1 | 2 | 3 | 4 | 5 | 7 | 10 | 12 | 14 |
| 1 | 3 | 4 | 5 | 7 | 8 | 10 | 12 | 14 |
| 1 | 4 | 5 | 7 | 8 | 10 | 11 | 12 | 14 |
| 0 | 1 | 3 | 5 | 6 | 7 | 8 | 14 | 15 |
| 0 | 1 | 3 | 5 | 6 | 8 | 13 | 14 | 15 |
| 1 | 2 | 3 | 5 | 6 | 8 | 12 | 14 | 15 |
| 1 | 2 | 3 | 4 | 5 | 6 | 8 | 14 | 15 |
| 1 | 2 | 3 | 4 | 5 | 6 | 8 | 12 | 15 |
| 1 | 2 | 3 | 4 | 5 | 10 | 12 | 13 | 15 |
| 1 | 2 | 3 | 4 | 5 | 7 | 10 | 12 | 15 |
| 1 | 2 | 3 | 4 | 5 | 7 | 12 | 14 | 15 |
| 1 | 2 | 3 | 4 | 5 | 8 | 12 | 14 | 15 |
| 0 | 1 | 2 | 3 | 4 | 5 | 6 | 7 | 15 |
| 0 | 1 | 2 | 4 | 5 | 6 | 7 | 9 | 15 |
| 1 | 2 | 3 | 4 | 5 | 6 | 7 | 8 | 15 |
| 1 | 2 | 4 | 5 | 6 | 7 | 12 | 14 | 15 |
| 1 | 2 | 3 | 5 | 7 | 8 | 12 | 14 | 15 |
| 1 | 2 | 4 | 5 | 7 | 8 | 12 | 14 | 15 |
| 1 | 3 | 5 | 7 | 8 | 10 | 12 | 14 | 15 |
| 1 | 4 | 5 | 6 | 7 | 8 | 12 | 14 | 15 |
| 0 | 1 | 2 | 3 | 4 | 6 | 7 | 9 | 15 |
| 0 | 1 | 2 | 3 | 4 | 5 | 6 | 7 | 9 |
| 0 | 1 | 2 | 3 | 6 | 7 | 8 | 9 | 10 |
| 0 | 1 | 2 | 3 | 6 | 7 | 8 | 9 | 15 |
| 0 | 1 | 2 | 3 | 5 | 6 | 7 | 9 | 12 |
| 0 | 1 | 2 | 3 | 6 | 7 | 9 | 10 | 12 |
| 0 | 1 | 2 | 3 | 7 | 8 | 9 | 10 | 12 |
| 0 | 1 | 3 | 6 | 7 | 8 | 9 | 10 | 12 |
| 1 | 2 | 3 | 7 | 8 | 9 | 10 | 12 | 15 |
| 1 | 3 | 5 | 7 | 8 | 9 | 10 | 12 | 14 |
| 0 | 1 | 2 | 3 | 6 | 8 | 9 | 10 | 15 |
| 0 | 1 | 3 | 6 | 7 | 8 | 9 | 10 | 15 |
| 1 | 2 | 3 | 6 | 8 | 9 | 10 | 12 | 15 |
| 1 | 3 | 4 | 6 | 8 | 9 | 10 | 13 | 15 |
| 0 | 1 | 2 | 3 | 4 | 6 | 8 | 9 | 15 |
| 0 | 1 | 2 | 4 | 6 | 7 | 8 | 9 | 15 |
| 1 | 2 | 3 | 4 | 5 | 6 | 8 | 9 | 15 |
| 0 | 1 | 4 | 7 | 8 | 9 | 10 | 11 | 14 |
| 0 | 1 | 2 | 4 | 7 | 8 | 9 | 10 | 11 |
| 1 | 2 | 4 | 7 | 8 | 9 | 10 | 11 | 15 |
| 1 | 4 | 5 | 7 | 8 | 9 | 10 | 11 | 14 |
| 1 | 4 | 6 | 7 | 8 | 9 | 10 | 11 | 15 |
| 0 | 1 | 2 | 4 | 5 | 7 | 9 | 11 | 14 |
| 0 | 1 | 2 | 4 | 7 | 9 | 10 | 11 | 14 |
| 0 | 1 | 2 | 7 | 8 | 9 | 10 | 11 | 14 |
| 0 | 1 | 2 | 7 | 8 | 9 | 11 | 14 | 15 |
| 0 | 1 | 2 | 5 | 7 | 9 | 11 | 12 | 14 |
| 0 | 1 | 2 | 7 | 9 | 11 | 12 | 14 | 15 |
| 0 | 1 | 2 | 7 | 8 | 9 | 12 | 14 | 15 |
| 0 | 1 | 2 | 8 | 9 | 11 | 12 | 14 | 15 |
| 1 | 2 | 3 | 5 | 8 | 9 | 12 | 14 | 15 |
| 1 | 2 | 3 | 8 | 9 | 10 | 12 | 14 | 15 |
| 0 | 1 | 2 | 8 | 9 | 10 | 11 | 14 | 15 |
| 0 | 1 | 7 | 8 | 9 | 10 | 11 | 14 | 15 |
| 1 | 2 | 4 | 8 | 9 | 10 | 11 | 14 | 15 |
| 1 | 4 | 8 | 9 | 10 | 11 | 13 | 14 | 15 |
| 1 | 7 | 8 | 9 | 10 | 11 | 12 | 14 | 15 |
| 0 | 1 | 2 | 3 | 4 | 10 | 11 | 13 | 15 |
| 0 | 1 | 2 | 4 | 10 | 11 | 13 | 14 | 15 |
| 1 | 2 | 3 | 4 | 10 | 11 | 12 | 13 | 15 |
| 1 | 2 | 4 | 8 | 9 | 10 | 11 | 13 | 15 |
| 0 | 1 | 2 | 3 | 4 | 11 | 13 | 14 | 15 |
| 0 | 1 | 2 | 3 | 4 | 10 | 11 | 13 | 14 |

| 0 | 1 | 2 | 3 | 8  | 10 | 11 | 13 | 14 |
| 0 | 1 | 2 | 3 | 5  | 8  | 11 | 13 | 14 |
| 0 | 1 | 2 | 3 | 5  | 11 | 12 | 13 | 14 |
| 0 | 1 | 2 | 3 | 11 | 12 | 13 | 14 | 15 |
| 0 | 1 | 2 | 3 | 4  | 12 | 13 | 14 | 15 |
| 0 | 1 | 2 | 4 | 11 | 12 | 13 | 14 | 15 |
| 1 | 2 | 3 | 4 | 10 | 12 | 13 | 14 | 15 |
| 1 | 2 | 4 | 5 | 7  | 12 | 13 | 14 | 15 |
| 0 | 1 | 3 | 4 | 6  | 7  | 13 | 14 | 15 |
| 0 | 1 | 3 | 4 | 6  | 7  | 10 | 13 | 14 |
| 0 | 1 | 3 | 6 | 7  | 8  | 13 | 14 | 15 |
| 0 | 1 | 3 | 5 | 6  | 7  | 8  | 13 | 14 |
| 0 | 1 | 3 | 5 | 6  | 7  | 12 | 13 | 14 |
| 0 | 1 | 3 | 6 | 7  | 10 | 12 | 13 | 14 |
| 0 | 1 | 3 | 4 | 7  | 10 | 12 | 13 | 14 |
| 0 | 1 | 3 | 4 | 6  | 7  | 10 | 12 | 13 |
| 1 | 2 | 3 | 4 | 5  | 7  | 10 | 12 | 13 |
| 1 | 2 | 3 | 4 | 7  | 10 | 12 | 13 | 15 |
| 0 | 1 | 3 | 4 | 6  | 7  | 10 | 13 | 15 |
| 0 | 1 | 4 | 6 | 7  | 10 | 13 | 14 | 15 |
| 1 | 3 | 4 | 6 | 7  | 8  | 10 | 13 | 15 |
| 1 | 4 | 6 | 7 | 8  | 9  | 10 | 13 | 15 |
| 1 | 4 | 6 | 7 | 10 | 12 | 13 | 14 | 15 |
| 0 | 1 | 3 | 4 | 6  | 8  | 13 | 14 | 15 |
| 0 | 1 | 4 | 6 | 7  | 8  | 13 | 14 | 15 |
| 1 | 3 | 4 | 6 | 8  | 10 | 13 | 14 | 15 |
| 1 | 4 | 5 | 6 | 7  | 8  | 13 | 14 | 15 |
| 0 | 1 | 3 | 4 | 8  | 10 | 11 | 13 | 14 |
| 0 | 1 | 2 | 3 | 4  | 8  | 10 | 11 | 13 |
| 1 | 2 | 3 | 4 | 5  | 8  | 10 | 11 | 13 |
| 1 | 3 | 4 | 6 | 8  | 10 | 11 | 13 | 15 |
| 0 | 2 | 3 | 4 | 5  | 6  | 8  | 9  | 11 |
| 0 | 2 | 3 | 4 | 5  | 6  | 8  | 11 | 13 |
| 1 | 2 | 3 | 4 | 5  | 6  | 8  | 10 | 11 |
| 1 | 2 | 3 | 4 | 5  | 6  | 8  | 11 | 15 |
| 0 | 2 | 3 | 4 | 6  | 9  | 11 | 13 | 15 |
| 0 | 2 | 3 | 4 | 5  | 6  | 9  | 11 | 13 |
| 0 | 2 | 3 | 6 | 8  | 9  | 10 | 11 | 13 |
| 0 | 2 | 3 | 5 | 6  | 8  | 9  | 11 | 13 |
| 0 | 2 | 3 | 6 | 9  | 10 | 11 | 12 | 13 |
| 0 | 2 | 3 | 6 | 9  | 11 | 12 | 13 | 15 |
| 0 | 2 | 3 | 4 | 5  | 6  | 11 | 13 | 15 |
| 0 | 2 | 4 | 5 | 6  | 9  | 11 | 13 | 15 |
| 2 | 3 | 4 | 5 | 6  | 11 | 12 | 13 | 15 |
| 2 | 4 | 5 | 6 | 8 | 9 | 11 | 13 | 15 |
| 2 | 4 | 5 | 6 | 11 | 12 | 13 | 14 | 15 |
| 0 | 2 | 3 | 4 | 5 | 6 | 7 | 9 | 12 |
| 0 | 1 | 2 | 3 | 4 | 5 | 6 | 7 | 12 |
| 1 | 2 | 3 | 4 | 5 | 6 | 7 | 10 | 12 |
| 2 | 3 | 4 | 5 | 6 | 7 | 12 | 13 | 15 |
| 0 | 2 | 3 | 4 | 6 | 9 | 12 | 13 | 15 |
| 0 | 2 | 4 | 6 | 9 | 11 | 12 | 13 | 15 |
| 2 | 3 | 4 | 5 | 6 | 9 | 12 | 13 | 15 |
| 2 | 4 | 5 | 6 | 7 | 9 | 12 | 13 | 15 |
| 0 | 1 | 2 | 3 | 5 | 7 | 9 | 10 | 12 |
| 0 | 2 | 3 | 5 | 6 | 7 | 9 | 10 | 12 |
| 1 | 2 | 3 | 4 | 5 | 7 | 9 | 10 | 12 |
| 1 | 2 | 4 | 5 | 8 | 9 | 10 | 11 | 15 |
| 1 | 2 | 3 | 4 | 5 | 8 | 9 | 10 | 11 |
| 2 | 3 | 4 | 5 | 8 | 9 | 10 | 11 | 12 |
| 2 | 4 | 5 | 6 | 8 | 9 | 10 | 11 | 15 |
| 2 | 4 | 5 | 7 | 8 | 9 | 10 | 11 | 12 |
| 1 | 2 | 3 | 4 | 5 | 6 | 8 | 10 | 15 |
| 1 | 2 | 3 | 4 | 5 | 8 | 10 | 11 | 15 |
| 1 | 2 | 3 | 4 | 8 | 9 | 10 | 11 | 15 |
| 1 | 2 | 3 | 4 | 8 | 9 | 10 | 12 | 15 |
| 1 | 2 | 3 | 4 | 6 | 8 | 10 | 13 | 15 |
| 1 | 2 | 3 | 4 | 8 | 10 | 12 | 13 | 15 |
| 0 | 2 | 3 | 6 | 9 | 10 | 12 | 13 | 15 |
| 0 | 2 | 3 | 9 | 10 | 11 | 12 | 13 | 15 |
| 1 | 2 | 3 | 4 | 9 | 10 | 12 | 13 | 15 |
| 1 | 2 | 3 | 8 | 9 | 10 | 12 | 13 | 15 |
| 1 | 2 | 3 | 8 | 9 | 10 | 11 | 12 | 15 |
| 1 | 2 | 4 | 8 | 9 | 10 | 11 | 12 | 15 |
| 2 | 3 | 5 | 8 | 9 | 10 | 11 | 12 | 15 |
| 2 | 4 | 8 | 9 | 10 | 11 | 12 | 13 | 15 |
| 2 | 5 | 8 | 9 | 10 | 11 | 12 | 14 | 15 |
| 0 | 2 | 3 | 5 | 8 | 11 | 12 | 13 | 14 |
| 0 | 1 | 2 | 3 | 5 | 8 | 11 | 12 | 14 |
| 1 | 2 | 3 | 5 | 8 | 11 | 12 | 14 | 15 |
| 2 | 3 | 5 | 8 | 9 | 10 | 11 | 12 | 14 |
| 0 | 1 | 2 | 3 | 5 | 12 | 13 | 14 | 15 |
| 0 | 2 | 3 | 5 | 11 | 12 | 13 | 14 | 15 |
| 1 | 2 | 3 | 5 | 8 | 12 | 13 | 14 | 15 |
| 2 | 3 | 4 | 5 | 6 | 12 | 13 | 14 | 15 |
| 0 | 2 | 4 | 5 | 7 | 9 | 11 | 12 | 14 |
| 0 | 1 | 2 | 4 | 5 | 7 | 11 | 12 | 14 |
| 1 | 2 | 4 | 5 | 7 | 11 | 12 | 14 | 15 |
| 2 | 4 | 5 | 7 | 9 | 10 | 11 | 12 | 14 |
| 2 | 4 | 5 | 7 | 11 | 12 | 13 | 14 | 15 |
| 0 | 1 | 2 | 5 | 7 | 9 | 12 | 14 | 15 |
| 0 | 2 | 5 | 7 | 9 | 11 | 12 | 14 | 15 |
| 1 | 2 | 4 | 5 | 7 | 9 | 12 | 14 | 15 |
| 2 | 4 | 5 | 6 | 7 | 9 | 12 | 14 | 15 |
| 0 | 3 | 5 | 6 | 7 | 8 | 12 | 13 | 14 |
| 0 | 1 | 3 | 5 | 6 | 7 | 8 | 12 | 14 |
| 1 | 3 | 5 | 6 | 7 | 8 | 10 | 12 | 14 |
| 3 | 5 | 6 | 7 | 8 | 9 | 10 | 12 | 14 |
| 3 | 5 | 6 | 7 | 8 | 12 | 13 | 14 | 15 |
| 0 | 1 | 3 | 5 | 7 | 10 | 12 | 13 | 14 |
| 0 | 3 | 5 | 6 | 7 | 10 | 12 | 13 | 14 |
| 1 | 3 | 5 | 7 | 8 | 10 | 12 | 13 | 14 |
| 3 | 4 | 5 | 6 | 7 | 10 | 12 | 13 | 14 |
| 0 | 2 | 3 | 5 | 6 | 8 | 9 | 10 | 11 |
| 0 | 3 | 5 | 6 | 8 | 9 | 10 | 11 | 13 |
| 2 | 3 | 5 | 6 | 8 | 9 | 10 | 11 | 12 |
| 3 | 4 | 5 | 6 | 8 | 9 | 10 | 11 | 13 |
| 3 | 5 | 6 | 7 | 8 | 9 | 10 | 11 | 12 |
| 0 | 2 | 3 | 8 | 9 | 10 | 11 | 12 | 13 |
| 0 | 3 | 6 | 8 | 9 | 10 | 11 | 12 | 13 |
| 2 | 3 | 5 | 8 | 9 | 10 | 11 | 12 | 13 |
| 3 | 5 | 8 | 9 | 10 | 11 | 12 | 13 | 14 |
| 3 | 6 | 8 | 9 | 10 | 11 | 12 | 13 | 15 |
| 1 | 3 | 4 | 5 | 6 | 8 | 10 | 13 | 15 |
| 1 | 2 | 3 | 4 | 5 | 6 | 8 | 13 | 15 |
| 2 | 3 | 4 | 5 | 6 | 8 | 12 | 13 | 15 |
| 3 | 4 | 5 | 6 | 8 | 10 | 11 | 13 | 15 |
| 1 | 2 | 3 | 4 | 6 | 10 | 12 | 13 | 15 |
| 1 | 3 | 4 | 6 | 8 | 10 | 12 | 13 | 15 |
| 2 | 3 | 4 | 5 | 6 | 10 | 12 | 13 | 15 |
| 3 | 4 | 5 | 6 | 7 | 10 | 12 | 13 | 15 |
| 0 | 4 | 5 | 6 | 7 | 8 | 9 | 11 | 14 |
| 0 | 4 | 5 | 6 | 7 | 8 | 11 | 13 | 14 |
| 1 | 4 | 5 | 6 | 7 | 8 | 10 | 11 | 14 |
| 1 | 4 | 5 | 6 | 7 | 8 | 11 | 14 | 15 |
| 4 | 5 | 6 | 7 | 8 | 9 | 10 | 11 | 14 |
| 4 | 5 | 6 | 7 | 8 | 11 | 13 | 14 | 15 |
| 1 | 4 | 5 | 6 | 7 | 8 | 9 | 10 | 15 |
| 1 | 4 | 5 | 6 | 7 | 8 | 9 | 14 | 15 |
| 2 | 4 | 5 | 6 | 7 | 8 | 9 | 11 | 15 |
| 2 | 4 | 5 | 6 | 7 | 8 | 9 | 12 | 15 |
| 4 | 5 | 6 | 7 | 8 | 9 | 10 | 11 | 15 |

| 1 | 4 | 5 | 6 | 7 | 8 | 10 | 14 | 15 |
| 1 | 4 | 5 | 7 | 8 | 10 | 11 | 14 | 15 |
| 1 | 4 | 7 | 8 | 9 | 10 | 12 | 14 | 15 |
| 1 | 4 | 6 | 7 | 8 | 9 | 10 | 14 | 15 |
| 1 | 4 | 7 | 8 | 10 | 11 | 13 | 14 | 15 |
| 1 | 4 | 7 | 8 | 10 | 12 | 13 | 14 | 15 |
| 1 | 4 | 5 | 8 | 10 | 11 | 13 | 14 | 15 |
| 1 | 4 | 5 | 7 | 8 | 10 | 11 | 13 | 14 |
| 3 | 4 | 5 | 6 | 8 | 10 | 11 | 13 | 14 |
| 3 | 4 | 5 | 8 | 10 | 11 | 12 | 13 | 14 |
| 4 | 5 | 6 | 8 | 10 | 11 | 13 | 14 | 15 |
| 4 | 5 | 7 | 8 | 10 | 11 | 12 | 13 | 14 |
| 2 | 4 | 5 | 6 | 7 | 9 | 10 | 11 | 12 |
| 2 | 4 | 5 | 6 | 7 | 9 | 10 | 12 | 15 |
| 3 | 4 | 5 | 6 | 7 | 8 | 9 | 10 | 12 |
| 3 | 4 | 5 | 6 | 7 | 9 | 10 | 12 | 13 |
| 4 | 5 | 6 | 7 | 8 | 9 | 10 | 11 | 12 |
| 0 | 4 | 7 | 9 | 10 | 11 | 12 | 13 | 14 |
| 0 | 4 | 6 | 7 | 9 | 10 | 11 | 12 | 13 |
| 2 | 4 | 5 | 7 | 9 | 10 | 11 | 12 | 13 |
| 2 | 4 | 7 | 9 | 10 | 11 | 12 | 13 | 15 |
| 4 | 5 | 7 | 9 | 10 | 11 | 12 | 13 | 14 |
| 4 | 6 | 7 | 9 | 10 | 11 | 12 | 13 | 15 |
| 2 | 4 | 5 | 6 | 7 | 9 | 11 | 12 | 15 |
| 2 | 4 | 5 | 6 | 8 | 9 | 11 | 12 | 15 |
| 2 | 4 | 5 | 9 | 10 | 11 | 12 | 13 | 15 |
| 2 | 4 | 5 | 7 | 9 | 10 | 11 | 12 | 15 |
| 2 | 4 | 5 | 8 | 9 | 11 | 12 | 14 | 15 |
| 2 | 4 | 5 | 9 | 11 | 12 | 13 | 14 | 15 |
| 1 | 4 | 7 | 10 | 11 | 12 | 13 | 14 | 15 |
| 1 | 4 | 8 | 10 | 11 | 12 | 13 | 14 | 15 |
| 2 | 4 | 5 | 10 | 11 | 12 | 13 | 14 | 15 |
| 2 | 4 | 9 | 10 | 11 | 12 | 13 | 14 | 15 |
| 4 | 5 | 7 | 10 | 11 | 12 | 13 | 14 | 15 |
| 0 | 5 | 6 | 7 | 8 | 9 | 11 | 14 | 15 |
| 0 | 5 | 6 | 8 | 9 | 11 | 13 | 14 | 15 |
| 2 | 5 | 6 | 8 | 9 | 11 | 12 | 14 | 15 |
| 2 | 4 | 5 | 6 | 8 | 9 | 11 | 14 | 15 |
| 4 | 5 | 6 | 8 | 9 | 11 | 13 | 14 | 15 |
| 5 | 6 | 7 | 8 | 9 | 11 | 12 | 14 | 15 |
| 0 | 4 | 5 | 6 | 7 | 9 | 10 | 11 | 13 |
| 0 | 4 | 5 | 6 | 7 | 10 | 11 | 13 | 14 |
| 3 | 4 | 5 | 6 | 7 | 8 | 10 | 11 | 13 |
| 3 | 4 | 5 | 6 | 7 | 10 | 11 | 12 | 13 |
| 4 | 5 | 6 | 7 | 8 | 9 | 10 | 11 | 13 |
| 4 | 5 | 6 | 7 | 10 | 11 | 12 | 13 | 14 |
| 3 | 4 | 5 | 6 | 7 | 8 | 10 | 12 | 13 |
| 3 | 5 | 6 | 7 | 8 | 9 | 10 | 12 | 13 |
| 3 | 5 | 6 | 8 | 10 | 11 | 12 | 13 | 14 |
| 3 | 4 | 5 | 6 | 8 | 10 | 11 | 12 | 13 |
| 3 | 5 | 6 | 8 | 9 | 10 | 12 | 13 | 15 |
| 3 | 5 | 6 | 8 | 10 | 12 | 13 | 14 | 15 |
| 2 | 4 | 5 | 8 | 11 | 12 | 13 | 14 | 15 |
| 2 | 5 | 8 | 9 | 11 | 12 | 13 | 14 | 15 |
| 3 | 5 | 6 | 8 | 11 | 12 | 13 | 14 | 15 |
| 3 | 5 | 8 | 10 | 11 | 12 | 13 | 14 | 15 |
| 4 | 5 | 6 | 8 | 11 | 12 | 13 | 14 | 15 |
| 1 | 4 | 6 | 7 | 8 | 9 | 10 | 12 | 15 |
| 1 | 6 | 7 | 8 | 9 | 10 | 12 | 14 | 15 |
| 3 | 6 | 7 | 8 | 9 | 10 | 12 | 13 | 15 |
| 3 | 5 | 6 | 7 | 8 | 9 | 10 | 12 | 15 |
| 4 | 6 | 7 | 8 | 9 | 10 | 12 | 13 | 15 |
| 5 | 6 | 7 | 8 | 9 | 10 | 12 | 14 | 15 |
| 0 | 4 | 5 | 6 | 7 | 9 | 11 | 13 | 14 |
| 0 | 4 | 6 | 7 | 9 | 10 | 11 | 13 | 14 |
| 0 | 6 | 7 | 8 | 9 | 11 | 13 | 14 | 15 |
| 0 | 5 | 6 | 7 | 8 | 9 | 11 | 13 | 14 |
| 0 | 6 | 7 | 9 | 10 | 11 | 12 | 13 | 14 |
| 0 | 6 | 7 | 9 | 11 | 12 | 13 | 14 | 15 |
| 0 | 6 | 7 | 8 | 9 | 12 | 13 | 14 | 15 |
| 0 | 6 | 8 | 9 | 11 | 12 | 13 | 14 | 15 |
| 3 | 5 | 6 | 8 | 9 | 12 | 13 | 14 | 15 |
| 3 | 6 | 8 | 9 | 10 | 12 | 13 | 14 | 15 |
| 5 | 6 | 7 | 8 | 9 | 12 | 13 | 14 | 15 |
| 0 | 6 | 7 | 9 | 10 | 12 | 13 | 14 | 15 |
| 0 | 7 | 9 | 10 | 11 | 12 | 13 | 14 | 15 |
| 1 | 4 | 7 | 9 | 10 | 12 | 13 | 14 | 15 |
| 1 | 7 | 8 | 9 | 10 | 12 | 13 | 14 | 15 |
| 4 | 6 | 7 | 9 | 10 | 12 | 13 | 14 | 15 |

The column indices contained in 96 A-(3,3,3) inter-connected cycles:

| | | | | | | | | |
|---|---|---|---|---|---|---|---|---|
| 0 | 3 | 4 | 5 | 6 | 7 | 10 | 11 | 13 |
| 0 | 1 | 3 | 4 | 5 | 6 | 7 | 13 | 14 |
| 0 | 2 | 3 | 4 | 5 | 6 | 7 | 12 | 13 |
| 0 | 1 | 2 | 3 | 4 | 5 | 7 | 10 | 11 |
| 0 | 1 | 2 | 3 | 4 | 6 | 7 | 9 | 10 |
| 0 | 1 | 2 | 3 | 4 | 7 | 10 | 12 | 13 |
| 0 | 1 | 2 | 3 | 4 | 5 | 11 | 13 | 14 |
| 0 | 1 | 2 | 3 | 4 | 5 | 6 | 8 | 11 |
| 0 | 2 | 3 | 4 | 6 | 9 | 10 | 11 | 13 |
| 0 | 1 | 3 | 4 | 6 | 8 | 10 | 11 | 13 |
| 0 | 1 | 4 | 6 | 7 | 9 | 10 | 13 | 15 |
| 0 | 1 | 4 | 6 | 7 | 8 | 9 | 10 | 11 |
| 0 | 1 | 2 | 4 | 9 | 10 | 11 | 13 | 15 |
| 0 | 1 | 2 | 4 | 7 | 9 | 11 | 14 | 15 |
| 0 | 1 | 2 | 4 | 6 | 8 | 9 | 11 | 15 |
| 0 | 2 | 4 | 7 | 9 | 10 | 11 | 12 | 13 |
| 0 | 4 | 6 | 7 | 9 | 11 | 13 | 14 | 15 |
| 0 | 2 | 4 | 6 | 7 | 9 | 12 | 13 | 15 |
| 0 | 2 | 4 | 5 | 6 | 11 | 13 | 14 | 15 |
| 0 | 2 | 4 | 5 | 7 | 11 | 12 | 13 | 14 |
| 0 | 1 | 2 | 4 | 5 | 6 | 7 | 14 | 15 |
| 0 | 1 | 2 | 4 | 7 | 12 | 13 | 14 | 15 |
| 0 | 1 | 4 | 5 | 6 | 7 | 8 | 11 | 14 |
| 0 | 1 | 4 | 6 | 8 | 11 | 13 | 14 | 15 |
| 0 | 1 | 2 | 3 | 5 | 6 | 8 | 14 | 15 |
| 0 | 1 | 2 | 3 | 5 | 6 | 7 | 8 | 9 |
| 0 | 2 | 5 | 6 | 8 | 9 | 11 | 14 | 15 |
| 0 | 2 | 3 | 6 | 8 | 9 | 11 | 13 | 15 |
| 0 | 1 | 2 | 5 | 7 | 8 | 9 | 11 | 14 |
| 0 | 2 | 3 | 5 | 8 | 9 | 11 | 12 | 14 |
| 0 | 1 | 2 | 3 | 8 | 11 | 13 | 14 | 15 |
| 0 | 1 | 2 | 3 | 8 | 9 | 12 | 14 | 15 |
| 0 | 3 | 5 | 6 | 7 | 8 | 9 | 10 | 11 |
| 0 | 3 | 5 | 6 | 7 | 8 | 9 | 12 | 14 |
| 0 | 1 | 3 | 5 | 7 | 8 | 10 | 11 | 14 |
| 0 | 1 | 3 | 6 | 7 | 8 | 10 | 13 | 14 |
| 0 | 1 | 3 | 7 | 8 | 9 | 10 | 12 | 14 |
| 0 | 6 | 7 | 8 | 9 | 10 | 11 | 13 | 14 |
| 0 | 1 | 3 | 6 | 8 | 9 | 10 | 13 | 15 |
| 0 | 1 | 8 | 9 | 10 | 11 | 13 | 14 | 15 |
| 0 | 3 | 8 | 9 | 10 | 11 | 12 | 13 | 14 |
| 0 | 3 | 6 | 8 | 9 | 12 | 13 | 14 | 15 |
| 0 | 1 | 7 | 9 | 10 | 12 | 13 | 14 | 15 |
| 0 | 1 | 2 | 7 | 9 | 10 | 11 | 12 | 14 |
| 0 | 1 | 2 | 3 | 9 | 10 | 12 | 13 | 15 |
| 0 | 1 | 2 | 3 | 6 | 7 | 9 | 12 | 15 |
| 0 | 1 | 2 | 3 | 10 | 11 | 12 | 13 | 14 |
| 0 | 1 | 3 | 6 | 7 | 12 | 13 | 14 | 15 |
| 0 | 2 | 3 | 5 | 6 | 12 | 13 | 14 | 15 |
| 0 | 2 | 3 | 5 | 6 | 9 | 11 | 12 | 13 |
| 0 | 2 | 5 | 6 | 7 | 9 | 12 | 14 | 15 |

| 0 | 5 | 6 | 7 | 9 | 11 | 12 | 13 | 14 |
|---|---|---|---|---|----|----|----|----|
| 0 | 2 | 3 | 5 | 7 | 9 | 10 | 11 | 12 |
| 0 | 3 | 5 | 7 | 10 | 11 | 12 | 13 | 14 |
| 1 | 2 | 4 | 5 | 7 | 10 | 11 | 12 | 14 |
| 1 | 2 | 4 | 5 | 7 | 8 | 9 | 10 | 11 |
| 1 | 2 | 3 | 5 | 8 | 10 | 11 | 12 | 14 |
| 1 | 2 | 3 | 4 | 5 | 8 | 10 | 12 | 15 |
| 1 | 2 | 3 | 5 | 7 | 8 | 9 | 10 | 12 |
| 1 | 3 | 4 | 5 | 8 | 10 | 11 | 13 | 14 |
| 1 | 4 | 5 | 7 | 8 | 10 | 12 | 14 | 15 |
| 1 | 3 | 4 | 5 | 7 | 10 | 12 | 13 | 14 |
| 1 | 3 | 5 | 6 | 7 | 8 | 12 | 14 | 15 |
| 1 | 3 | 4 | 5 | 6 | 8 | 13 | 14 | 15 |
| 1 | 2 | 3 | 4 | 5 | 6 | 7 | 12 | 15 |
| 1 | 2 | 3 | 4 | 5 | 12 | 13 | 14 | 15 |
| 1 | 2 | 4 | 5 | 6 | 7 | 8 | 9 | 15 |
| 1 | 2 | 5 | 7 | 8 | 9 | 12 | 14 | 15 |
| 1 | 3 | 6 | 7 | 8 | 9 | 10 | 12 | 15 |
| 1 | 2 | 3 | 4 | 6 | 8 | 9 | 10 | 15 |
| 1 | 4 | 7 | 8 | 9 | 10 | 11 | 14 | 15 |
| 1 | 2 | 8 | 9 | 10 | 11 | 12 | 14 | 15 |
| 1 | 2 | 4 | 10 | 11 | 12 | 13 | 14 | 15 |
| 1 | 2 | 3 | 4 | 8 | 10 | 11 | 13 | 15 |
| 1 | 3 | 4 | 6 | 7 | 10 | 12 | 13 | 15 |
| 1 | 4 | 6 | 7 | 8 | 10 | 13 | 14 | 15 |
| 2 | 3 | 4 | 5 | 6 | 8 | 11 | 13 | 15 |
| 2 | 3 | 4 | 5 | 6 | 8 | 9 | 10 | 11 |
| 2 | 4 | 5 | 6 | 9 | 11 | 12 | 13 | 15 |
| 2 | 3 | 4 | 5 | 6 | 7 | 9 | 10 | 12 |
| 2 | 3 | 4 | 6 | 9 | 10 | 12 | 13 | 15 |
| 2 | 4 | 5 | 8 | 9 | 10 | 11 | 12 | 15 |
| 2 | 3 | 8 | 9 | 10 | 11 | 12 | 13 | 15 |
| 2 | 3 | 5 | 8 | 11 | 12 | 13 | 14 | 15 |
| 2 | 4 | 5 | 7 | 9 | 11 | 12 | 14 | 15 |
| 3 | 5 | 6 | 7 | 8 | 10 | 12 | 13 | 14 |
| 3 | 5 | 6 | 8 | 9 | 10 | 11 | 12 | 13 |
| 3 | 4 | 5 | 6 | 8 | 10 | 12 | 13 | 15 |
| 4 | 5 | 6 | 7 | 8 | 9 | 11 | 14 | 15 |
| 4 | 5 | 6 | 7 | 8 | 10 | 11 | 13 | 14 |
| 4 | 5 | 6 | 7 | 8 | 9 | 10 | 12 | 15 |
| 4 | 5 | 8 | 10 | 11 | 12 | 13 | 14 | 15 |
| 4 | 5 | 6 | 7 | 9 | 10 | 11 | 12 | 13 |
| 4 | 7 | 9 | 10 | 11 | 12 | 13 | 14 | 15 |
| 5 | 6 | 8 | 9 | 11 | 12 | 13 | 14 | 15 |

|   |   | 6 | 7 | 8 | 9 | 10 | 12 | 13 | 14 | 15 |
|---|---|---|---|---|---|----|----|----|----|----|

The column indices contained in 192 C-(1,1,2,2,2,2,1) inter-connected cycles:

| 0 | 2 | 3 | 4 | 6 | 8 | 10 | 11 | 15 |
|---|---|---|---|---|---|----|----|----|
| 0 | 2 | 3 | 4 | 8 | 9 | 10 | 13 | 15 |
| 0 | 1 | 3 | 4 | 5 | 8 | 11 | 13 | 15 |
| 0 | 2 | 3 | 4 | 5 | 8 | 13 | 14 | 15 |
| 0 | 1 | 3 | 4 | 5 | 6 | 12 | 14 | 15 |
| 0 | 1 | 3 | 4 | 5 | 7 | 12 | 13 | 15 |
| 0 | 1 | 3 | 4 | 6 | 9 | 10 | 12 | 15 |
| 0 | 2 | 3 | 4 | 6 | 7 | 10 | 12 | 15 |
| 0 | 1 | 3 | 4 | 5 | 6 | 11 | 14 | 15 |
| 0 | 1 | 3 | 4 | 6 | 9 | 10 | 11 | 15 |
| 0 | 1 | 3 | 4 | 8 | 9 | 11 | 13 | 15 |
| 0 | 1 | 3 | 4 | 7 | 9 | 12 | 13 | 15 |
| 0 | 2 | 3 | 4 | 7 | 9 | 10 | 13 | 15 |
| 0 | 2 | 3 | 4 | 5 | 7 | 13 | 14 | 15 |
| 0 | 2 | 3 | 4 | 6 | 7 | 12 | 14 | 15 |
| 0 | 2 | 3 | 4 | 6 | 8 | 11 | 14 | 15 |
| 0 | 1 | 2 | 4 | 5 | 6 | 8 | 9 | 10 |
| 0 | 2 | 3 | 4 | 5 | 7 | 8 | 9 | 10 |
| 0 | 1 | 4 | 5 | 7 | 8 | 9 | 11 | 15 |
| 0 | 2 | 4 | 5 | 7 | 8 | 9 | 14 | 15 |
| 0 | 2 | 4 | 5 | 6 | 9 | 10 | 12 | 13 |
| 0 | 3 | 4 | 5 | 6 | 9 | 10 | 11 | 12 |
| 0 | 4 | 5 | 7 | 9 | 11 | 12 | 13 | 15 |
| 0 | 4 | 5 | 6 | 9 | 11 | 12 | 14 | 15 |
| 0 | 1 | 3 | 4 | 5 | 6 | 9 | 10 | 11 |
| 0 | 1 | 4 | 5 | 6 | 9 | 11 | 14 | 15 |
| 0 | 1 | 3 | 4 | 5 | 7 | 8 | 9 | 11 |
| 0 | 1 | 2 | 4 | 5 | 6 | 8 | 9 | 14 |
| 0 | 2 | 3 | 4 | 5 | 7 | 9 | 10 | 13 |
| 0 | 2 | 4 | 5 | 7 | 9 | 13 | 14 | 15 |
| 0 | 2 | 4 | 5 | 6 | 9 | 12 | 13 | 14 |
| 0 | 3 | 4 | 5 | 7 | 9 | 11 | 12 | 13 |
| 0 | 4 | 6 | 7 | 8 | 10 | 11 | 14 | 15 |
| 0 | 4 | 7 | 8 | 9 | 10 | 13 | 14 | 15 |
| 0 | 1 | 4 | 5 | 6 | 8 | 10 | 13 | 14 |
| 0 | 3 | 4 | 5 | 7 | 8 | 10 | 13 | 14 |
| 0 | 1 | 3 | 4 | 5 | 10 | 11 | 12 | 14 |
| 0 | 1 | 2 | 4 | 5 | 10 | 12 | 13 | 14 |
| 0 | 1 | 4 | 9 | 10 | 11 | 12 | 14 | 15 |
| 0 | 2 | 4 | 7 | 10 | 11 | 12 | 14 | 15 |
| 0 | 1 | 3 | 4 | 5 | 6 | 10 | 11 | 14 |

| 0 | 1 | 2 | 4 | 9 | 10 | 12 | 13 | 14 |
|---|---|---|---|---|----|----|----|----|
| 0 | 1 | 4 | 6 | 9 | 10 | 11 | 14 | 15 |
| 0 | 1 | 4 | 6 | 8 | 9 | 10 | 13 | 14 |
| 0 | 2 | 3 | 4 | 5 | 7 | 10 | 13 | 14 |
| 0 | 2 | 4 | 7 | 9 | 10 | 13 | 14 | 15 |
| 0 | 2 | 3 | 4 | 7 | 10 | 11 | 12 | 14 |
| 0 | 3 | 4 | 6 | 7 | 8 | 10 | 11 | 14 |
| 0 | 1 | 2 | 5 | 7 | 8 | 10 | 11 | 12 |
| 0 | 1 | 2 | 5 | 8 | 9 | 10 | 12 | 14 |
| 0 | 1 | 2 | 8 | 10 | 11 | 12 | 13 | 15 |
| 0 | 2 | 3 | 8 | 10 | 11 | 12 | 14 | 15 |
| 0 | 1 | 2 | 5 | 6 | 7 | 8 | 10 | 11 |
| 0 | 1 | 2 | 4 | 6 | 8 | 9 | 10 | 13 |
| 0 | 1 | 2 | 6 | 8 | 10 | 11 | 13 | 15 |
| 0 | 1 | 2 | 8 | 9 | 10 | 12 | 13 | 14 |
| 0 | 2 | 3 | 4 | 6 | 7 | 8 | 10 | 11 |
| 0 | 2 | 3 | 5 | 7 | 8 | 9 | 10 | 14 |
| 0 | 2 | 3 | 7 | 8 | 10 | 11 | 12 | 14 |
| 0 | 2 | 3 | 8 | 9 | 10 | 13 | 14 | 15 |
| 0 | 1 | 2 | 5 | 6 | 7 | 8 | 12 | 15 |
| 0 | 1 | 3 | 5 | 7 | 8 | 9 | 12 | 15 |
| 0 | 1 | 6 | 7 | 8 | 10 | 12 | 13 | 15 |
| 0 | 3 | 6 | 7 | 8 | 10 | 12 | 14 | 15 |
| 0 | 1 | 2 | 5 | 6 | 7 | 8 | 11 | 15 |
| 0 | 1 | 4 | 7 | 8 | 9 | 11 | 13 | 15 |
| 0 | 1 | 6 | 7 | 8 | 10 | 11 | 13 | 15 |
| 0 | 1 | 3 | 7 | 8 | 9 | 12 | 13 | 15 |
| 0 | 2 | 4 | 6 | 7 | 8 | 11 | 14 | 15 |
| 0 | 2 | 3 | 6 | 7 | 8 | 12 | 14 | 15 |
| 0 | 2 | 3 | 5 | 7 | 8 | 9 | 14 | 15 |
| 0 | 3 | 7 | 8 | 9 | 10 | 13 | 14 | 15 |
| 0 | 5 | 6 | 7 | 8 | 10 | 11 | 12 | 13 |
| 0 | 5 | 6 | 8 | 9 | 10 | 12 | 13 | 14 |
| 0 | 2 | 5 | 6 | 8 | 11 | 12 | 13 | 15 |
| 0 | 3 | 5 | 8 | 9 | 11 | 12 | 13 | 15 |
| 0 | 1 | 5 | 6 | 7 | 8 | 10 | 11 | 13 |
| 0 | 1 | 2 | 5 | 6 | 8 | 11 | 13 | 15 |
| 0 | 1 | 2 | 4 | 5 | 6 | 8 | 13 | 14 |
| 0 | 1 | 3 | 4 | 5 | 7 | 8 | 11 | 13 |
| 0 | 2 | 5 | 6 | 8 | 9 | 12 | 13 | 14 |
| 0 | 2 | 3 | 5 | 8 | 9 | 13 | 14 | 15 |
| 0 | 3 | 5 | 7 | 8 | 9 | 10 | 13 | 14 |
| 0 | 3 | 5 | 7 | 8 | 9 | 11 | 12 | 13 |
| 0 | 1 | 3 | 5 | 7 | 8 | 9 | 11 | 12 |

| 0 | 1 | 3 | 5 | 9 | 10 | 11 | 12 | 14 |
| 0 | 1 | 2 | 5 | 6 | 8 | 9 | 12 | 14 |
| 0 | 1 | 3 | 5 | 6 | 9 | 12 | 14 | 15 |
| 0 | 1 | 2 | 4 | 5 | 6 | 12 | 13 | 14 |
| 0 | 1 | 2 | 5 | 6 | 7 | 12 | 13 | 15 |
| 0 | 1 | 3 | 4 | 5 | 7 | 11 | 12 | 13 |
| 0 | 1 | 2 | 5 | 7 | 10 | 11 | 12 | 13 |
| 0 | 1 | 2 | 4 | 6 | 9 | 10 | 12 | 13 |
| 0 | 1 | 6 | 8 | 9 | 10 | 12 | 13 | 14 |
| 0 | 1 | 3 | 6 | 9 | 10 | 12 | 14 | 15 |
| 0 | 1 | 2 | 6 | 7 | 10 | 12 | 13 | 15 |
| 0 | 2 | 3 | 4 | 6 | 7 | 10 | 11 | 12 |
| 0 | 2 | 5 | 6 | 7 | 10 | 11 | 12 | 13 |
| 0 | 3 | 6 | 7 | 8 | 10 | 11 | 12 | 14 |
| 0 | 3 | 5 | 6 | 9 | 10 | 11 | 12 | 14 |
| 0 | 1 | 3 | 8 | 9 | 11 | 12 | 13 | 15 |
| 0 | 1 | 3 | 9 | 10 | 11 | 12 | 14 | 15 |
| 0 | 1 | 4 | 7 | 9 | 11 | 12 | 13 | 15 |
| 0 | 1 | 2 | 7 | 10 | 11 | 12 | 13 | 15 |
| 0 | 2 | 4 | 6 | 7 | 11 | 12 | 14 | 15 |
| 0 | 2 | 5 | 6 | 7 | 11 | 12 | 13 | 15 |
| 0 | 2 | 3 | 6 | 8 | 11 | 12 | 14 | 15 |
| 0 | 3 | 5 | 6 | 9 | 11 | 12 | 14 | 15 |
| 1 | 2 | 3 | 5 | 6 | 7 | 9 | 10 | 11 |
| 1 | 2 | 4 | 5 | 6 | 8 | 9 | 10 | 12 |
| 1 | 3 | 4 | 5 | 6 | 9 | 10 | 12 | 15 |
| 1 | 3 | 5 | 6 | 7 | 10 | 11 | 13 | 14 |
| 1 | 4 | 5 | 6 | 8 | 10 | 12 | 13 | 14 |
| 1 | 5 | 6 | 7 | 8 | 10 | 12 | 13 | 15 |
| 1 | 2 | 5 | 6 | 7 | 8 | 10 | 12 | 15 |
| 1 | 2 | 3 | 5 | 6 | 7 | 9 | 10 | 15 |
| 1 | 3 | 4 | 5 | 6 | 10 | 12 | 14 | 15 |
| 1 | 3 | 5 | 6 | 7 | 10 | 13 | 14 | 15 |
| 1 | 4 | 5 | 7 | 8 | 9 | 11 | 12 | 15 |
| 1 | 4 | 5 | 9 | 10 | 11 | 12 | 14 | 15 |
| 1 | 2 | 5 | 6 | 7 | 9 | 11 | 14 | 15 |
| 1 | 2 | 3 | 5 | 6 | 11 | 13 | 14 | 15 |
| 1 | 2 | 5 | 8 | 10 | 11 | 12 | 13 | 15 |
| 1 | 3 | 4 | 5 | 8 | 11 | 12 | 13 | 15 |
| 1 | 2 | 3 | 5 | 10 | 11 | 13 | 14 | 15 |
| 1 | 2 | 5 | 7 | 8 | 10 | 11 | 12 | 15 |
| 1 | 2 | 5 | 7 | 9 | 10 | 11 | 14 | 15 |
| 1 | 3 | 4 | 5 | 10 | 11 | 12 | 14 | 15 |
| 1 | 2 | 3 | 4 | 6 | 8 | 9 | 11 | 13 |
| 1 | 2 | 3 | 6 | 9 | 10 | 11 | 13 | 15 |
| 1 | 2 | 3 | 8 | 9 | 11 | 12 | 13 | 14 |
| 1 | 2 | 3 | 4 | 6 | 7 | 8 | 9 | 11 |
| 1 | 2 | 3 | 5 | 7 | 9 | 10 | 11 | 14 |
| 1 | 2 | 3 | 7 | 8 | 9 | 11 | 12 | 14 |
| 1 | 2 | 3 | 9 | 10 | 11 | 13 | 14 | 15 |
| 1 | 2 | 3 | 4 | 6 | 7 | 9 | 12 | 13 |
| 1 | 2 | 4 | 7 | 9 | 11 | 12 | 13 | 14 |
| 1 | 2 | 3 | 4 | 6 | 7 | 8 | 9 | 12 |
| 1 | 2 | 4 | 5 | 8 | 9 | 10 | 12 | 14 |
| 1 | 2 | 4 | 7 | 8 | 9 | 11 | 12 | 14 |
| 1 | 3 | 4 | 5 | 7 | 8 | 9 | 12 | 15 |
| 1 | 4 | 6 | 7 | 8 | 9 | 11 | 13 | 14 |
| 1 | 6 | 7 | 9 | 10 | 11 | 13 | 14 | 15 |
| 1 | 3 | 6 | 7 | 8 | 9 | 12 | 13 | 14 |
| 1 | 2 | 4 | 6 | 7 | 8 | 9 | 11 | 14 |
| 1 | 2 | 3 | 6 | 7 | 8 | 9 | 12 | 14 |
| 1 | 2 | 3 | 5 | 6 | 7 | 9 | 14 | 15 |
| 1 | 3 | 6 | 7 | 9 | 10 | 13 | 14 | 15 |
| 1 | 2 | 3 | 6 | 7 | 9 | 10 | 13 | 15 |
| 1 | 2 | 3 | 5 | 6 | 7 | 13 | 14 | 15 |
| 1 | 2 | 3 | 4 | 6 | 7 | 12 | 13 | 14 |
| 1 | 2 | 3 | 4 | 6 | 8 | 11 | 13 | 14 |
| 1 | 2 | 3 | 5 | 7 | 10 | 11 | 13 | 14 |
| 1 | 2 | 7 | 9 | 10 | 11 | 13 | 14 | 15 |
| 1 | 2 | 3 | 4 | 7 | 11 | 12 | 13 | 14 |
| 1 | 3 | 4 | 6 | 7 | 8 | 11 | 13 | 14 |
| 1 | 2 | 4 | 5 | 8 | 10 | 12 | 13 | 14 |
| 1 | 2 | 3 | 4 | 8 | 11 | 12 | 13 | 14 |
| 1 | 3 | 4 | 5 | 7 | 8 | 12 | 13 | 15 |
| 1 | 3 | 4 | 6 | 7 | 8 | 12 | 13 | 14 |
| 2 | 3 | 5 | 6 | 7 | 9 | 10 | 11 | 13 |
| 2 | 4 | 6 | 7 | 9 | 11 | 12 | 13 | 14 |
| 2 | 5 | 6 | 7 | 9 | 11 | 13 | 14 | 15 |
| 2 | 3 | 4 | 6 | 7 | 9 | 11 | 12 | 13 |
| 2 | 4 | 5 | 6 | 8 | 9 | 10 | 12 | 13 |
| 2 | 5 | 6 | 8 | 10 | 11 | 12 | 13 | 15 |
| 2 | 3 | 4 | 6 | 7 | 8 | 10 | 12 | 15 |
| 2 | 3 | 6 | 8 | 9 | 11 | 12 | 13 | 14 |
| 2 | 3 | 4 | 6 | 8 | 9 | 11 | 12 | 13 |
| 2 | 3 | 4 | 6 | 8 | 10 | 11 | 12 | 15 |
| 2 | 3 | 4 | 5 | 8 | 10 | 13 | 14 | 15 |
| 2 | 3 | 5 | 6 | 9 | 10 | 11 | 13 | 15 |
| 2 | 3 | 4 | 5 | 8 | 9 | 10 | 13 | 15 |
| 2 | 4 | 5 | 7 | 8 | 9 | 10 | 14 | 15 |
| 2 | 4 | 7 | 8 | 10 | 11 | 12 | 14 | 15 |
| 2 | 3 | 4 | 5 | 7 | 8 | 9 | 10 | 15 |
| 2 | 3 | 4 | 8 | 10 | 11 | 12 | 14 | 15 |
| 2 | 3 | 5 | 6 | 9 | 11 | 13 | 14 | 15 |
| 3 | 4 | 5 | 7 | 8 | 10 | 13 | 14 | 15 |
| 3 | 4 | 6 | 7 | 8 | 10 | 12 | 14 | 15 |
| 3 | 5 | 6 | 7 | 9 | 10 | 11 | 13 | 14 |
| 3 | 6 | 7 | 8 | 9 | 11 | 12 | 13 | 14 |
| 3 | 4 | 5 | 6 | 9 | 10 | 11 | 12 | 15 |
| 3 | 4 | 5 | 8 | 9 | 11 | 12 | 13 | 15 |
| 4 | 5 | 6 | 8 | 9 | 10 | 12 | 13 | 14 |
| 4 | 6 | 7 | 8 | 9 | 11 | 12 | 13 | 14 |
| 4 | 5 | 7 | 8 | 9 | 10 | 13 | 14 | 15 |
| 4 | 5 | 7 | 8 | 9 | 11 | 12 | 13 | 15 |
| 4 | 5 | 6 | 9 | 10 | 11 | 12 | 14 | 15 |
| 4 | 6 | 7 | 8 | 10 | 11 | 12 | 14 | 15 |
| 5 | 6 | 7 | 9 | 10 | 11 | 13 | 14 | 15 |
| 5 | 6 | 7 | 8 | 10 | 11 | 12 | 13 | 15 |

The column indices contained in 144 E-(4,1,4) inter-connected cycles:

| | | | | | | | | |
|---|---|---|---|---|---|---|---|---|
| 0 | 1 | 2 | 3 | 4 | 5 | 6 | 13 | 15 |
| 0 | 1 | 2 | 3 | 4 | 6 | 7 | 13 | 15 |
| 0 | 1 | 2 | 3 | 4 | 6 | 8 | 13 | 15 |
| 0 | 1 | 2 | 3 | 4 | 6 | 9 | 13 | 15 |
| 0 | 1 | 2 | 3 | 4 | 6 | 10 | 13 | 15 |
| 0 | 1 | 2 | 3 | 4 | 6 | 11 | 13 | 15 |
| 0 | 1 | 2 | 3 | 4 | 6 | 12 | 13 | 15 |
| 0 | 1 | 2 | 3 | 4 | 6 | 13 | 14 | 15 |
| 0 | 1 | 2 | 4 | 5 | 6 | 7 | 9 | 11 |
| 0 | 2 | 3 | 4 | 5 | 6 | 7 | 9 | 11 |
| 0 | 2 | 4 | 5 | 6 | 7 | 8 | 9 | 11 |
| 0 | 2 | 4 | 5 | 6 | 7 | 9 | 10 | 11 |
| 0 | 2 | 4 | 5 | 6 | 7 | 9 | 11 | 12 |
| 0 | 2 | 4 | 5 | 6 | 7 | 9 | 11 | 13 |
| 0 | 2 | 4 | 5 | 6 | 7 | 9 | 11 | 14 |
| 0 | 2 | 4 | 5 | 6 | 7 | 9 | 11 | 15 |
| 0 | 1 | 2 | 4 | 7 | 10 | 11 | 13 | 14 |
| 0 | 1 | 3 | 4 | 7 | 10 | 11 | 13 | 14 |
| 0 | 1 | 4 | 5 | 7 | 10 | 11 | 13 | 14 |
| 0 | 1 | 4 | 6 | 7 | 10 | 11 | 13 | 14 |
| 0 | 1 | 4 | 7 | 8 | 10 | 11 | 13 | 14 |
| 0 | 1 | 4 | 7 | 9 | 10 | 11 | 13 | 14 |
| 0 | 1 | 4 | 7 | 10 | 11 | 12 | 13 | 14 |
| 0 | 1 | 4 | 7 | 10 | 11 | 13 | 14 | 15 |
| 0 | 1 | 2 | 3 | 4 | 8 | 9 | 10 | 11 |
| 0 | 1 | 2 | 3 | 5 | 8 | 9 | 10 | 11 |
| 0 | 1 | 2 | 3 | 6 | 8 | 9 | 10 | 11 |
| 0 | 1 | 2 | 3 | 7 | 8 | 9 | 10 | 11 |
| 0 | 1 | 2 | 3 | 8 | 9 | 10 | 11 | 12 |
| 0 | 1 | 2 | 3 | 8 | 9 | 10 | 11 | 13 |
| 0 | 1 | 2 | 3 | 8 | 9 | 10 | 11 | 14 |
| 0 | 1 | 2 | 3 | 8 | 9 | 10 | 11 | 15 |
| 0 | 1 | 2 | 6 | 7 | 8 | 9 | 14 | 15 |
| 0 | 1 | 3 | 6 | 7 | 8 | 9 | 14 | 15 |
| 0 | 1 | 4 | 6 | 7 | 8 | 9 | 14 | 15 |
| 0 | 1 | 5 | 6 | 7 | 8 | 9 | 14 | 15 |
| 0 | 1 | 6 | 7 | 8 | 9 | 10 | 14 | 15 |
| 0 | 1 | 6 | 7 | 8 | 9 | 11 | 14 | 15 |
| 0 | 1 | 6 | 7 | 8 | 9 | 12 | 14 | 15 |
| 0 | 1 | 6 | 7 | 8 | 9 | 13 | 14 | 15 |
| 0 | 1 | 3 | 5 | 6 | 8 | 11 | 13 | 14 |
| 0 | 2 | 3 | 5 | 6 | 8 | 11 | 13 | 14 |
| 0 | 3 | 4 | 5 | 6 | 8 | 11 | 13 | 14 |
| 0 | 3 | 5 | 6 | 7 | 8 | 11 | 13 | 14 |
| 0 | 3 | 5 | 6 | 8 | 9 | 11 | 13 | 14 |
| 0 | 3 | 5 | 6 | 8 | 10 | 11 | 13 | 14 |
| 0 | 3 | 5 | 6 | 8 | 11 | 12 | 13 | 14 |
| 0 | 3 | 5 | 6 | 8 | 11 | 13 | 14 | 15 |
| 0 | 1 | 2 | 3 | 4 | 5 | 7 | 12 | 14 |
| 0 | 1 | 2 | 3 | 5 | 6 | 7 | 12 | 14 |
| 0 | 1 | 2 | 3 | 5 | 7 | 8 | 12 | 14 |
| 0 | 1 | 2 | 3 | 5 | 7 | 9 | 12 | 14 |
| 0 | 1 | 2 | 3 | 5 | 7 | 10 | 12 | 14 |
| 0 | 1 | 2 | 3 | 5 | 7 | 11 | 12 | 14 |
| 0 | 1 | 2 | 3 | 5 | 7 | 12 | 13 | 14 |
| 0 | 1 | 2 | 3 | 5 | 7 | 12 | 14 | 15 |
| 0 | 1 | 3 | 6 | 7 | 9 | 10 | 12 | 13 |
| 0 | 2 | 3 | 6 | 7 | 9 | 10 | 12 | 13 |
| 0 | 3 | 4 | 6 | 7 | 9 | 10 | 12 | 13 |
| 0 | 3 | 5 | 6 | 7 | 9 | 10 | 12 | 13 |
| 0 | 3 | 6 | 7 | 8 | 9 | 10 | 12 | 13 |
| 0 | 3 | 6 | 7 | 9 | 10 | 11 | 12 | 13 |
| 0 | 3 | 6 | 7 | 9 | 10 | 12 | 13 | 14 |
| 0 | 3 | 6 | 7 | 9 | 10 | 12 | 13 | 15 |
| 0 | 1 | 2 | 9 | 11 | 12 | 13 | 14 | 15 |
| 0 | 2 | 3 | 9 | 11 | 12 | 13 | 14 | 15 |
| 0 | 2 | 4 | 9 | 11 | 12 | 13 | 14 | 15 |
| 0 | 2 | 5 | 9 | 11 | 12 | 13 | 14 | 15 |
| 0 | 2 | 6 | 9 | 11 | 12 | 13 | 14 | 15 |
| 0 | 2 | 7 | 9 | 11 | 12 | 13 | 14 | 15 |
| 0 | 2 | 8 | 9 | 11 | 12 | 13 | 14 | 15 |
| 0 | 2 | 9 | 10 | 11 | 12 | 13 | 14 | 15 |
| 0 | 1 | 3 | 4 | 5 | 6 | 7 | 8 | 10 |
| 1 | 2 | 3 | 4 | 5 | 6 | 7 | 8 | 10 |
| 1 | 3 | 4 | 5 | 6 | 7 | 8 | 9 | 10 |
| 1 | 3 | 4 | 5 | 6 | 7 | 8 | 10 | 11 |
| 1 | 3 | 4 | 5 | 6 | 7 | 8 | 10 | 12 |
| 1 | 3 | 4 | 5 | 6 | 7 | 8 | 10 | 13 |
| 1 | 3 | 4 | 5 | 6 | 7 | 8 | 10 | 14 |
| 1 | 3 | 4 | 5 | 6 | 7 | 8 | 10 | 15 |
| 0 | 1 | 2 | 4 | 5 | 8 | 11 | 14 | 15 |
| 1 | 2 | 3 | 4 | 5 | 8 | 11 | 14 | 15 |
| 1 | 2 | 4 | 5 | 6 | 8 | 11 | 14 | 15 |
| 1 | 2 | 4 | 5 | 7 | 8 | 11 | 14 | 15 |
| 1 | 2 | 4 | 5 | 8 | 9 | 11 | 14 | 15 |
| 1 | 2 | 4 | 5 | 8 | 10 | 11 | 14 | 15 |
| 1 | 2 | 4 | 5 | 8 | 11 | 12 | 14 | 15 |
| 1 | 2 | 4 | 5 | 8 | 11 | 13 | 14 | 15 |
| 0 | 1 | 2 | 4 | 7 | 9 | 10 | 12 | 15 |
| 1 | 2 | 3 | 4 | 7 | 9 | 10 | 12 | 15 |
| 1 | 2 | 4 | 5 | 7 | 9 | 10 | 12 | 15 |
| 1 | 2 | 4 | 6 | 7 | 9 | 10 | 12 | 15 |
| 1 | 2 | 4 | 7 | 8 | 9 | 10 | 12 | 15 |
| 1 | 2 | 4 | 7 | 9 | 10 | 11 | 12 | 15 |
| 1 | 2 | 4 | 7 | 9 | 10 | 12 | 13 | 15 |
| 1 | 2 | 4 | 7 | 9 | 10 | 12 | 14 | 15 |
| 0 | 1 | 3 | 8 | 10 | 12 | 13 | 14 | 15 |
| 1 | 2 | 3 | 8 | 10 | 12 | 13 | 14 | 15 |
| 1 | 3 | 4 | 8 | 10 | 12 | 13 | 14 | 15 |
| 1 | 3 | 5 | 8 | 10 | 12 | 13 | 14 | 15 |
| 1 | 3 | 6 | 8 | 10 | 12 | 13 | 14 | 15 |
| 1 | 3 | 7 | 8 | 10 | 12 | 13 | 14 | 15 |
| 1 | 3 | 8 | 9 | 10 | 12 | 13 | 14 | 15 |
| 1 | 3 | 8 | 10 | 11 | 12 | 13 | 14 | 15 |
| 0 | 2 | 3 | 5 | 6 | 8 | 9 | 12 | 15 |
| 1 | 2 | 3 | 5 | 6 | 8 | 9 | 12 | 15 |
| 2 | 3 | 4 | 5 | 6 | 8 | 9 | 12 | 15 |
| 2 | 3 | 5 | 6 | 7 | 8 | 9 | 12 | 15 |
| 2 | 3 | 5 | 6 | 8 | 9 | 10 | 12 | 15 |
| 2 | 3 | 5 | 6 | 8 | 9 | 11 | 12 | 15 |
| 2 | 3 | 5 | 6 | 8 | 9 | 12 | 13 | 15 |
| 2 | 3 | 5 | 6 | 8 | 9 | 12 | 14 | 15 |
| 0 | 2 | 3 | 4 | 5 | 10 | 11 | 12 | 13 |
| 1 | 2 | 3 | 4 | 5 | 10 | 11 | 12 | 13 |
| 2 | 3 | 4 | 5 | 6 | 10 | 11 | 12 | 13 |
| 2 | 3 | 4 | 5 | 7 | 10 | 11 | 12 | 13 |
| 2 | 3 | 4 | 5 | 8 | 10 | 11 | 12 | 13 |
| 2 | 3 | 4 | 5 | 9 | 10 | 11 | 12 | 13 |
| 2 | 3 | 4 | 5 | 10 | 11 | 12 | 13 | 14 |
| 2 | 3 | 4 | 5 | 10 | 11 | 12 | 13 | 15 |
| 0 | 4 | 6 | 8 | 9 | 10 | 11 | 13 | 15 |
| 1 | 4 | 6 | 8 | 9 | 10 | 11 | 13 | 15 |
| 2 | 4 | 6 | 8 | 9 | 10 | 11 | 13 | 15 |
| 3 | 4 | 6 | 8 | 9 | 10 | 11 | 13 | 15 |
| 4 | 5 | 6 | 8 | 9 | 10 | 11 | 13 | 15 |
| 4 | 6 | 7 | 8 | 9 | 10 | 11 | 13 | 15 |
| 4 | 6 | 8 | 9 | 10 | 11 | 12 | 13 | 15 |
| 4 | 6 | 8 | 9 | 10 | 11 | 13 | 14 | 15 |
| 0 | 4 | 5 | 6 | 7 | 12 | 13 | 14 | 15 |
| 1 | 4 | 5 | 6 | 7 | 12 | 13 | 14 | 15 |
| 2 | 4 | 5 | 6 | 7 | 12 | 13 | 14 | 15 |
| 3 | 4 | 5 | 6 | 7 | 12 | 13 | 14 | 15 |
| 4 | 5 | 6 | 7 | 8 | 12 | 13 | 14 | 15 |
| 4 | 5 | 6 | 7 | 9 | 12 | 13 | 14 | 15 |
| 4 | 5 | 6 | 7 | 10 | 12 | 13 | 14 | 15 |
| 4 | 5 | 6 | 7 | 11 | 12 | 13 | 14 | 15 |
| 0 | 5 | 7 | 8 | 9 | 10 | 11 | 12 | 14 |
| 1 | 5 | 7 | 8 | 9 | 10 | 11 | 12 | 14 |
| 2 | 5 | 7 | 8 | 9 | 10 | 11 | 12 | 14 |
| 3 | 5 | 7 | 8 | 9 | 10 | 11 | 12 | 14 |
| 4 | 5 | 7 | 8 | 9 | 10 | 11 | 12 | 14 |
| 5 | 6 | 7 | 8 | 9 | 10 | 11 | 12 | 14 |
| 5 | 7 | 8 | 9 | 10 | 11 | 12 | 13 | 14 |
| 5 | 7 | 8 | 9 | 10 | 11 | 12 | 14 | 15 |

The column indices contained in 576 F-(2,3,1,2,1) inter-connected cycles:

| | | | | | | | | |
|---|---|---|---|---|---|---|---|---|
| 0 | 2 | 3 | 4 | 5 | 8 | 9 | 10 | 15 |
| 0 | 2 | 3 | 4 | 8 | 10 | 11 | 12 | 15 |
| 0 | 2 | 3 | 4 | 5 | 8 | 10 | 13 | 15 |
| 0 | 2 | 3 | 4 | 6 | 8 | 10 | 12 | 15 |
| 0 | 1 | 3 | 4 | 5 | 8 | 12 | 13 | 15 |
| 0 | 3 | 4 | 5 | 8 | 10 | 13 | 14 | 15 |
| 0 | 3 | 4 | 5 | 8 | 11 | 12 | 13 | 15 |
| 0 | 1 | 3 | 4 | 5 | 10 | 12 | 14 | 15 |
| 0 | 1 | 3 | 4 | 5 | 7 | 8 | 12 | 15 |
| 0 | 1 | 3 | 4 | 5 | 6 | 10 | 12 | 15 |
| 0 | 3 | 4 | 5 | 6 | 9 | 10 | 12 | 15 |
| 0 | 3 | 4 | 6 | 7 | 8 | 10 | 12 | 15 |
| 0 | 1 | 3 | 4 | 5 | 7 | 11 | 13 | 15 |
| 0 | 1 | 3 | 4 | 5 | 6 | 10 | 11 | 15 |
| 0 | 1 | 3 | 4 | 5 | 7 | 8 | 11 | 15 |
| 0 | 1 | 3 | 4 | 5 | 10 | 11 | 14 | 15 |
| 0 | 1 | 3 | 4 | 7 | 9 | 11 | 13 | 15 |
| 0 | 1 | 3 | 4 | 6 | 9 | 11 | 14 | 15 |
| 0 | 1 | 3 | 4 | 7 | 8 | 9 | 11 | 15 |
| 0 | 1 | 3 | 4 | 9 | 10 | 11 | 14 | 15 |
| 0 | 1 | 3 | 4 | 8 | 9 | 12 | 13 | 15 |
| 0 | 1 | 3 | 4 | 9 | 10 | 12 | 14 | 15 |
| 0 | 1 | 3 | 4 | 6 | 9 | 12 | 14 | 15 |
| 0 | 1 | 3 | 4 | 7 | 8 | 9 | 12 | 15 |
| 0 | 2 | 3 | 4 | 5 | 7 | 10 | 13 | 15 |
| 0 | 2 | 3 | 4 | 6 | 7 | 10 | 11 | 15 |
| 0 | 2 | 3 | 4 | 7 | 10 | 11 | 12 | 15 |
| 0 | 2 | 3 | 4 | 5 | 7 | 9 | 10 | 15 |
| 0 | 2 | 3 | 4 | 7 | 9 | 13 | 14 | 15 |
| 0 | 2 | 3 | 4 | 6 | 7 | 11 | 14 | 15 |
| 0 | 2 | 3 | 4 | 7 | 11 | 12 | 14 | 15 |
| 0 | 2 | 3 | 4 | 5 | 7 | 9 | 14 | 15 |
| 0 | 2 | 3 | 4 | 5 | 8 | 9 | 14 | 15 |
| 0 | 2 | 3 | 4 | 8 | 9 | 13 | 14 | 15 |
| 0 | 2 | 3 | 4 | 6 | 8 | 12 | 14 | 15 |
| 0 | 2 | 3 | 4 | 8 | 11 | 12 | 14 | 15 |
| 0 | 3 | 4 | 7 | 9 | 11 | 12 | 13 | 15 |
| 0 | 3 | 4 | 7 | 9 | 10 | 13 | 14 | 15 |
| 0 | 3 | 4 | 6 | 7 | 8 | 11 | 14 | 15 |
| 0 | 3 | 4 | 5 | 6 | 9 | 11 | 14 | 15 |
| 0 | 3 | 4 | 8 | 9 | 10 | 13 | 14 | 15 |
| 0 | 3 | 4 | 8 | 9 | 11 | 12 | 13 | 15 |
| 0 | 3 | 4 | 5 | 6 | 9 | 12 | 14 | 15 |
| 0 | 3 | 4 | 6 | 7 | 8 | 12 | 14 | 15 |
| 0 | 3 | 4 | 5 | 7 | 11 | 12 | 13 | 15 |
| 0 | 3 | 4 | 5 | 7 | 10 | 13 | 14 | 15 |
| 0 | 3 | 4 | 6 | 7 | 8 | 10 | 11 | 15 |
| 0 | 3 | 4 | 5 | 6 | 9 | 10 | 11 | 15 |
| 0 | 1 | 2 | 4 | 5 | 8 | 9 | 10 | 12 |
| 0 | 2 | 4 | 5 | 7 | 8 | 9 | 10 | 15 |
| 0 | 2 | 4 | 5 | 6 | 8 | 9 | 10 | 12 |
| 0 | 1 | 4 | 5 | 7 | 8 | 9 | 12 | 15 |
| 0 | 4 | 5 | 7 | 8 | 9 | 10 | 14 | 15 |
| 0 | 4 | 5 | 7 | 8 | 9 | 11 | 12 | 15 |
| 0 | 4 | 5 | 6 | 9 | 10 | 11 | 12 | 15 |
| 0 | 4 | 5 | 6 | 8 | 9 | 10 | 12 | 13 |
| 0 | 4 | 5 | 9 | 10 | 11 | 12 | 14 | 15 |
| 0 | 4 | 5 | 8 | 9 | 11 | 12 | 13 | 15 |
| 0 | 1 | 4 | 5 | 6 | 9 | 10 | 11 | 15 |
| 0 | 1 | 2 | 4 | 5 | 6 | 9 | 10 | 13 |
| 0 | 1 | 4 | 5 | 6 | 8 | 9 | 10 | 13 |
| 0 | 1 | 3 | 4 | 5 | 6 | 9 | 10 | 15 |
| 0 | 1 | 4 | 5 | 7 | 9 | 11 | 13 | 15 |
| 0 | 1 | 4 | 5 | 8 | 9 | 11 | 13 | 15 |
| 0 | 1 | 4 | 5 | 9 | 10 | 11 | 14 | 15 |
| 0 | 1 | 3 | 4 | 5 | 7 | 9 | 11 | 13 |
| 0 | 1 | 3 | 4 | 5 | 6 | 9 | 11 | 14 |
| 0 | 1 | 3 | 4 | 5 | 9 | 10 | 11 | 14 |
| 0 | 1 | 3 | 4 | 5 | 8 | 9 | 11 | 13 |
| 0 | 1 | 2 | 4 | 5 | 6 | 9 | 13 | 14 |
| 0 | 1 | 3 | 4 | 5 | 6 | 9 | 14 | 15 |
| 0 | 1 | 4 | 5 | 6 | 8 | 9 | 13 | 14 |
| 0 | 2 | 4 | 5 | 7 | 9 | 10 | 13 | 15 |
| 0 | 1 | 2 | 4 | 5 | 9 | 10 | 12 | 13 |
| 0 | 2 | 3 | 4 | 5 | 9 | 10 | 13 | 15 |
| 0 | 2 | 3 | 4 | 5 | 7 | 8 | 9 | 14 |
| 0 | 2 | 4 | 5 | 6 | 8 | 9 | 12 | 14 |
| 0 | 1 | 2 | 4 | 5 | 8 | 9 | 12 | 14 |
| 0 | 2 | 3 | 4 | 5 | 7 | 9 | 13 | 14 |
| 0 | 1 | 2 | 4 | 5 | 9 | 12 | 13 | 14 |
| 0 | 2 | 3 | 4 | 5 | 9 | 13 | 14 | 15 |
| 0 | 3 | 4 | 5 | 7 | 8 | 9 | 10 | 14 |
| 0 | 3 | 4 | 5 | 7 | 8 | 9 | 11 | 12 |
| 0 | 1 | 3 | 4 | 5 | 7 | 8 | 9 | 12 |
| 0 | 1 | 3 | 4 | 5 | 7 | 9 | 12 | 13 |
| 0 | 3 | 4 | 5 | 7 | 9 | 10 | 13 | 14 |
| 0 | 3 | 4 | 5 | 6 | 9 | 11 | 12 | 14 |
| 0 | 3 | 4 | 5 | 9 | 10 | 11 | 12 | 14 |
| 0 | 3 | 4 | 5 | 8 | 9 | 11 | 12 | 13 |
| 0 | 4 | 5 | 6 | 8 | 9 | 12 | 13 | 14 |
| 0 | 1 | 4 | 5 | 7 | 9 | 12 | 13 | 15 |
| 0 | 4 | 5 | 7 | 9 | 10 | 13 | 14 | 15 |
| 0 | 4 | 5 | 7 | 8 | 10 | 13 | 14 | 15 |
| 0 | 4 | 7 | 8 | 10 | 11 | 12 | 14 | 15 |
| 0 | 4 | 6 | 7 | 8 | 10 | 12 | 14 | 15 |
| 0 | 1 | 4 | 5 | 8 | 10 | 12 | 13 | 14 |
| 0 | 4 | 5 | 6 | 8 | 10 | 12 | 13 | 14 |

| 0 | 1 | 2 | 4 | 5 | 8 | 10 | 12 | 14 |
|---|---|---|---|---|---|----|----|----|
| 0 | 1 | 4 | 5 | 10 | 11 | 12 | 14 | 15 |
| 0 | 2 | 4 | 8 | 10 | 11 | 12 | 14 | 15 |
| 0 | 1 | 2 | 4 | 5 | 6 | 10 | 13 | 14 |
| 0 | 1 | 4 | 5 | 6 | 10 | 11 | 14 | 15 |
| 0 | 1 | 2 | 4 | 5 | 6 | 8 | 10 | 14 |
| 0 | 1 | 3 | 4 | 5 | 6 | 10 | 14 | 15 |
| 0 | 1 | 4 | 8 | 9 | 10 | 12 | 13 | 14 |
| 0 | 1 | 3 | 4 | 9 | 10 | 11 | 12 | 14 |
| 0 | 1 | 2 | 4 | 8 | 9 | 10 | 12 | 14 |
| 0 | 1 | 3 | 4 | 6 | 9 | 10 | 11 | 14 |
| 0 | 1 | 2 | 4 | 6 | 9 | 10 | 13 | 14 |
| 0 | 1 | 3 | 4 | 6 | 9 | 10 | 14 | 15 |
| 0 | 1 | 2 | 4 | 6 | 8 | 9 | 10 | 14 |
| 0 | 2 | 4 | 5 | 7 | 10 | 13 | 14 | 15 |
| 0 | 2 | 4 | 5 | 6 | 10 | 12 | 13 | 14 |
| 0 | 2 | 3 | 4 | 5 | 10 | 13 | 14 | 15 |
| 0 | 2 | 4 | 6 | 7 | 10 | 11 | 14 | 15 |
| 0 | 2 | 4 | 6 | 7 | 10 | 12 | 14 | 15 |
| 0 | 2 | 4 | 5 | 7 | 9 | 10 | 14 | 15 |
| 0 | 2 | 3 | 4 | 6 | 7 | 10 | 11 | 14 |
| 0 | 2 | 3 | 4 | 7 | 9 | 10 | 13 | 14 |
| 0 | 2 | 3 | 4 | 6 | 7 | 10 | 12 | 14 |
| 0 | 2 | 3 | 4 | 5 | 7 | 9 | 10 | 14 |
| 0 | 2 | 4 | 6 | 9 | 10 | 12 | 13 | 14 |
| 0 | 2 | 3 | 4 | 9 | 10 | 13 | 14 | 15 |
| 0 | 3 | 4 | 7 | 8 | 9 | 10 | 13 | 14 |
| 0 | 3 | 4 | 7 | 8 | 10 | 11 | 12 | 14 |
| 0 | 3 | 4 | 6 | 7 | 8 | 10 | 12 | 14 |
| 0 | 2 | 3 | 4 | 6 | 8 | 10 | 11 | 14 |
| 0 | 3 | 4 | 5 | 6 | 9 | 10 | 11 | 14 |
| 0 | 2 | 3 | 4 | 8 | 10 | 11 | 12 | 14 |
| 0 | 4 | 6 | 8 | 9 | 10 | 12 | 13 | 14 |
| 0 | 2 | 4 | 6 | 8 | 10 | 11 | 14 | 15 |
| 0 | 4 | 5 | 6 | 9 | 10 | 11 | 14 | 15 |
| 0 | 1 | 2 | 5 | 7 | 8 | 10 | 12 | 15 |
| 0 | 1 | 2 | 5 | 8 | 10 | 11 | 12 | 15 |
| 0 | 2 | 5 | 8 | 10 | 11 | 12 | 13 | 15 |
| 0 | 1 | 2 | 5 | 6 | 8 | 10 | 11 | 15 |
| 0 | 1 | 2 | 5 | 6 | 8 | 9 | 10 | 14 |
| 0 | 1 | 2 | 5 | 6 | 7 | 8 | 10 | 15 |
| 0 | 1 | 2 | 6 | 7 | 8 | 10 | 11 | 13 |
| 0 | 1 | 2 | 6 | 8 | 9 | 10 | 13 | 14 |
| 0 | 1 | 2 | 6 | 7 | 8 | 10 | 13 | 15 |
| 0 | 1 | 2 | 4 | 6 | 8 | 10 | 13 | 14 |
| 0 | 1 | 2 | 4 | 8 | 9 | 10 | 12 | 13 |
| 0 | 1 | 2 | 7 | 8 | 10 | 12 | 13 | 15 |
| 0 | 1 | 2 | 7 | 8 | 10 | 11 | 12 | 13 |
| 0 | 1 | 2 | 4 | 8 | 10 | 12 | 13 | 14 |
| 0 | 2 | 3 | 6 | 7 | 8 | 10 | 11 | 14 |
| 0 | 2 | 4 | 6 | 7 | 8 | 10 | 11 | 14 |
| 0 | 2 | 5 | 6 | 7 | 8 | 10 | 11 | 13 |
| 0 | 2 | 3 | 7 | 8 | 9 | 10 | 13 | 14 |
| 0 | 2 | 3 | 6 | 7 | 8 | 10 | 12 | 14 |
| 0 | 2 | 3 | 5 | 7 | 8 | 10 | 13 | 14 |
| 0 | 2 | 6 | 8 | 9 | 10 | 12 | 13 | 14 |
| 0 | 2 | 7 | 8 | 9 | 10 | 13 | 14 | 15 |
| 0 | 2 | 3 | 4 | 5 | 7 | 8 | 10 | 13 |
| 0 | 2 | 3 | 4 | 7 | 8 | 9 | 10 | 13 |
| 0 | 2 | 3 | 4 | 7 | 8 | 10 | 11 | 12 |
| 0 | 2 | 3 | 4 | 6 | 7 | 8 | 10 | 12 |
| 0 | 2 | 3 | 6 | 8 | 10 | 11 | 14 | 15 |
| 0 | 2 | 3 | 5 | 8 | 9 | 10 | 14 | 15 |
| 0 | 2 | 3 | 6 | 8 | 10 | 12 | 14 | 15 |
| 0 | 2 | 3 | 5 | 8 | 10 | 13 | 14 | 15 |
| 0 | 2 | 4 | 7 | 8 | 9 | 10 | 13 | 15 |
| 0 | 2 | 4 | 6 | 8 | 9 | 10 | 12 | 13 |
| 0 | 2 | 5 | 6 | 8 | 9 | 10 | 12 | 14 |
| 0 | 2 | 5 | 7 | 8 | 9 | 10 | 14 | 15 |
| 0 | 2 | 5 | 6 | 8 | 10 | 11 | 13 | 15 |
| 0 | 2 | 5 | 7 | 8 | 10 | 11 | 12 | 13 |
| 0 | 2 | 4 | 7 | 8 | 10 | 11 | 12 | 14 |
| 0 | 1 | 5 | 6 | 7 | 8 | 10 | 12 | 15 |
| 0 | 5 | 6 | 7 | 8 | 10 | 12 | 13 | 15 |
| 0 | 1 | 3 | 5 | 7 | 8 | 9 | 11 | 15 |
| 0 | 1 | 5 | 6 | 7 | 8 | 10 | 11 | 15 |
| 0 | 1 | 2 | 5 | 7 | 8 | 10 | 11 | 15 |
| 0 | 1 | 2 | 6 | 7 | 8 | 11 | 13 | 15 |
| 0 | 1 | 3 | 7 | 8 | 9 | 11 | 13 | 15 |
| 0 | 1 | 2 | 7 | 8 | 10 | 11 | 13 | 15 |
| 0 | 1 | 3 | 4 | 7 | 8 | 11 | 13 | 15 |
| 0 | 1 | 4 | 7 | 8 | 9 | 12 | 13 | 15 |
| 0 | 1 | 2 | 6 | 7 | 8 | 12 | 13 | 15 |
| 0 | 1 | 3 | 4 | 7 | 8 | 12 | 13 | 15 |
| 0 | 2 | 3 | 6 | 7 | 8 | 11 | 14 | 15 |
| 0 | 2 | 3 | 4 | 6 | 7 | 8 | 11 | 15 |
| 0 | 2 | 5 | 6 | 7 | 8 | 11 | 13 | 15 |
| 0 | 2 | 4 | 6 | 7 | 8 | 12 | 14 | 15 |

| 0 | 2 | 5 | 6 | 7 | 8 | 12 | 13 | 15 |
|---|---|---|---|---|---|----|----|----|
| 0 | 2 | 3 | 4 | 6 | 7 | 8 | 12 | 15 |
| 0 | 2 | 3 | 7 | 8 | 9 | 13 | 14 | 15 |
| 0 | 2 | 3 | 7 | 8 | 11 | 12 | 14 | 15 |
| 0 | 2 | 3 | 5 | 7 | 8 | 13 | 14 | 15 |
| 0 | 2 | 4 | 7 | 8 | 9 | 13 | 14 | 15 |
| 0 | 2 | 4 | 5 | 7 | 8 | 13 | 14 | 15 |
| 0 | 2 | 4 | 7 | 8 | 11 | 12 | 14 | 15 |
| 0 | 2 | 3 | 7 | 8 | 9 | 10 | 13 | 15 |
| 0 | 3 | 7 | 8 | 9 | 11 | 12 | 13 | 15 |
| 0 | 3 | 5 | 7 | 8 | 9 | 10 | 14 | 15 |
| 0 | 3 | 5 | 7 | 8 | 9 | 11 | 12 | 15 |
| 0 | 2 | 3 | 5 | 7 | 8 | 9 | 10 | 15 |
| 0 | 3 | 6 | 7 | 8 | 10 | 11 | 14 | 15 |
| 0 | 3 | 7 | 8 | 10 | 11 | 12 | 14 | 15 |
| 0 | 3 | 5 | 7 | 8 | 10 | 13 | 14 | 15 |
| 0 | 4 | 7 | 8 | 9 | 11 | 12 | 13 | 15 |
| 0 | 5 | 6 | 7 | 8 | 10 | 11 | 13 | 15 |
| 0 | 5 | 6 | 8 | 10 | 11 | 12 | 13 | 15 |
| 0 | 1 | 5 | 6 | 8 | 10 | 11 | 13 | 15 |
| 0 | 1 | 5 | 6 | 8 | 9 | 10 | 13 | 14 |
| 0 | 1 | 5 | 6 | 7 | 8 | 10 | 13 | 15 |
| 0 | 1 | 3 | 5 | 8 | 9 | 11 | 13 | 15 |
| 0 | 1 | 2 | 5 | 8 | 10 | 11 | 13 | 15 |
| 0 | 1 | 2 | 5 | 6 | 8 | 9 | 13 | 14 |
| 0 | 1 | 2 | 5 | 6 | 7 | 8 | 11 | 13 |
| 0 | 1 | 2 | 5 | 6 | 7 | 8 | 13 | 15 |
| 0 | 1 | 2 | 4 | 5 | 6 | 8 | 9 | 13 |
| 0 | 1 | 3 | 5 | 7 | 8 | 9 | 11 | 13 |
| 0 | 1 | 2 | 5 | 7 | 8 | 10 | 11 | 13 |
| 0 | 1 | 4 | 5 | 7 | 8 | 9 | 11 | 13 |
| 0 | 2 | 4 | 5 | 6 | 8 | 12 | 13 | 14 |
| 0 | 2 | 5 | 6 | 7 | 8 | 11 | 12 | 13 |
| 0 | 2 | 4 | 5 | 6 | 8 | 9 | 12 | 13 |
| 0 | 2 | 3 | 4 | 5 | 7 | 8 | 13 | 14 |
| 0 | 1 | 2 | 4 | 5 | 8 | 12 | 13 | 14 |
| 0 | 2 | 3 | 5 | 7 | 8 | 9 | 13 | 14 |
| 0 | 1 | 2 | 5 | 8 | 9 | 12 | 13 | 14 |
| 0 | 2 | 5 | 7 | 8 | 9 | 13 | 14 | 15 |
| 0 | 3 | 4 | 5 | 7 | 8 | 11 | 12 | 13 |
| 0 | 1 | 3 | 4 | 5 | 7 | 8 | 12 | 13 |
| 0 | 1 | 3 | 5 | 7 | 8 | 9 | 12 | 13 |
| 0 | 2 | 3 | 5 | 7 | 8 | 9 | 10 | 13 |
| 0 | 3 | 5 | 8 | 9 | 10 | 13 | 14 | 15 |

| 0 | 1 | 3 | 5 | 8 | 9 | 12 | 13 | 15 |
|---|---|---|---|---|---|---|---|---|
| 0 | 2 | 3 | 5 | 8 | 9 | 10 | 13 | 15 |
| 0 | 1 | 5 | 8 | 9 | 10 | 12 | 13 | 14 |
| 0 | 5 | 7 | 8 | 9 | 10 | 13 | 14 | 15 |
| 0 | 4 | 5 | 7 | 8 | 9 | 11 | 12 | 13 |
| 0 | 1 | 3 | 5 | 6 | 9 | 11 | 12 | 14 |
| 0 | 1 | 3 | 5 | 7 | 9 | 11 | 12 | 13 |
| 0 | 1 | 3 | 5 | 6 | 9 | 10 | 11 | 12 |
| 0 | 1 | 3 | 5 | 8 | 9 | 11 | 12 | 13 |
| 0 | 1 | 2 | 5 | 6 | 9 | 12 | 13 | 14 |
| 0 | 1 | 5 | 6 | 9 | 11 | 12 | 14 | 15 |
| 0 | 1 | 5 | 6 | 8 | 9 | 12 | 13 | 14 |
| 0 | 1 | 2 | 5 | 6 | 7 | 11 | 12 | 13 |
| 0 | 1 | 2 | 5 | 6 | 11 | 12 | 13 | 15 |
| 0 | 1 | 2 | 4 | 5 | 6 | 9 | 12 | 13 |
| 0 | 1 | 4 | 5 | 7 | 9 | 11 | 12 | 13 |
| 0 | 1 | 5 | 6 | 7 | 10 | 11 | 12 | 13 |
| 0 | 1 | 2 | 4 | 5 | 6 | 8 | 12 | 14 |
| 0 | 1 | 2 | 5 | 6 | 8 | 11 | 12 | 15 |
| 0 | 1 | 2 | 5 | 6 | 7 | 8 | 11 | 12 |
| 0 | 1 | 2 | 4 | 5 | 6 | 8 | 9 | 12 |
| 0 | 1 | 2 | 5 | 9 | 10 | 12 | 13 | 14 |
| 0 | 1 | 2 | 5 | 7 | 10 | 12 | 13 | 15 |
| 0 | 1 | 2 | 5 | 10 | 11 | 12 | 13 | 15 |
| 0 | 1 | 3 | 4 | 5 | 6 | 10 | 11 | 12 |
| 0 | 1 | 3 | 4 | 5 | 6 | 11 | 12 | 14 |
| 0 | 1 | 3 | 4 | 5 | 7 | 8 | 11 | 12 |
| 0 | 1 | 3 | 4 | 5 | 8 | 11 | 12 | 13 |
| 0 | 1 | 3 | 5 | 9 | 10 | 12 | 14 | 15 |
| 0 | 1 | 3 | 5 | 7 | 9 | 12 | 13 | 15 |
| 0 | 1 | 3 | 5 | 6 | 9 | 10 | 12 | 15 |
| 0 | 1 | 4 | 5 | 6 | 11 | 12 | 14 | 15 |
| 0 | 1 | 4 | 5 | 6 | 8 | 12 | 13 | 14 |
| 0 | 1 | 5 | 9 | 10 | 11 | 12 | 14 | 15 |
| 0 | 1 | 5 | 6 | 7 | 10 | 12 | 13 | 15 |
| 0 | 1 | 5 | 6 | 7 | 8 | 10 | 11 | 12 |
| 0 | 1 | 4 | 5 | 7 | 8 | 9 | 11 | 12 |
| 0 | 1 | 3 | 4 | 6 | 9 | 10 | 11 | 12 |
| 0 | 1 | 4 | 6 | 9 | 10 | 11 | 12 | 15 |
| 0 | 1 | 4 | 6 | 8 | 9 | 10 | 12 | 13 |
| 0 | 1 | 2 | 4 | 6 | 8 | 9 | 10 | 12 |
| 0 | 1 | 2 | 6 | 9 | 10 | 12 | 13 | 14 |
| 0 | 1 | 3 | 6 | 9 | 10 | 11 | 12 | 14 |
| 0 | 1 | 2 | 6 | 8 | 9 | 10 | 12 | 14 |
| 0 | 1 | 6 | 9 | 10 | 11 | 12 | 14 | 15 |
| 0 | 1 | 2 | 6 | 7 | 10 | 11 | 12 | 13 |
| 0 | 1 | 2 | 6 | 10 | 11 | 12 | 13 | 15 |
| 0 | 1 | 2 | 4 | 6 | 10 | 12 | 13 | 14 |
| 0 | 1 | 6 | 7 | 8 | 10 | 11 | 12 | 13 |
| 0 | 1 | 6 | 8 | 10 | 11 | 12 | 13 | 15 |
| 0 | 1 | 4 | 6 | 8 | 10 | 12 | 13 | 14 |
| 0 | 2 | 3 | 6 | 7 | 10 | 11 | 12 | 14 |
| 0 | 1 | 2 | 5 | 6 | 7 | 10 | 11 | 12 |
| 0 | 2 | 4 | 6 | 7 | 10 | 11 | 12 | 14 |
| 0 | 2 | 5 | 6 | 7 | 10 | 12 | 13 | 15 |
| 0 | 2 | 5 | 6 | 9 | 10 | 12 | 13 | 14 |
| 0 | 2 | 5 | 6 | 10 | 11 | 12 | 13 | 15 |
| 0 | 2 | 3 | 6 | 7 | 10 | 12 | 14 | 15 |
| 0 | 1 | 2 | 5 | 6 | 7 | 10 | 12 | 15 |
| 0 | 1 | 3 | 5 | 6 | 10 | 11 | 12 | 14 |
| 0 | 2 | 3 | 6 | 8 | 10 | 11 | 12 | 14 |
| 0 | 3 | 4 | 6 | 7 | 8 | 10 | 11 | 12 |
| 0 | 2 | 3 | 4 | 6 | 8 | 10 | 11 | 12 |
| 0 | 3 | 5 | 6 | 9 | 10 | 12 | 14 | 15 |
| 0 | 1 | 3 | 5 | 6 | 10 | 12 | 14 | 15 |
| 0 | 5 | 6 | 9 | 10 | 11 | 12 | 14 | 15 |
| 0 | 4 | 6 | 7 | 8 | 10 | 11 | 12 | 14 |
| 0 | 1 | 3 | 7 | 9 | 11 | 12 | 13 | 15 |
| 0 | 1 | 3 | 6 | 9 | 11 | 12 | 14 | 15 |
| 0 | 1 | 3 | 7 | 8 | 9 | 11 | 12 | 15 |
| 0 | 1 | 3 | 6 | 9 | 10 | 11 | 12 | 15 |
| 0 | 1 | 4 | 6 | 9 | 11 | 12 | 14 | 15 |
| 0 | 1 | 4 | 8 | 9 | 11 | 12 | 13 | 15 |
| 0 | 1 | 4 | 7 | 8 | 9 | 11 | 12 | 15 |
| 0 | 1 | 2 | 6 | 7 | 11 | 12 | 13 | 15 |
| 0 | 1 | 3 | 4 | 7 | 11 | 12 | 13 | 15 |
| 0 | 1 | 6 | 7 | 10 | 11 | 12 | 13 | 15 |
| 0 | 1 | 2 | 6 | 8 | 11 | 12 | 13 | 15 |
| 0 | 1 | 3 | 4 | 8 | 11 | 12 | 13 | 15 |
| 0 | 2 | 3 | 6 | 7 | 11 | 12 | 14 | 15 |
| 0 | 1 | 2 | 5 | 6 | 7 | 11 | 12 | 15 |
| 0 | 2 | 3 | 4 | 6 | 7 | 11 | 12 | 15 |
| 0 | 2 | 4 | 6 | 8 | 11 | 12 | 14 | 15 |
| 0 | 2 | 3 | 4 | 6 | 8 | 11 | 12 | 15 |
| 0 | 2 | 3 | 7 | 10 | 11 | 12 | 14 | 15 |
| 0 | 2 | 5 | 7 | 10 | 11 | 12 | 13 | 15 |
| 0 | 1 | 2 | 5 | 7 | 10 | 11 | 12 | 15 |
| 0 | 1 | 3 | 5 | 6 | 11 | 12 | 14 | 15 |
| 0 | 3 | 6 | 7 | 8 | 11 | 12 | 14 | 15 |
| 0 | 3 | 5 | 9 | 10 | 11 | 12 | 14 | 15 |
| 0 | 3 | 5 | 7 | 9 | 11 | 12 | 13 | 15 |
| 0 | 3 | 5 | 6 | 9 | 10 | 11 | 12 | 15 |
| 0 | 1 | 3 | 5 | 10 | 11 | 12 | 14 | 15 |
| 0 | 4 | 6 | 7 | 8 | 11 | 12 | 14 | 15 |
| 0 | 5 | 6 | 7 | 10 | 11 | 12 | 13 | 15 |
| 1 | 2 | 3 | 5 | 6 | 9 | 10 | 11 | 13 |
| 1 | 3 | 4 | 5 | 6 | 9 | 10 | 11 | 12 |
| 1 | 3 | 5 | 6 | 7 | 9 | 10 | 11 | 13 |
| 1 | 2 | 4 | 5 | 6 | 9 | 10 | 12 | 13 |
| 1 | 4 | 5 | 6 | 9 | 10 | 11 | 12 | 15 |
| 1 | 4 | 5 | 6 | 8 | 9 | 10 | 12 | 13 |
| 1 | 5 | 6 | 7 | 8 | 10 | 11 | 12 | 13 |
| 1 | 5 | 6 | 7 | 9 | 10 | 11 | 13 | 14 |
| 1 | 5 | 6 | 8 | 10 | 11 | 12 | 13 | 15 |
| 1 | 5 | 6 | 8 | 9 | 10 | 12 | 13 | 14 |
| 1 | 2 | 5 | 6 | 7 | 8 | 10 | 11 | 12 |
| 1 | 2 | 3 | 5 | 6 | 7 | 10 | 11 | 14 |
| 1 | 2 | 5 | 6 | 7 | 9 | 10 | 11 | 14 |
| 1 | 2 | 4 | 5 | 6 | 8 | 10 | 12 | 14 |
| 1 | 2 | 5 | 6 | 8 | 9 | 10 | 12 | 14 |
| 1 | 2 | 5 | 6 | 8 | 10 | 11 | 12 | 15 |
| 1 | 2 | 3 | 5 | 6 | 7 | 10 | 14 | 15 |
| 1 | 2 | 5 | 6 | 7 | 9 | 10 | 14 | 15 |
| 1 | 3 | 4 | 5 | 6 | 10 | 11 | 12 | 14 |
| 1 | 2 | 3 | 5 | 6 | 10 | 11 | 13 | 14 |
| 1 | 3 | 5 | 6 | 7 | 9 | 10 | 13 | 15 |
| 1 | 2 | 3 | 5 | 6 | 9 | 10 | 13 | 15 |
| 1 | 2 | 3 | 5 | 6 | 10 | 13 | 14 | 15 |
| 1 | 2 | 4 | 5 | 6 | 10 | 12 | 13 | 14 |
| 1 | 4 | 5 | 6 | 10 | 11 | 12 | 14 | 15 |
| 1 | 5 | 6 | 7 | 9 | 10 | 13 | 14 | 15 |
| 1 | 4 | 5 | 7 | 9 | 11 | 12 | 13 | 15 |
| 1 | 4 | 5 | 6 | 9 | 11 | 12 | 14 | 15 |
| 1 | 4 | 5 | 8 | 9 | 11 | 12 | 13 | 15 |
| 1 | 2 | 5 | 6 | 9 | 11 | 13 | 14 | 15 |
| 1 | 5 | 6 | 7 | 9 | 11 | 13 | 14 | 15 |
| 1 | 2 | 3 | 5 | 6 | 9 | 11 | 13 | 15 |
| 1 | 2 | 5 | 6 | 8 | 11 | 12 | 13 | 15 |
| 1 | 3 | 5 | 8 | 9 | 11 | 12 | 13 | 15 |
| 1 | 2 | 3 | 5 | 6 | 7 | 11 | 14 | 15 |
| 1 | 2 | 5 | 6 | 7 | 8 | 11 | 12 | 15 |
| 1 | 2 | 3 | 5 | 6 | 7 | 9 | 11 | 15 |

| 1 | 2 | 5 | 9 | 10 | 11 | 13 | 14 | 15 |
| 1 | 2 | 3 | 5 | 9 | 10 | 11 | 13 | 15 |
| 1 | 2 | 3 | 5 | 7 | 10 | 11 | 14 | 15 |
| 1 | 2 | 3 | 5 | 7 | 9 | 10 | 11 | 15 |
| 1 | 3 | 4 | 5 | 6 | 11 | 12 | 14 | 15 |
| 1 | 3 | 5 | 6 | 7 | 11 | 13 | 14 | 15 |
| 1 | 3 | 4 | 5 | 7 | 8 | 11 | 12 | 15 |
| 1 | 3 | 4 | 5 | 7 | 11 | 12 | 13 | 15 |
| 1 | 3 | 4 | 5 | 6 | 10 | 11 | 12 | 15 |
| 1 | 3 | 5 | 7 | 10 | 11 | 13 | 14 | 15 |
| 1 | 5 | 7 | 9 | 10 | 11 | 13 | 14 | 15 |
| 1 | 3 | 5 | 7 | 8 | 9 | 11 | 12 | 15 |
| 1 | 5 | 6 | 7 | 8 | 10 | 11 | 12 | 15 |
| 1 | 2 | 3 | 4 | 6 | 9 | 11 | 12 | 13 |
| 1 | 2 | 3 | 6 | 8 | 9 | 11 | 12 | 13 |
| 1 | 3 | 6 | 8 | 9 | 11 | 12 | 13 | 14 |
| 1 | 2 | 3 | 6 | 7 | 8 | 9 | 11 | 12 |
| 1 | 2 | 3 | 6 | 7 | 9 | 10 | 11 | 15 |
| 1 | 2 | 3 | 4 | 6 | 7 | 9 | 11 | 12 |
| 1 | 2 | 3 | 4 | 7 | 8 | 9 | 11 | 14 |
| 1 | 2 | 3 | 7 | 9 | 10 | 11 | 14 | 15 |
| 1 | 2 | 3 | 4 | 7 | 9 | 11 | 12 | 14 |
| 1 | 2 | 3 | 5 | 7 | 9 | 11 | 14 | 15 |
| 1 | 2 | 3 | 5 | 9 | 10 | 11 | 13 | 14 |
| 1 | 2 | 3 | 4 | 9 | 11 | 12 | 13 | 14 |
| 1 | 2 | 3 | 4 | 8 | 9 | 11 | 13 | 14 |
| 1 | 2 | 3 | 5 | 9 | 11 | 13 | 14 | 15 |
| 1 | 3 | 4 | 6 | 7 | 8 | 9 | 11 | 14 |
| 1 | 3 | 4 | 5 | 7 | 8 | 9 | 11 | 15 |
| 1 | 3 | 4 | 9 | 10 | 11 | 12 | 14 | 15 |
| 1 | 3 | 7 | 9 | 10 | 11 | 13 | 14 | 15 |
| 1 | 3 | 4 | 6 | 8 | 9 | 11 | 13 | 14 |
| 1 | 3 | 4 | 5 | 8 | 9 | 11 | 13 | 15 |
| 1 | 3 | 4 | 5 | 9 | 10 | 11 | 12 | 14 |
| 1 | 3 | 5 | 7 | 9 | 10 | 11 | 13 | 14 |
| 1 | 3 | 6 | 7 | 9 | 10 | 11 | 13 | 15 |
| 1 | 3 | 4 | 6 | 9 | 10 | 11 | 12 | 15 |
| 1 | 3 | 6 | 7 | 8 | 9 | 11 | 12 | 14 |
| 1 | 2 | 4 | 6 | 7 | 9 | 11 | 12 | 13 |
| 1 | 4 | 6 | 7 | 9 | 11 | 12 | 13 | 14 |
| 1 | 2 | 4 | 6 | 7 | 8 | 9 | 11 | 12 |
| 1 | 2 | 3 | 4 | 6 | 8 | 9 | 11 | 12 |
| 1 | 2 | 3 | 4 | 7 | 8 | 9 | 12 | 14 |
| 1 | 2 | 3 | 4 | 8 | 9 | 11 | 12 | 14 |

| 1 | 2 | 4 | 5 | 9 | 10 | 12 | 13 | 14 |
|---|---|---|---|---|----|----|----|----|
| 1 | 2 | 3 | 4 | 7 | 9 | 12 | 13 | 14 |
| 1 | 3 | 4 | 6 | 7 | 8 | 9 | 12 | 14 |
| 1 | 3 | 4 | 5 | 7 | 9 | 12 | 13 | 15 |
| 1 | 3 | 4 | 6 | 7 | 9 | 12 | 13 | 14 |
| 1 | 3 | 4 | 5 | 9 | 10 | 12 | 14 | 15 |
| 1 | 3 | 4 | 5 | 6 | 9 | 12 | 14 | 15 |
| 1 | 3 | 4 | 5 | 8 | 9 | 12 | 13 | 15 |
| 1 | 4 | 5 | 8 | 9 | 10 | 12 | 13 | 14 |
| 1 | 4 | 6 | 7 | 8 | 9 | 11 | 12 | 14 |
| 1 | 6 | 7 | 8 | 9 | 11 | 12 | 13 | 14 |
| 1 | 2 | 6 | 7 | 8 | 9 | 11 | 12 | 14 |
| 1 | 2 | 6 | 7 | 9 | 10 | 11 | 14 | 15 |
| 1 | 2 | 4 | 6 | 7 | 9 | 11 | 12 | 14 |
| 1 | 2 | 3 | 6 | 8 | 9 | 11 | 12 | 14 |
| 1 | 2 | 3 | 6 | 7 | 9 | 10 | 14 | 15 |
| 1 | 2 | 3 | 4 | 6 | 7 | 8 | 9 | 14 |
| 1 | 2 | 3 | 4 | 6 | 7 | 9 | 12 | 14 |
| 1 | 2 | 3 | 5 | 6 | 7 | 9 | 10 | 14 |
| 1 | 2 | 3 | 4 | 6 | 8 | 9 | 11 | 14 |
| 1 | 2 | 4 | 5 | 6 | 8 | 9 | 10 | 14 |
| 1 | 3 | 5 | 6 | 7 | 9 | 13 | 14 | 15 |
| 1 | 3 | 4 | 6 | 7 | 8 | 9 | 13 | 14 |
| 1 | 3 | 5 | 6 | 7 | 9 | 10 | 13 | 14 |
| 1 | 2 | 3 | 5 | 6 | 9 | 13 | 14 | 15 |
| 1 | 2 | 3 | 6 | 9 | 10 | 13 | 14 | 15 |
| 1 | 3 | 4 | 6 | 9 | 10 | 12 | 14 | 15 |
| 1 | 4 | 5 | 6 | 8 | 9 | 10 | 13 | 14 |
| 1 | 2 | 6 | 9 | 10 | 11 | 13 | 14 | 15 |
| 1 | 4 | 6 | 9 | 10 | 11 | 12 | 14 | 15 |
| 1 | 2 | 3 | 6 | 7 | 10 | 13 | 14 | 15 |
| 1 | 2 | 6 | 7 | 8 | 10 | 12 | 13 | 15 |
| 1 | 2 | 6 | 7 | 9 | 10 | 13 | 14 | 15 |
| 1 | 2 | 3 | 4 | 6 | 7 | 8 | 13 | 14 |
| 1 | 2 | 3 | 6 | 7 | 8 | 12 | 13 | 14 |
| 1 | 2 | 3 | 5 | 6 | 7 | 10 | 13 | 14 |
| 1 | 2 | 4 | 6 | 7 | 8 | 11 | 13 | 14 |
| 1 | 2 | 4 | 5 | 6 | 8 | 10 | 13 | 14 |
| 1 | 2 | 3 | 5 | 6 | 7 | 9 | 13 | 15 |
| 1 | 2 | 3 | 6 | 7 | 8 | 9 | 12 | 13 |
| 1 | 2 | 3 | 4 | 6 | 7 | 8 | 9 | 13 |
| 1 | 2 | 3 | 5 | 6 | 7 | 9 | 10 | 13 |
| 1 | 2 | 3 | 6 | 10 | 11 | 13 | 14 | 15 |
| 1 | 2 | 3 | 4 | 6 | 11 | 12 | 13 | 14 |
| 1 | 2 | 3 | 6 | 8 | 11 | 12 | 13 | 14 |
| 1 | 2 | 4 | 6 | 7 | 8 | 9 | 11 | 13 |
| 1 | 2 | 4 | 5 | 6 | 8 | 9 | 10 | 13 |
| 1 | 2 | 4 | 6 | 7 | 11 | 12 | 13 | 14 |
| 1 | 2 | 5 | 6 | 7 | 8 | 12 | 13 | 15 |
| 1 | 2 | 5 | 6 | 7 | 9 | 13 | 14 | 15 |
| 1 | 2 | 6 | 8 | 10 | 11 | 12 | 13 | 15 |
| 1 | 2 | 5 | 7 | 8 | 10 | 11 | 12 | 13 |
| 1 | 2 | 5 | 7 | 9 | 10 | 11 | 13 | 14 |
| 1 | 2 | 3 | 5 | 7 | 9 | 10 | 11 | 13 |
| 1 | 2 | 3 | 7 | 10 | 11 | 13 | 14 | 15 |
| 1 | 2 | 3 | 7 | 9 | 10 | 11 | 13 | 15 |
| 1 | 2 | 7 | 8 | 10 | 11 | 12 | 13 | 15 |
| 1 | 2 | 3 | 4 | 7 | 8 | 11 | 13 | 14 |
| 1 | 2 | 3 | 7 | 8 | 11 | 12 | 13 | 14 |
| 1 | 2 | 3 | 5 | 7 | 11 | 13 | 14 | 15 |
| 1 | 2 | 4 | 7 | 8 | 9 | 11 | 13 | 14 |
| 1 | 2 | 7 | 8 | 9 | 11 | 12 | 13 | 14 |
| 1 | 2 | 5 | 7 | 9 | 11 | 13 | 14 | 15 |
| 1 | 2 | 3 | 4 | 6 | 7 | 8 | 11 | 13 |
| 1 | 3 | 4 | 5 | 7 | 8 | 11 | 13 | 15 |
| 1 | 3 | 4 | 6 | 7 | 11 | 12 | 13 | 14 |
| 1 | 3 | 6 | 7 | 10 | 11 | 13 | 14 | 15 |
| 1 | 3 | 6 | 7 | 8 | 11 | 12 | 13 | 14 |
| 1 | 2 | 3 | 4 | 6 | 7 | 11 | 12 | 13 |
| 1 | 4 | 5 | 7 | 8 | 9 | 11 | 13 | 15 |
| 1 | 6 | 7 | 8 | 10 | 11 | 12 | 13 | 15 |
| 1 | 2 | 5 | 7 | 8 | 10 | 12 | 13 | 15 |
| 1 | 2 | 5 | 8 | 9 | 10 | 12 | 13 | 14 |
| 1 | 2 | 4 | 5 | 8 | 9 | 10 | 12 | 13 |
| 1 | 2 | 3 | 4 | 7 | 8 | 12 | 13 | 14 |
| 1 | 2 | 4 | 7 | 8 | 11 | 12 | 13 | 14 |
| 1 | 2 | 3 | 7 | 8 | 9 | 12 | 13 | 14 |
| 1 | 2 | 3 | 4 | 6 | 7 | 8 | 12 | 13 |
| 1 | 3 | 5 | 7 | 8 | 9 | 12 | 13 | 15 |
| 1 | 3 | 4 | 6 | 8 | 11 | 12 | 13 | 14 |
| 1 | 2 | 3 | 4 | 6 | 8 | 11 | 12 | 13 |
| 1 | 4 | 6 | 7 | 8 | 11 | 12 | 13 | 14 |
| 1 | 4 | 5 | 7 | 8 | 9 | 12 | 13 | 15 |
| 2 | 3 | 5 | 6 | 7 | 10 | 11 | 13 | 14 |
| 2 | 5 | 6 | 7 | 8 | 10 | 11 | 12 | 13 |
| 2 | 5 | 6 | 7 | 9 | 10 | 11 | 13 | 14 |
| 2 | 4 | 6 | 7 | 8 | 10 | 11 | 14 | 15 |
| 2 | 4 | 6 | 7 | 8 | 9 | 11 | 13 | 14 |
| 2 | 6 | 7 | 8 | 9 | 11 | 12 | 13 | 14 |
| 2 | 6 | 7 | 9 | 10 | 11 | 13 | 14 | 15 |
| 2 | 3 | 4 | 6 | 7 | 8 | 9 | 11 | 13 |
| 2 | 3 | 4 | 6 | 7 | 8 | 10 | 11 | 15 |
| 2 | 3 | 5 | 6 | 7 | 9 | 11 | 13 | 15 |
| 2 | 3 | 6 | 7 | 9 | 10 | 11 | 13 | 15 |
| 2 | 3 | 6 | 7 | 8 | 9 | 11 | 12 | 13 |
| 2 | 3 | 4 | 6 | 7 | 10 | 11 | 12 | 15 |
| 2 | 4 | 6 | 7 | 10 | 11 | 12 | 14 | 15 |
| 2 | 3 | 5 | 6 | 7 | 11 | 13 | 14 | 15 |
| 2 | 5 | 6 | 7 | 8 | 11 | 12 | 13 | 15 |
| 2 | 4 | 5 | 6 | 8 | 10 | 12 | 13 | 14 |
| 2 | 5 | 6 | 7 | 8 | 10 | 12 | 13 | 15 |
| 2 | 5 | 6 | 8 | 9 | 10 | 12 | 13 | 14 |
| 2 | 3 | 6 | 7 | 8 | 10 | 12 | 14 | 15 |
| 2 | 4 | 6 | 7 | 8 | 10 | 12 | 14 | 15 |
| 2 | 3 | 6 | 7 | 8 | 9 | 12 | 13 | 14 |
| 2 | 3 | 4 | 6 | 7 | 8 | 9 | 12 | 13 |
| 2 | 3 | 6 | 8 | 10 | 11 | 12 | 14 | 15 |
| 2 | 4 | 6 | 7 | 8 | 9 | 11 | 12 | 13 |
| 2 | 4 | 6 | 8 | 10 | 11 | 12 | 14 | 15 |
| 2 | 3 | 4 | 5 | 7 | 8 | 10 | 13 | 14 |
| 2 | 4 | 5 | 7 | 8 | 10 | 13 | 14 | 15 |
| 2 | 3 | 4 | 5 | 7 | 8 | 9 | 10 | 13 |
| 2 | 3 | 5 | 6 | 10 | 11 | 13 | 14 | 15 |
| 2 | 4 | 5 | 7 | 8 | 9 | 10 | 13 | 15 |
| 2 | 5 | 6 | 9 | 10 | 11 | 13 | 14 | 15 |
| 2 | 4 | 7 | 8 | 9 | 10 | 13 | 14 | 15 |
| 2 | 3 | 4 | 7 | 8 | 9 | 10 | 13 | 15 |
| 2 | 3 | 4 | 7 | 8 | 10 | 11 | 12 | 15 |
| 2 | 3 | 4 | 5 | 7 | 8 | 10 | 13 | 15 |
| 2 | 3 | 5 | 6 | 7 | 9 | 10 | 11 | 15 |
| 2 | 5 | 6 | 7 | 9 | 10 | 11 | 14 | 15 |
| 2 | 3 | 7 | 8 | 10 | 11 | 12 | 14 | 15 |
| 2 | 5 | 7 | 8 | 10 | 11 | 12 | 13 | 15 |
| 2 | 3 | 4 | 7 | 9 | 11 | 12 | 13 | 14 |
| 2 | 3 | 4 | 7 | 10 | 11 | 12 | 14 | 15 |
| 2 | 3 | 4 | 5 | 7 | 8 | 9 | 14 | 15 |
| 2 | 3 | 5 | 6 | 7 | 9 | 11 | 14 | 15 |
| 2 | 3 | 4 | 6 | 7 | 9 | 12 | 13 | 14 |
| 2 | 3 | 4 | 6 | 7 | 10 | 12 | 14 | 15 |
| 2 | 3 | 4 | 5 | 7 | 8 | 9 | 10 | 14 |
| 2 | 3 | 5 | 6 | 7 | 9 | 10 | 11 | 14 |
| 2 | 3 | 4 | 5 | 7 | 8 | 13 | 14 | 15 |
| 2 | 3 | 7 | 8 | 9 | 11 | 12 | 13 | 14 |
| 2 | 3 | 4 | 6 | 8 | 9 | 11 | 13 | 14 |
| 2 | 3 | 4 | 6 | 8 | 10 | 11 | 14 | 15 |
| 2 | 3 | 4 | 8 | 9 | 11 | 12 | 13 | 14 |
| 2 | 3 | 4 | 5 | 8 | 9 | 10 | 14 | 15 |
| 2 | 3 | 4 | 8 | 9 | 10 | 13 | 14 | 15 |
| 2 | 3 | 4 | 6 | 8 | 10 | 12 | 14 | 15 |
| 2 | 4 | 5 | 6 | 8 | 9 | 10 | 12 | 14 |
| 2 | 4 | 7 | 8 | 9 | 11 | 12 | 13 | 14 |
| 2 | 3 | 4 | 6 | 9 | 11 | 12 | 13 | 14 |
| 2 | 3 | 6 | 9 | 10 | 11 | 13 | 14 | 15 |
| 2 | 3 | 5 | 6 | 9 | 10 | 11 | 13 | 14 |
| 2 | 3 | 4 | 5 | 8 | 9 | 13 | 14 | 15 |
| 2 | 4 | 5 | 6 | 9 | 10 | 12 | 13 | 14 |
| 2 | 4 | 5 | 7 | 8 | 9 | 13 | 14 | 15 |
| 3 | 4 | 6 | 7 | 8 | 10 | 11 | 14 | 15 |
| 3 | 4 | 6 | 7 | 8 | 9 | 11 | 13 | 14 |
| 3 | 4 | 5 | 7 | 8 | 9 | 10 | 14 | 15 |
| 3 | 4 | 5 | 7 | 8 | 9 | 11 | 12 | 15 |
| 3 | 4 | 7 | 8 | 10 | 11 | 12 | 14 | 15 |
| 3 | 4 | 7 | 8 | 9 | 10 | 13 | 14 | 15 |
| 3 | 4 | 6 | 7 | 8 | 9 | 12 | 13 | 14 |
| 3 | 4 | 5 | 7 | 8 | 11 | 12 | 13 | 15 |
| 3 | 4 | 6 | 7 | 9 | 11 | 12 | 13 | 14 |
| 3 | 5 | 6 | 7 | 9 | 11 | 13 | 14 | 15 |
| 3 | 6 | 7 | 9 | 10 | 11 | 13 | 14 | 15 |
| 3 | 4 | 5 | 7 | 9 | 11 | 12 | 13 | 15 |
| 3 | 4 | 5 | 7 | 8 | 9 | 10 | 13 | 14 |
| 3 | 5 | 7 | 8 | 9 | 11 | 12 | 13 | 15 |
| 3 | 4 | 5 | 6 | 9 | 11 | 12 | 14 | 15 |
| 3 | 4 | 5 | 6 | 9 | 10 | 11 | 12 | 14 |
| 3 | 6 | 7 | 8 | 10 | 11 | 12 | 14 | 15 |
| 3 | 4 | 5 | 9 | 10 | 11 | 12 | 14 | 15 |
| 3 | 4 | 6 | 8 | 9 | 11 | 12 | 13 | 14 |
| 3 | 4 | 6 | 7 | 8 | 10 | 11 | 12 | 15 |
| 3 | 4 | 5 | 8 | 9 | 10 | 13 | 14 | 15 |
| 3 | 5 | 6 | 7 | 9 | 10 | 11 | 13 | 15 |
| 3 | 4 | 5 | 6 | 9 | 10 | 12 | 14 | 15 |
| 3 | 5 | 6 | 7 | 10 | 11 | 13 | 14 | 15 |

The column indices contained in 16 G-(1,1,1,1,1,1,1,1,1) inter-connected cycles:

| | | | | | | | | |
|---|---|---|---|---|---|---|---|---|
| 0 | 3 | 4 | 5 | 7 | 8 | 9 | 13 | 15 |
| 0 | 1 | 3 | 4 | 5 | 9 | 11 | 12 | 15 |
| 0 | 2 | 3 | 4 | 7 | 8 | 10 | 14 | 15 |

| 0 | 3 | 4 | 6  | 10 | 11 | 12 | 14 | 15 |
| 0 | 2 | 4 | 5  | 8  | 9  | 10 | 13 | 14 |
| 0 | 1 | 4 | 5  | 6  | 9  | 10 | 12 | 14 |
| 0 | 2 | 6 | 7  | 8  | 10 | 11 | 12 | 15 |
| 0 | 1 | 2 | 5  | 6  | 8  | 10 | 12 | 13 |
| 0 | 1 | 5 | 7  | 8  | 11 | 12 | 13 | 15 |
| 1 | 3 | 5 | 6  | 9  | 10 | 11 | 14 | 15 |
| 1 | 2 | 5 | 6  | 7  | 10 | 11 | 13 | 15 |
| 1 | 3 | 4 | 7  | 8  | 9  | 11 | 12 | 13 |
| 1 | 2 | 3 | 6  | 7  | 9  | 11 | 13 | 14 |
| 1 | 2 | 4 | 6  | 8  | 9  | 12 | 13 | 14 |
| 2 | 3 | 4 | 6  | 7  | 8  | 11 | 12 | 14 |
| 2 | 3 | 5 | 7  | 9  | 10 | 13 | 14 | 15 |